\documentclass[preprint,review,12pt]{elsarticle}

\usepackage[numbers]{natbib}

\usepackage{amssymb}
\usepackage{amsmath}

\usepackage{bm}
\usepackage{url}
\usepackage{siunitx}
\usepackage{multirow}
\usepackage{graphicx}
\usepackage{subcaption}
\usepackage{lineno}
\usepackage{xspace}
\usepackage{duckuments}
\usepackage{tikz}
\usetikzlibrary{fit, positioning}
\usepackage{soul}
\usepackage{derivative}
\usepackage{amsmath}
\usepackage{calc}

\usepackage{lscape}
\usepackage{array}
\usepackage{ragged2e}

\usepackage{footnote}
\makesavenoteenv{tabular}
\makesavenoteenv{table}

\usepackage[export]{adjustbox}
\usepackage{threeparttable}
\usepackage{hyperref}

\newcommand{\bra}[1]{\left[#1\right]}
\newcommand{\pp}[1]{\left(#1\right)}
\newcommand{\norm}[1]{\left\|#1\right\|}

\newcommand{\vm}[1]{\ensuremath{\bm{#1}}}
\newcommand{\tm}[1]{\bm{#1}}

\newcommand{\ltwo}[1]{\ensuremath{\mathcal{L}^2(#1)}}

\newcommand{\commentblue}[1]{{\color{black} #1}}
\newcommand{\commentred}[1]{{\color{black} #1}}

\newcommand{\len}[0]{{\mathbb{L}}}
\newcommand{\m}[0]{{\mathbb{M}}}
\newcommand{\ti}[0]{{\mathbb{T}}}

\newcommand{\transpose}{^\mathsf{T}}

\newcommand{\lethe}[0]{\mbox{\texttt{Lethe}}\xspace}

\newcommand*\justified{%
  \fontdimen2\font=0.4em
  \fontdimen3\font=0.2em
  \fontdimen4\font=0.1em
  \fontdimen7\font=0.1em
  \hyphenchar\font=`\-
}

\renewcommand{\texttt}[1]{%
  \begingroup
  \ttfamily
  \begingroup\lccode`~=`/\lowercase{\endgroup\def~}{/\discretionary{}{}{}}%
  \begingroup\lccode`~=`[\lowercase{\endgroup\def~}{[\discretionary{}{}{}}%
  \begingroup\lccode`~=`.\lowercase{\endgroup\def~}{.\discretionary{}{}{}}%
  \catcode`/=\active\catcode`[=\active\catcode`.=\active
  \justified\scantokens{#1\noexpand}%
  \endgroup
}

\newcolumntype{P}[1]{>{\RaggedRight\hspace{0pt}}p{#1}}
\newcolumntype{M}[1]{>{\RaggedRight\hspace{0pt}}m{#1}}
\newcolumntype{R}[1]{>{\RaggedLeft\hspace{0pt}}p{#1}}
\newcolumntype{C}[1]{>{\centering\arraybackslash}p{#1}}

\journal{CMAME}

\begin{document}

\begin{frontmatter}



\title{} 


\title{A Geometric Reinitialization for Conservative Level-Set Methods in the Context of Capillary Flows}


\author[a]{Hélène Papillon-Laroche}
\author[a]{Amishga Alphonius}
\author[b]{Magdalena Schreter-Fleischhacker}
\author[c]{Jean-Philippe Harvey}
\author[a]{Bruno Blais} 

\affiliation[a]{organization={CHAOS Laboratory},
            addressline={2500 Chemin de Polytechnique},
            city={Montréal},
            postcode={H3T 1J4},
            state={Québec},
            country={Canada}}

\affiliation[b]{organization={Professorship of Simulation for Additive Manufacturing, Technical University of Munich},
            addressline={Freisinger Landstraße 52},
            city={ Garching},
            postcode={85748},
            country={Germany}}
\affiliation[c]{organization={Centre de recherche en calcul thermochimique (CRCT), Polytechnique Montréal},
            addressline={2500 Chemin de Polytechnique},
            city={Montréal},
            postcode={H3T 1J4},
            state={Québec},
            country={Canada}}           
\begin{abstract}
Simulations of immiscible two-phase flows involving surface tension (ST) require a robust high-fidelity framework. State-of-the-art multi-phase models, such as the Conservative Level-Set (CLS) approach, use Eulerian representations of the fluids and their interface, and require reinitialization methods to ensure volume conservation and accurate ST force modeling. \commentblue{This work proposes a new geometric reinitialization strategy for CLS formulations as an alternative to the typical PDE-based reinitialization methods. Building on the geometric reinitialization developed in context of level-set frameworks, the proposed method is an extension to the CLS approach within an open-source finite element framework that allows for fully distributed and adaptively refined, two- and three-dimensional meshes. This extension enables application of the geometric reinitialization method to large-scale three-dimensional problems with strongly deforming interfaces, including breakup.} The method is quantitatively compared to two established reinitialization approaches: the PDE-based reinitialization proposed in the original CLS method and a simple projection-based approach. The assessment comprises three 3D application cases of increasing complexity:  the capillary migration of a droplet, the rise of a bubble, and the Rayleigh-Plateau instability development in a capillary jet. The geometric method leads to high-quality, spatially-converged results in good agreement with benchmark and reference solutions. \commentblue{It robustly and accurately recovers expected interface topology-based metrics (e.g., the volume, surface area and sphericity), and relevant flow quantities. In addition, the proposed method has only two parameters and is robust across parameter choices and cases.}
\end{abstract}

\begin{graphicalabstract}
\begin{figure}
    \centering
        \makebox[\textwidth][c]{
        \includegraphics[width=1.3\linewidth]{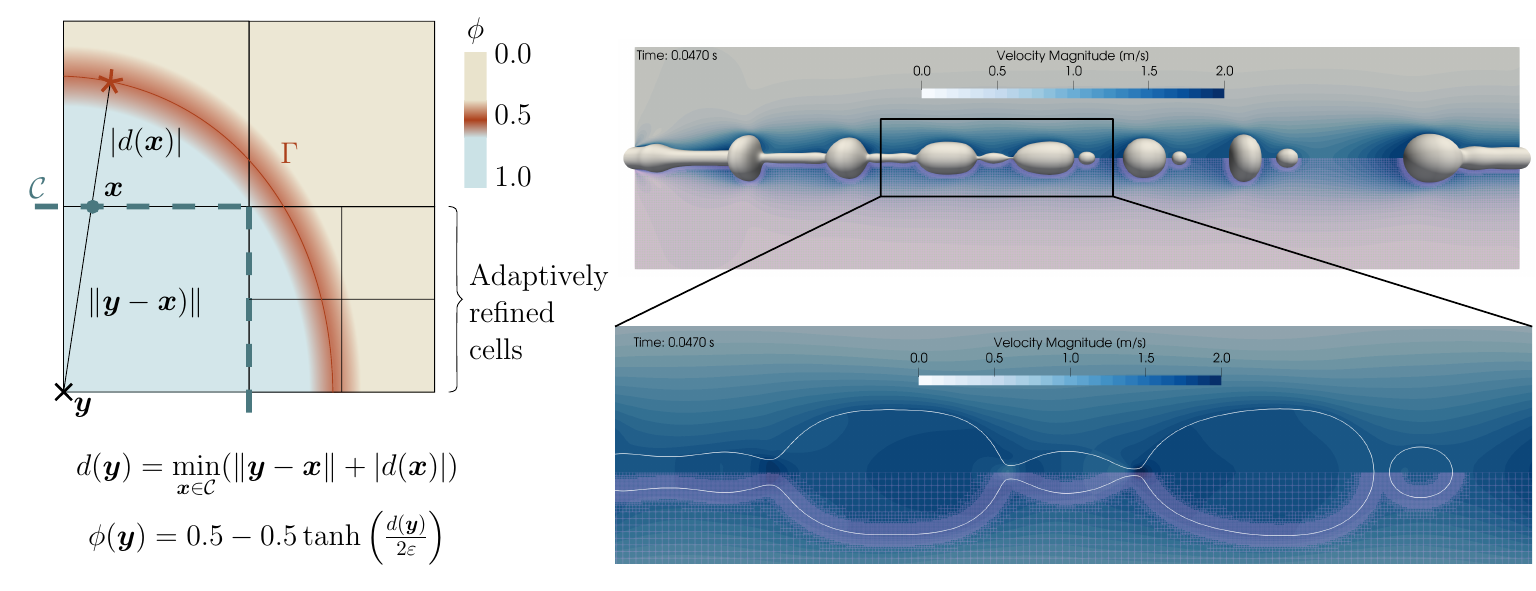}}
\end{figure}

\end{graphicalabstract}

\begin{highlights}
\item \textcolor{black}{Proposition of a geometric reinitialization for Conservative Level-Set frameworks}
\item \textcolor{black}{Extension to quad-based, adaptively refined, parallel-distributed, 2D and 3D meshes}
\item \textcolor{black}{Two-parameters framework, robust across parameters choice and cases}
\item \textcolor{black}{High-quality results on practical capillary flows, including breakup}
\end{highlights}


\end{frontmatter}



\section{Introduction} \label{sec:intro}
At length scales approaching and below the capillary length, surface tension (ST) drives immiscible flows \cite{deGennes2005}. This commonly occurs in many multiphase applications including droplets and bubbles \cite{Popinet2018} or metal additive manufacturing processes~\cite{Cook2020}. The current work focuses on incompressible interfacial flows.

\subsection{Modeling of Incompressible Interfacial Flows}

The incompressible Navier-Stokes (NS) equations govern the dynamics of each fluid. The ST force results in a jump in the stress tensor at the interface between the two fluids. The magnitude of this jump depends on the interfacial properties of the material (i.e., the ST coefficient and its derivatives with respect to temperature and/or composition) as well as the curvature of the interface.

In most capillary-driven flows, the interface moves and deforms, making a boundary (interface) conforming approach challenging, in particular due to re-meshing requirements when the interface presents large deformations, breaks-up or coalesces \cite{Popinet2018}. Considering this, one-fluid approaches are usually preferred to two-fluid models. 

One-fluid models treat the jump boundary condition as a singular force on the right-hand side of the NS momentum equation \cite{Tryggvason2011}. However, the position and geometric properties of the moving interface are unknowns. Hence, simulations require an additional interface solver coupled to the NS solver. Current available methods range from Lagrangian to Eulerian descriptions of the interface~\cite{Popinet2018}. 

The Eulerian interface representation uses a phase indicator $\phi$ to capture the transport of the interface with the flow. The definition of the phase indicator varies among the scientific literature. The Level-Set (LS) method employs a signed-distance function with the interface corresponding to $\phi=0$ \cite{Tryggvason2011}. The Volume-of-Fluid (VOF) method uses the phase fraction, typically ranging from $0$ to $1$. It changes from one extremum to the other at the interface, represented by the iso-contour $\phi=0.5$. The Conservative Level-Set (CLS) approach \cite{Olsson2007} considers a similar description as the VOF, however, the phase indicator represents a pseudo-signed-distance instead of a fluid fraction. 

The advection of the phase indicator with the velocity field models the transport of the interface \cite{Mirjalili2017,Garcia-Villalba2025}. Coupling with the NS momentum equation happens through the ST force and the change in physical properties (density and viscosity) at the interface \cite{Mirjalili2017}.

On a discrete level, simple conservative advection schemes using the phase indicator description of the VOF or CLS methods conserve the global volume defined as:
\begin{equation*}
    V_\mathrm{global} = \int_\Omega \phi \mathrm{d}\Omega
\end{equation*}
over the domain $\Omega$. However, when considering the volume of the domain enclosed by the $\phi=0.5$ iso-contour, denoted $\Omega_1$,
\begin{equation*}
    V = \int_{\Omega_1} 1\mathrm{d}\Omega
\end{equation*}
non-conservative behavior can be observed \cite{Olsson2007}. This leads to artificial volume transfer between the two phases \cite{Aliabadi2000}. \commentblue{Indeed, the space and time discretizations of typical numerical schemes as well as stabilization methods (or limiters) add artificial diffusion in the problem. Without any self-sharpening term or specific advection schemes, it results in volume changes and anisotropic smearing of the interface as the simulation time progresses \cite{Hirt1981,Tryggvason2011}. In the case of the LS method, the phase indicator losses its distance property after advection \cite{Shakoor2025}.}

\commentblue{The loss of a well-defined phase indicator is problematic for the computation of accurate geometric properties of the interface such as the normal and curvature.} 
Therefore, Eulerian frameworks need mechanisms to preserve the interface and the enclosed volume. Available strategies include: 
\begin{itemize}
    \item \textbf{Improved discretizations.} Simulations of single-phase flows can benefit from high-order methods~\cite{lethe2025}. However, in the case of fluid--fluid flows with ST, the discontinuities at the interface reduce the convergence rate of the solution~\cite{Tryggvason2011, Fuster2025, Ausas2010}. A smaller cell size increases the accuracy of the interface position and helps reduce numerical diffusion~\cite{Olsson2007}, but uniform refinement has a significant computational cost. Dynamic mesh adaptation offers a suitable alternative \cite{Mirjalili2017} which increases the cell density in high error areas, while applying coarsening to the cells in low error regions. This process requires adequate error estimation to identify where the interface should be refined.
    
    \item \textbf{Specialized advection schemes.} Conservation of the phase fluxes across cell boundaries reduces excessive diffusion in the original VOF method by \citet{Hirt1981} and later developments. \commentblue{The resulting advection schemes have a built-in volume-conservation property, which is the main advantage compared to LS and CLS approaches. However, they rely on an algebraic representation or geometric reconstruction of the interface \cite{Mirjalili2017, Tryggvason2011}, which become challenging and computationally expensive when the problem at hand requires more accuracy}. Additionally, standard schemes found in the Continuous Galerkin (CG) Finite Element Method (FEM) are unsuitable for such a specialized framework~\cite{Olsson2007}.
    
    \item \textbf{Reinitialization methods.} For LS and CLS methods, there is not built-in volume conservation in the advection scheme. Most authors turn to reinitialization procedures to preserve the interface quality and volume conservation while keeping a simple advection scheme \cite{Olsson2007, Aliabadi2000, Ausas2011, Henri2022}. In a LS framework, they aim at regaining a signed distance function either by a direct geometric computation of the latter \cite{Ausas2011,Henri2022} or by the resolution of a PDE enforcing $\|\nabla\phi\|=1$~\cite{Tryggvason2011}. \commentblue{The literature review of \citet{Shakoor2025} offers a great overview of the current state-of-the-art methods for LS reinitialization.} For the CLS method, the original works of \citet{Olsson2005} and \citet{Olsson2007} uses a PDE-based reinitialization which balances compressive and diffusive terms in the normal direction of the interface. \commentblue{More recent CLS reinitialization methods, e.g. \citet{McCaslin2014}, \citet{Chiodi2017}, and \citet{Janodet2022}), are reformulations and/or extensions of the PDE-based reinitialization of \citet{Olsson2007}}.
\end{itemize}

In the context of a FEM framework, dynamic mesh adaptation and reinitialization procedures are suitable methods to mitigate the excessive interface diffusion and non-conservative behavior. \commentblue{This work focuses on the reinitialization method in the context of CLS frameworks.}

\subsection{Reinitialization in CLS Methods} \label{sec:reinitialization}
\commentblue{For the reinitialization method in CLS frameworks, the original works of \citet{Olsson2005} and \citet{Olsson2007} propose a phase indicator that approximates smoothly the Heaviside function stepping at the interface $\Gamma$: $\phi(\vm{x}) \approx H_\Gamma (\vm{x})$. In \cite{Olsson2007}, they use the definition:} 
\begin{equation}
    \phi(\vm{x}) = 0.5-0.5\tanh\pp{\frac{d(\vm{x})}{2\varepsilon}}\label{eq:heaviside_phi}
\end{equation}
\commentblue{where $d(\vm{x})$ is the signed distance from the interface and $\varepsilon$ is a measure of the thickness of the transition region between the two phases. The authors approximate the volume $V$ enclosed by interface using the integral of the phase indicator over the domain, i.e.,
\begin{equation}
    V = \int_{\Omega_1} 1d\Omega  \approx V_\text{global} = \int_\Omega \phi(\vm{x})d\Omega \label{eq:heaviside_approx}
\end{equation}
stating that in the limit of $\varepsilon \rightarrow 0$, in which case $\phi(\vm{x})=H_\Gamma(\vm{x})$, the approximation \eqref{eq:heaviside_approx} is exact. As a result, a method that conserves $V_\text{global}$ should minimize the error on $V$ \cite{Olsson2005,Janodet2022}; the remaining error on $V$ comes from the approximation \eqref{eq:heaviside_approx} when $\varepsilon$ does not tend to 0.

Hence, to reduce this error on $V$, \citet{Olsson2005} propose to limit $\varepsilon$ to the smallest resolvable value, i.e. $\varepsilon \approx \Delta x/2$, where $\Delta x$ is the mesh size. To achieve this throughout the simulation, the authors mitigate the smearing of the transition region with the reinitialization of $\phi$ applied at each time step. It takes the form of the following PDE, which balances at steady-state a compressive and diffusive term in the direction normal to the interface:}
\begin{equation*}
    \underbrace{\frac{\partial \phi}{\partial \tau} }_\text{transient} + \underbrace{\nabla \cdot \left[ \phi (1-\phi ) \vm{n}\right]}_\text{compression} - \underbrace{\varepsilon \nabla \cdot \left[ (\nabla \phi  \cdot \vm{n}) \vm{n} \right]}_\text{diffusion} = 0
\end{equation*}
\commentblue{where $\tau$ is an {artificial} time and $\vm{n}$ is the normal vector to the interface. This results in the displacement of the interface in its normal direction to preserve both $V_\text{global}$ \cite{Chiodi2017} and the thickness of the interface profile \cite{Olsson2005, Desjardins2008}.} 

\commentblue{With the approximation \eqref{eq:heaviside_approx}, \citet{Olsson2005} and \citet{Olsson2007} put a strong restriction on the size of $\varepsilon$. This leads to a conflicting choice of the value of $\varepsilon$. On the one hand, the phase indicator has to be sufficiently smooth for the numerical solution of the advection equation, and the computation of the normal vector $\vm{n}$ and curvature $\kappa$, defined respectively by:}
\begin{align}
    \vm{n} &= \frac{\nabla \phi}{\|\nabla \phi\|} \label{eq::n_intro}\\
    \kappa &= -\nabla\cdot\vm{n} \label{eq::kappa_intro}
\end{align}
\commentblue{This smoothness is naturally given by choosing a large value for $\varepsilon$. On the other hand, choosing it small to minimize the volume conservation error leads to spurious oscillations of the normal, resulting in a balance of the compressive and diffusive terms in the wrong direction. Over time, these errors accumulate and lead to spurious deformation of the interface \cite{Desjardins2008}.} 

\commentblue{\citet{Desjardins2008} propose the Accurate CLS (ACLS) which addresses the spurious oscillations of the normal by decoupling of the computation of $\vm{n}$ from $\phi$. Instead of using Equation \eqref{eq::n_intro} to obtain $\vm{n}$, they introduce the computation of the signed distance $d(\vm{x})$ to the interface using a Fast Marching Method (FMM) and define $\vm{n} = \frac{\nabla d}{\|\nabla d\|}$. Then, they solve the same reinitialization PDE. Later, \citet{McCaslin2014} and \citet{Chiodi2017} propose reformulations of the PDE reinitialization in the ACLS to reduce the accumulation of error in static regions of the interface, which is relevant for many applications such as metal additive manufacturing processes, where e.g., implicitly represented powder particles in the far-field of the laser are static. \citet{Janodet2022} use the approach of \citet{Chiodi2017} and, to reduce the computational time, they compute the distance only in a narrow band around the interface, a technique already proposed in LS frameworks \citep{Chopp2001,Shakoor2025}. They also adapt the ACLS to unstructured, adaptively refined, tetrahedral grids.}

\commentblue{Despite the improvements of the reinitialization procedure introduced by the ACLS of \citet{Desjardins2008} and its extensions \cite{Chiodi2017, McCaslin2014}, the premise stays the same: to conserve the volume approximation described in Equation \eqref{eq:heaviside_approx}, $\varepsilon$ should tend to 0. The resulting sharpness of the transition makes the numerical advection and the computation of the interfacial quantities $\vm{n}$ and $\kappa$ challenging, complexifies the solvers and adds computational cost. Additionally, this premise comes at the cost of artificial displacement of the interface, even for the most recent formulation of the ACLS (e.g. \cite{Chiodi2017}).}

\subsection{Contribution of the Present Work}
\commentblue{This work proposes a new geometric reinitialization method for CLS frameworks that uses the same hyperbolic tangent definition of the phase indicator as in Equation \eqref{eq:heaviside_phi}. This definition of $\phi$ does not alter the position of the interface defined by $d(\vm{x})=0$, hence on a continuous level, it conserves the volume $V$. Yet, the new method does do not assume that $\phi$ should approximate $H_\Gamma(\vm{x})$, making the approximation $V \approx V_\mathrm{global}$ of \citet{Olsson2005}, and extensions, unnecessary.  

This lifts the strong restriction on $\varepsilon$, easing the numerical advection and computations of the normal and curvature. It increases the regularity of the solution, which is important for spatial convergence behavior in FEM. 
However, the numerical advection still brings artificial diffusion that needs mitigation to ensure the accuracy of the interface quantities and conservation of the volume. We propose to reinitialize the phase indicator with Equation \eqref{eq:heaviside_phi} by recomputing the signed distance $d(\vm{x})$ from a reconstruction of the interface, borrowing from geometric reinitialization methods developed in the context of LS frameworks \cite{Shakoor2025, Ausas2011}.

Hence, this work presents the new geometric reinitialization method for CLS frameworks, along with the description of a simple and robust 2D and 3D, two-phase flow solver in an open-source FEM framework that is capable of dynamic mesh adaptation. 

Building on the geometric reinitialization of \citet{Ausas2011} developed in context of level-set frameworks, the current work extends this approach to:
\begin{itemize}
    \item CLS frameworks;
    \item quadrilateral and hexahedral, adaptively refined, two- and three-dimensional meshes;
    \item fully parallel-distributed simulations.
\end{itemize}
The resulting two-parameter reinitialization method is robust across parameter choices and cases. Hence, it handles large-scale, three-dimensional, practical problems with strongly deforming interfaces, including breakup, which were outside the scope of the original work of \citet{Ausas2011}. This work provides also a quantitative comparison of the proposed reinitialization to the PDE-based method of \citet{Olsson2007} and a simple projection-based approach proposed by \citet{Aliabadi2000} to assess the suitability and robustness of the new approach. }

The paper takes the following structure: first, Section \ref{sec:problem-formulation} presents the problem formulation, along with the treatment of interfacial physical property changes, the ST force model, and the description of the general solver. The presentation of the geometric reinitialization method follows in Section \ref{sec:geometric_reinit}. Then, Section \ref{sec:mitigation} introduces briefly the reference methods used to compare the proposed approach. Finally, Section \ref{sec:results} presents the results and quantitative comparisons of the reinitialization methods for three 3D application cases: the capillary migration of a droplet \cite{Buscaglia2011}, the 3D rising bubble benchmark \cite{Hysing2009,Turek2019}, and the Rayleigh-Plateau instability \cite{Denner2022}. 
\section{Problem formulation} \label{sec:problem-formulation}
The problem formulation described in this work considers a one-fluid framework with an Eulerian interface representation. The domain of interest is $\Omega = \Omega_0 \cup \Omega_1$, where each subscript $i \in \{0,1\}$ indicates one fluid. The fluid-fluid interface is denoted by $\Gamma$ and the domain boundary by $\partial\Omega$. We consider the time $t\in[0,t_\text{end}]$, where $t_\text{end}$ is the end time. 

For simplicity, the following finite element formulations use the inner product notations in the weak forms~\cite{lethe2025}:
\begin{align*}
    \pp{a,b}_\Omega &= \int_\Omega a\cdot b \odif{\Omega}\\
    \pp{\nabla \vm{a}, \nabla \vm{b}}_{\Omega} &= \int_\Omega \nabla \vm{a}: \nabla \vm{b} \ \mathrm{d}\Omega \\
    \pp{a,b}_{\Omega_k} &= \sum_k\int_{\Omega_k} a\cdot b \odif{\Omega_k}
\end{align*}
where $\sum_k\int_{\Omega_k}\cdot\odif{\Omega_k}$ represents the sum of the integral over all $k$ elements ($\Omega_k$) of the mesh.

\subsection{Navier-Stokes Equations}
The \emph{strong formulation} of the incompressible NS equations considered in this work is:
\begin{subequations}
\begin{align}
    \nabla \cdot \vm{u} &= 0 &\quad \forall \vm{x}\in\Omega\times(0,t_\text{end}]\\
    \rho\pp{\pdv{\vm{u}}{t} + (\vm{u} \cdot \nabla)\vm{u}} &= \nabla \cdot \tm{\sigma} + \vm{F_\Gamma} + \rho\vm{g} & 
\quad \forall \vm{x}\in\Omega\times(0,t_\text{end}] \label{eq:ns_momentum_strong}
\end{align}
\end{subequations}
where $\vm{u}$ is the velocity vector, $\rho$ is the density and $\tm{\sigma} = -p\tm{I} + \mu\pp{\nabla\vm{u} + \pp{\nabla\vm{u}}\transpose}$ is the stress tensor, with the pressure $p$ and the dynamic viscosity $\mu$. The only forces considered are the ST force at the interface denoted $\vm{F_\Gamma}$ and the body force due the gravitational acceleration $\vm{g}$. Section \ref{sec:sft} describes the formulation of the ST force. Without loss of generality, the imposed boundary conditions on $\partial\Omega$ are no-slip conditions and the initial condition considers the fluids at rest.

The \emph{weak form} for $p$ and $\vm{u}$ is obtained using the corresponding test functions $q$ and $\vm{v}$. The formulation includes stabilization terms to have a well-posed problem. Uniqueness of the solution is ensured using Pressure-Stabilizing/Petrov-Galerkin (PSPG) stabilization, which enable the use of same order elements for $\vm{u}$ and $p$. Additionally, the weak form includes the Streamline-Upwind/Petrov-Galerkin (SUPG) stabilization \cite{lethe2025} to avoid oscillations in advection-dominated problems. 

The solution and test spaces are \cite{lethe2025}:
\begin{subequations}
\begin{align} 
    &\mathcal{U} := \left\{\vm{u} \in H^1(\Omega)^D : \vm{u} = \vm{0} \ \ \textrm{on} \ \partial \Omega \right\} \\ 
    &\mathcal{V} := \left\{ \vm{v} \in H^1(\Omega)^D: \vm{v} = \bm{0} \ \ \textrm{on} \ \partial \Omega \right\}\\ 
    &\mathcal{P} := \left\{ p \in L^2(\Omega) \right\} \\
    &\mathcal{Q} := \left\{ q \in L^2(\Omega) \right\}
\end{align}
\end{subequations}
where $D$ denotes the spatial dimension. The complete weak problem is: 


\noindent Find $p\in \mathcal{P}\times (0,t_\text{end}]$ and $\vm{u}\in \mathcal{U}\times (0,t_\text{end}],$ such that:
\begin{subequations}
\begin{align}
    \pp{q,\nabla\cdot\vm{u}}_\Omega + \underbrace{\pp{\tau_{u,k}\nabla q,\vm{\mathcal{R}_u}}_{\Omega_k}}_{\text{PSPG}}&= 0 \hphantom{\pp{\vm{v},\vm{F_\Gamma} + \rho\vm{g}}_\Omega} \quad \forall q\in\mathcal{Q}\times(0,t_\text{end}] \label{eq:mass_weak_form}\\
    \rho\pp{\vm{v},\pdv{\vm{u}}{t}}_\Omega + \rho\pp{\vm{v},(\vm{u} \cdot \nabla)\vm{u}}_\Omega &- \pp{\nabla\cdot\vm{v}, p}_\Omega + \mu\pp{\nabla\vm{v},\pp{\nabla\vm{u} + \pp{\nabla\vm{u}}\transpose}}_\Omega \nonumber \\
    + \underbrace{\pp{\tau_{u,k}\pp{\vm{u}\cdot\nabla}\vm{v},\vm{\mathcal{R}_u}}_{\Omega_k}}_{\text{SUPG}}&= \pp{\vm{v},\vm{F_\Gamma} + \rho\vm{g}}_\Omega \hphantom{0} \quad \forall \vm{v}\in\mathcal{V}\times(0,t_\text{end}]\label{eq:ns_weak_form}
\end{align}
\end{subequations}
holds, where $\vm{\mathcal{R}_u}$ is the strong residual of the momentum equation,
\begin{equation}
    \vm{\mathcal{R}_u}=\rho\pp{\pdv{\vm{u}}{t} + (\vm{u} \cdot \nabla)\vm{u}} - \nabla \cdot \tm{\sigma} - \vm{F_\Gamma} - \rho\vm{g}
\end{equation} 
and $\tau_{u,k}$ is a stabilization parameter. For transient problems, it takes the form \cite{ lethe2025, Tezduyar1992}:
\begin{align} \label{eq::tau_stabilization}
  \tau_{u,k} = \left[ \left( \frac{1}{\Delta t}\right)^2 + \left( \frac{2 \lVert \vm{u}_k \rVert n_p}{h_k} \right)^2 + 9 \left( \frac{4\nu n_p^2}{h_k^2} \right)^{2}\right]^{-1/2}
\end{align}
where $\Delta t$ is the time step, $\nu=\mu/\rho$ is the kinematic viscosity, $\lVert \vm{u}_k \rVert$ is the Euclidean norm of the cell velocity vector $\vm{u}_k$, and $n_p$ is the degree of the polynomial approximation. $h_k$ denotes the diameter of a disk (2D) or a sphere (3D) of the same area (2D) or volume (3D) as the element $k$:
\begin{align} \label{eq::h_stabilization}
    h_{k,\mathrm{2D}} = \sqrt{\frac{4 V_{k}}{\pi}} \ \ , \ h_{k,\mathrm{3D}} = \left( \frac{6 V_{k}}{\pi} \right)^{\frac{1}{3}}
\end{align}
where $V_k$ is the area in 2D or volume in 3D of the element $k$. The current work uses Lagrange tensor product elements $Q_m$ of degree $m=1$ for $\vm{u}$ and $p$.


\subsection{Phase Indicator Transport}
The present framework defines the phase indicator as: 
\begin{equation}
    \phi(\vm{x}) = \begin{cases}
        0 &\;\forall \vm{x} \in \Omega_0 \\
        1 &\;\forall \vm{x} \in \Omega_1
    \end{cases}
\end{equation}
with the interface $\Gamma := \{\vm{x} \in \Omega \mid \phi\pp{\vm{x}}=0.5\}$. The \textit{strong form} of the problem for the transport of $\phi$ is:
\begin{equation}
    \pdv{\phi}{t} + \vm{u} \cdot \nabla \phi = 0 \quad \forall \vm{x} \in \Omega\times(0,t_\text{end}]\label{eq:advection}
\end{equation}
Without loss of generality, the following boundary conditions are considered:\begin{subequations}
\begin{align}
    \phi(\vm{x})&=0\quad\forall \vm{x} \in \partial\Omega_D\times(0,t_\text{end}]\\
    \nabla\phi\cdot \vm{n} &= 0 \quad\forall \vm{x} \in \partial\Omega_N\times(0,t_\text{end}] 
\end{align}
\end{subequations}
where $\partial\Omega_D$ and $\partial\Omega_N$ correspond to Dirichlet and Neumann boundary conditions, respectively.
The initial condition is:
\begin{equation}
    \phi=\phi_0  \quad \forall \vm{x} \in \Omega, t=0
\end{equation}
\paragraph{Weak form} 
The weak form uses the scalar test function $s$ and includes SUPG stabilization due to the pure-advective form of Equation \eqref{eq:advection}. 

The solution and test spaces are:
\begin{subequations}
\begin{align} 
    &\Phi := \{{\phi}  \in H^1(\Omega) : {\phi} =0 \ \ \textrm{on} \ \partial \Omega_D \} \\ 
    &\mathcal{S}:= \{ {s}\;  \in H^1(\Omega): s =0 \ \ \textrm{on} \ \partial \Omega_D \}
\end{align}
\end{subequations}

respectively. The problem reads:

\noindent Find $\phi\in \Phi\times (0,t_\text{end}]$ such that
\begin{equation}
    \pp{s,\pdv{\phi}{t}  + \vm{u} \cdot \nabla \phi}_\Omega + \underbrace{\pp{\tau_{\phi,k}\pp{\vm{u}\cdot\nabla s},\mathcal{R}_\phi}_{\Omega_k}}_{\text{SUPG}}= 0\quad \forall s\in\mathcal{S}\times(0,t_\text{end}]
    \label{eq:vof_weak_form}
\end{equation}
where $\mathcal{R}_\phi$ is the strong residual of Equation \eqref{eq:advection}, $\mathcal{R}_\phi=\pdv{\phi}{t} + \vm{u}\cdot\nabla \phi$, and $\tau_{\phi,k}$ is a stabilization parameter defined by \citet{Tezduyar1992}:
\begin{equation}
    \tau_{\phi,k} =  \bra{\pp{\frac{1}{\Delta t}}^2 + \pp{\frac{2 \lVert \vm{u}_k\rVert}{h_k}}^2 }^{-1/2}
\end{equation}
The current work uses $Q_m$ elements of degree $m=1$ for $\phi$. 


\subsection{Time Integration and Equations Coupling}

An implicit Backward Differentiation Formula of order 2 (BDF2) discretizes the transient terms of Equations \eqref{eq:ns_weak_form} and \eqref{eq:vof_weak_form}.

The framework explicitly treats the coupling between the NS equations and the phase indicator transport, which occurs through the velocity $\vm{u}$, the physical properties ($\rho(\phi)$ and $\mu(\phi)$), and the ST term ($\vm{F_\Gamma}(\phi)$). For a given time iteration $n+1$, corresponding to the time $t+\Delta t$, the framework:
\begin{enumerate}
    \item solves the phase indicator advection \eqref{eq:vof_weak_form} for $\phi^{n+1}$ using an approximation of the velocity $\vm{\hat{u}}^{n+1}$ computed from a linear extrapolation of the two previous solutions, $\vm{u}^{n}$ and $\vm{u}^{n-1}$,
    \item performs the reinitialization of $\phi^{n+1}$ if the time iteration $n+1$ is a reinitialization time step, and,
    \item {solves the monolithic formulation of the fluid dynamics system formed by Equations \eqref{eq:mass_weak_form} and \eqref{eq:ns_weak_form} for $\vm{u}^{n+1}$ and $p^{n+1}$, with $\rho(\phi^{n+1})$, $\mu(\phi^{n+1})$, and $\vm{F_\Gamma}(\phi^{n+1})$.}
\end{enumerate}

\subsection{Physical Properties} \label{sec:properties}\label{sec:property_reg}
The change of any given physical property $\lambda$ at the interface follows~\cite{Garcia-Villalba2025}:
\begin{equation}
    \lambda = \lambda_0 (1-\hat{H}_\Gamma(\vm{x})) + \lambda_1\hat{H}_\Gamma(\vm{x})
\end{equation}
where $\lambda_i$, with $i \in \{0,1\}$, is the value of the property in $\Omega_i$. $\hat{H}_\Gamma(\vm{x})$ is the approximation of the Heaviside function stepping at the interface:
\begin{equation}
    \hat{H}_\Gamma(\vm{x}) = 0.5\tanh \bra{\beta\pp{\phi(\vm{x}) - 0.5}} +0.5 \label{eq:heaviside}
\end{equation}
where $\beta$ is a model parameter that controls the sharpness of the phase transition. \commentblue{This smooth Heaviside field models the discontinuous physical property changes more sharply than a direct usage of the phase indicator field.}

\subsection{Surface Tension Force} \label{sec:sft}
The volumetric formulation of the ST force in Equation \eqref{eq:ns_momentum_strong} is:
\begin{equation}
    \vm{F_\Gamma} = (\underbrace{-\gamma \kappa \vm{n}}_{\substack{\text{\tiny Normal} \\\text{\tiny force}}} + \underbrace{\nabla_\Gamma \gamma}_{\substack{\text{\tiny Tangential} \\\text{\tiny force}}} )\|\nabla \hat{H}_\Gamma\| \label{eq:surface_tension_force}
\end{equation}
where $\gamma$ is the ST coefficient, $\kappa$ is the curvature, $\vm{n}$ is the unit normal vector of the interface pointing away from $\Omega_0$. $\nabla_\Gamma$ represents the orthogonal projection of $\nabla$ on $\Gamma$. $\|\nabla \hat{H}_\Gamma\|$ is the approximation of the Dirac Delta distribution with a support only on the interface, as proposed in the Continuous Surface Force (CSF) model by \citet{Brackbill1992}. \commentblue{The introduction of the additional field $\hat{H}_\Gamma$ keeps a smooth transition of the phase indicator in the CLS solver, while ensuring a sharper definition of the interfacial forces and consequently lowering the modeling error.}


This work considers a linear variation of the ST coefficient $\gamma$ to a given scalar field $G(\vm{x})$:
\begin{equation}
    \gamma(\vm{x}) = \gamma_0 + \gamma' \pp{G(\vm{x}) - G_0}
\end{equation}
where $\gamma_0$ is the reference ST coefficient at the reference scalar value $G_0$ and $\gamma'$ is the constant rate of change of $\gamma$ with respect to $G(\vm{x})$. For example, if thermo-capillarity is considered, $G(\vm{x})$ would represent the temperature field.

With this definition of $\gamma(\vm{x})$, Equation \eqref{eq:surface_tension_force} takes the form:
\begin{equation}
    \vm{F_\Gamma} = \left[-\gamma \kappa \vm{n} +  \gamma'\pp{\nabla G - \vm{n}\pp{\vm{n}\cdot\nabla G}} \right]\|\nabla \hat{H}_\Gamma\|
\end{equation}

\subsection{Normal and Curvature Computations}
The normal $\vm{n}$ and curvature $\kappa$ terms in Equation \eqref{eq:surface_tension_force} are respectively:
\begin{align}
    \vm{n} &= \frac{\nabla\phi}{\|\nabla\phi\|}\label{eq:normal} \\ 
    \kappa &= -\nabla\cdot\vm{n} \label{eq:curvature}
\end{align}

The finite element approximation of $\phi$ is in the approximation space constructed using the Lagrange $Q_m$ polynomials. The current framework typically uses $Q_1$ elements. It results in $\kappa$ not belonging to the approximation space since it involves the second derivatives of the approximation of $\phi$. 

For this reason, instead of computing the interface normal vector field with Equation \eqref{eq:normal} and the curvature with Equation \eqref{eq:curvature}, the solver computes their $L^2$ projections to ensure $\kappa$ belongs to the approximation space using the approach proposed by \citet{Zahedi2012}

Considering an arbitrary vector function on the domain $\vm{\nu} \in L^2\pp{\Omega}^D$  as the test function, the following problem is considered for the projection of the regularized phase fraction gradient $\vm{\zeta}$:

Find $\vm{\zeta} \in \ltwo{\Omega}^D$ such that
\begin{equation}
    \pp{\vm{\nu},\vm{\zeta}}_\Omega + \pp{\nabla\vm{\nu},\eta_\mathrm{\zeta}\nabla\vm{\zeta}}_\Omega = \pp{\vm{\nu},\nabla {\phi}}_\Omega  \quad \forall \vm{\nu} \in \ltwo{\Omega}^D \label{eq:phase_gradient_projection}
\end{equation}
where $\eta_\mathrm{\zeta} = a h^2$ is the normal filter with $a$ a constant usually set to $4$. Following Equation \eqref{eq:normal}, the normal vector approximation is:
\begin{equation}
    \vm{n} \approx \vm{\hat{n}}= \frac{\vm{\zeta}}{\norm{\vm{\zeta}}}
    \label{eq:projected_normal}
\end{equation}
The projected curvature $\hat{\kappa}$ is computed by solving the following problem:

Find ${\hat{\kappa}} \in \ltwo{\Omega}$ such that
\begin{equation}
    \pp{K,\hat{\kappa}}_\Omega + \pp{\nabla K, \eta_\mathrm{{\hat{\kappa}}}\nabla\hat{\kappa}}_\Omega = \pp{\nabla K, \vm{\hat{n}}}_\Omega \quad \forall K \in \ltwo{\Omega}
    \label{eq:curvature_projection}
\end{equation}
Here, $K \in \ltwo{\Omega}$ is the arbitrary scalar test function and $\eta_\mathrm{{\hat{\kappa}}} = b h^2$ is the curvature filter with $b$ a constant usually set to $1$. The framework uses $Q_m$ elements with the same degree as the phase indicator to approximate the normal and curvature.

\subsection{Software} \label{sec:lethe}
\lethe{}, an open-source CFD software framework based on the \texttt{deal.II} finite element library \cite{dealII97}, implements the current work. It is a fully-parallel distributed, multi-phase and multi-physics solver and uses FEM with a Continuous Galerkin (CG) formulation. It considers Lagrange tensor product elements ($Q_m$) of arbitrary degree $m$ (quadrilaterals in 2D and hexahedra in 3D). It features adaptive mesh refinement capabilities using the \texttt{p4est} library \cite{p4est}. A thorough description of \lethe{} is available in \cite{lethe2025}. The version of \lethe{} used to generate the results in this work is openly available \cite{lethe_zenodo}.

\lethe estimates the error for the mesh adaptation using Kelly's error estimator \cite{Kelly1983}. It performs refinement and coarsening by estimating the error via the jump in the gradient between the faces for a selected variable. For the interfacial problems targeted in the current work, selecting the variable $\phi$ enables local refinement around the interface and coarsening away from it.

\section{Geometric reinitialization for CLS frameworks} \label{sec:geometric_reinit}
The proposed geometric reinitialization approach builds on the method of \citet{Mut2006} and \citet{Ausas2011} developed in the context of LS frameworks.  \commentblue{This work is an extension of the latter to conservative level-set method and fully distributed, adaptively refined, two- and three-dimensional meshes.}

It uses the following property of signed distance functions \cite{Mut2006}, illustrated in Figure \ref{fig:distance_properties}:

Let $\mathcal{C}$ be a surface of $\mathbb{R}^D$ dividing the latter into two open domains $\Omega^+$ and $\Omega^-$ such that $\Gamma\subset\Omega^-$. Then,
\begin{equation}
    |d(\vm{y})| = \min_{\vm{x}\in \mathcal{C}}\pp{\|\vm{y}-\vm{x}\|+|d(\vm{x})|} \quad \forall \vm{y}\in\Omega^+ \label{eq:d_abs}
\end{equation}
where $d(\vm{x})$ is a signed distance function:
\begin{equation}
    d(\vm{x}) = S(\vm{x})\min_{\vm{c}\in\Gamma}\|\vm{x}-\vm{c}\| \label{eq:d}
\end{equation}
with $\vm{c}$, the closest point of $\vm{x}$ on $\Gamma$, and
\begin{equation}
    S(\vm{x}) = 
    \begin{cases}
        -1 \quad \forall \vm{x} \in \Omega_1 \\
        \hphantom{-}1 \quad \forall \vm{x} \in \Omega_0
    \end{cases}
\end{equation}
\begin{figure}[h]
    \centering
    \def\svgwidth{0.9\textwidth}
          \input{Figures/distance_property103}
    \caption{Distance property described in Equation \eqref{eq:d_abs}.}
    \label{fig:distance_properties}
\end{figure}

In other words, the property states that the minimal distance of a given point $\vm{y}$ to $\Gamma$ can be computed knowing the distance function on an intermediary surface $\mathcal{C}$. 

In a discrete frame,  the implementation of the geometric method consists of three steps, as illustrated in the flow diagram of Figure \ref{fig:geometric-reintialization-process-chart}:

\begin{figure}[h!]
    \centering
    \usetikzlibrary{shapes.geometric, arrows}
\definecolor{beige}{HTML}{e9e3cc}
\definecolor{dbeige}{HTML}{7e5b31}

\definecolor{blue}{HTML}{cbe2e6}
\definecolor{dblue}{HTML}{22525b}

\definecolor{interface}{HTML}{ac401b}
\definecolor{dark}{HTML}{464646}

\tikzstyle{startstop} = [rectangle, 
rounded corners=0.75cm, 
minimum width=3cm, 
minimum height=1cm,
text width=5cm,
text centered, 
draw=dbeige, 
fill=beige]
\tikzstyle{process} = [rectangle,
minimum width=3.5cm, 
minimum height=1cm,
text width=6.5cm, 
text centered, 
draw=dblue, 
fill=blue]
\tikzstyle{decision} = [diamond, 
minimum width=1cm, 
minimum height=1cm, 
text width=3cm,
text centered, 
draw=black, 
fill=cyan!30]
\tikzstyle{arrow} = [thick,
->,
>=stealth]

\begin{tikzpicture}[node distance=2.5cm]

\node (start) [startstop] {
\begin{tabular}[t]{c}Solve for $\phi^{n+1}$ \\Eq. \eqref{eq:vof_weak_form}\end{tabular}};

\node (pro1) [process, below of=start, yshift=-0.75cm] {
\begin{tabular}[t]{c}Identify cells intersected by \\$\Gamma$ with a marching cube method\end{tabular}};

\node (pro2) [process, below of=pro1, yshift=-1.5cm] {
\begin{tabular}[t]{c}Compute $d(\vm{x})$ of the\\ intersected cells DoFs to \\construct the 1st intermediate \\surfaces $\mathcal{C}$ on both sides of $\Gamma$\end{tabular}};

\node (pro3) [process, below of=pro2, yshift=-1.85cm, text width=5cm] {
\begin{tabular}[t]{c}Apply local and global \\volume corrections\\(\citet{Mut2006})\end{tabular}};

\node (step1) [above right,inner sep=0mm, outer sep=0, xshift=-0.5cm, yshift=0.25cm] at (pro1.north west) {\renewcommand{\arraystretch}{0.5} \begin{tabular}[t]{l}Step 1:\end{tabular}};

\node[draw, dashed, thick, fit=(pro1) (pro2) (pro3),inner sep=2mm,outer sep=0] (Box1) {};

\node (pro4) [process, right of=pro2, xshift=5.4cm, yshift=0.35cm, text width=5.5cm] {
\begin{tabular}[t]{c}Iteratively compute $d(\vm{x})$\\for the rest of the mesh\\ up to $d_\text{max}$\\Eq. \eqref{eq:d_face}\end{tabular}};

\node (step2) [above right,inner sep=0mm, outer sep=0, xshift=-0.4cm] at (pro4.north west) {\renewcommand{\arraystretch}{0.5} \begin{tabular}[t]{l}Step 2:\end{tabular}};

\node (pro5) [process, below of=pro4, yshift=-2cm, text width=5.5cm] {
\begin{tabular}[t]{c}Compute $\phi^{n+1}_\mathrm{reinit}$ \\Eq. \eqref{eq:geo-reinit-tanh}\end{tabular}};

\node (step3) [above right,inner sep=0mm, outer sep=0, xshift=-0.4cm] at (pro5.north west) {\renewcommand{\arraystretch}{0.5} \begin{tabular}[t]{l}Step 3:\end{tabular}};

\node (stop) [startstop, below of=pro5, yshift=-0.75cm] {
\begin{tabular}[t]{c}Set\\$\phi^{n+1} = \phi_\mathrm{reinit}^{n+1}$\end{tabular}};

\draw [arrow] (start) -- (pro1);
\draw [arrow] (pro1) -- (pro2);
\draw [arrow] (pro2) -- (pro3);
\draw [arrow] (pro3.east) -- ++ (1.5,0) |- (pro4.west);
\draw [arrow] (pro4) -- (pro5);
\draw [arrow] (pro5) -- (stop);

\end{tikzpicture}
    \caption{Geometric reinitialization flow diagram}
    \label{fig:geometric-reintialization-process-chart}
\end{figure}

\paragraph{Step 1: First Intermediary Surface}
The method reconstructs the interface $\Gamma$ in each intersected cell into linear segments in 2D and planar elements in 3D from the iso-contour $\phi=0.5$ using a marching cube method. 

It builds the first intermediary surface $\mathcal{C}$ on each side of the interface by computing the distance between the interface reconstruction and the DoFs of the intersected cells. In 2D, it corresponds to point-to-line-segment distance computation, and in 3D, to point-to-plane distance computation. 

The method of \citet{Mut2006} also includes a volume correction step to account for volume loss/gain introduced by the signed distance function not being part of the Lagrange polynomial space ($d(\vm{x}) \notin Q_m$). It corrects the DoF values of the intersected cells to conserve cell-wise, i.e. local, and global volumes; the reader can refer to \cite{Mut2006} for more details on the volume conservation algorithm. 

\paragraph{Step 2: Distance Computation of the Rest of the Mesh}
With the knowledge of $d(\vm{x})$ on the first surface $\mathcal{C}$, the approach computes the distance for the rest of the mesh up to a user-defined maximum distance $d_\text{max}$. It builds successive surfaces $\mathcal{C}$ by resolving the following minimization problem for each DoF $i$ of a cell, as illustrated in Figure \ref{fig:simplex_quad}:
\begin{equation}
    |d(\vm{y_i})| = \min_{\vm{x}\in \mathcal{F}_j}\pp{\|\vm{y_i}-\vm{x}\|+|d(\vm{x})|} \label{eq:d_face}
\end{equation}
where $\vm{y_i}$ is the coordinate of the DoF $i$, $\mathcal{F}_j$ is a face opposite to the DoF $i$, and \vm{x} is a point on $\mathcal{F}_j$. Equation \eqref{eq:d_face} corresponds to Equation \eqref{eq:d_abs} in a discrete frame, hence, $\mathcal{F}_j$ is the counterpart of $\mathcal{C}$. 

It is an iterative approach similar to a marching method: it solves equation \eqref{eq:d_face} for each DoF of each cell until $d(\vm{y_i})$ converges.

The main differences between the current framework and the original works of \citet{Mut2006} and \citet{Ausas2011} is highlighted in Figure \ref{fig:simplex_quad}: the proposed approach is suitable for quadrilateral/hexahedral elements and adaptively refined grids. The adaptation to quadrilateral/hexahedral lies in the treatment of multiple opposite faces instead of only one. For the adaptively refined grids, the method treats the coarse and refined cells in the same way. The only requirement is to constraint the distance value at the hanging nodes to ensure continuity of the distance field, as typically done in dynamic mesh adaptation \cite{Bangerth2009}. The proposed method applies this constrain upon convergence of the iterative loop. Finally, the method works in a distributed parallel framework, which \citet{Mut2006} and \citet{Ausas2011} do not discuss in their works. 

\begin{figure}[h]
    \centering
    \def\svgwidth{\textwidth}
          \input{Figures/simplex_quad}
    \caption{Geometric reinitialization for quadrilateral and adaptively refined (right) meshes compared to the method of \citet{Mut2006} and \citet{Ausas2011} for simplex meshes (left).}
    \label{fig:simplex_quad}
\end{figure}

\paragraph{Step 3: Phase Indicator Computation}
The last step is to convert the signed distance field to a phase indicator using a $\tanh$-based transformation:
\begin{equation}
    \phi_\text{reinit}(\vm{x}) = 0.5 - 0.5\tanh{\pp{\frac{d(\vm{x})}{2\varepsilon}}}
    \label{eq:geo-reinit-tanh}
\end{equation}
where $\varepsilon$ is a measure of the interface thickness. This transformation ensures a smooth transition of the phase indicator, which is numerically desirable to solve Equation \eqref{eq:vof_weak_form}. It imposes $\phi_\text{reinit}(\vm{x})=0.5$ where $d(\vm{x}) = 0$. However, it does not strongly impose the extrema $\phi_\text{reinit}(\vm{x}) = 0$ or $1$ away from $\Gamma$ and their values depend on $d_\text{max}$. It has no significant impact on the results when $d_{\max{}}$ is sufficiently large, i.e., $\sim 4\varepsilon$.

\section{Baseline Interface Reinitialization Methods} \label{sec:mitigation}


\commentblue{For completeness and to assess the advantages of the geometric method, this work benchmarks the latter against two established reinitialization approaches of the literature: the PDE-based reinitialization of \citet{Olsson2007} and a simple projection-based reinitialization proposed by \citet{Aliabadi2000}. }

\subsection{PDE-Based Reinitialization}

The PDE-based reinitialization method proposed by \citet{Olsson2007} solves the following PDE in an {artificial time} until reaching steady-state:  
\begin{equation}
    \underbrace{\frac{\partial \phi_\text{reinit}}{\partial \tau} }_\text{transient} + \underbrace{\nabla \cdot \left[ \phi_\text{reinit}  (1-\phi_\text{reinit} ) \vm{n}\right]}_\text{compression} - \underbrace{\varepsilon \nabla \cdot \left[ (\nabla \phi_\text{reinit}  \cdot \vm{n}) \vm{n} \right]}_\text{diffusion} = 0
    \label{eq:algebraic_reinitialization_pde}
\end{equation}
where $\tau$ is the {artificial} time and $\varepsilon$ is a diffusion coefficient. The solution corresponds to the quasi-steady balance between the compressive and diffusive terms and the interface takes the form of the hyperbolic tangent profile decribed in Equation \eqref{eq:heaviside_phi} along its thickness \cite{Mirjalili2017, Garcia-Villalba2025, Chiu2011}. The choice of the diffusion coefficient $\varepsilon$ determines the interface thickness.

\paragraph{Steady-state} 
In the current numerical framework, Equation \eqref{eq:algebraic_reinitialization_pde} reaches steady-state when one of the two following stopping criteria is met\footnote{Note that the solver uses an implicit BDF1 time-integration scheme for the PDE-based reinitialization since the objective is to reach a steady-state solution and the temporal accuracy of the artificial time-stepping method is not important \cite{Olsson2007}.}:

\paragraph{Criterion I}
    \begin{equation*}
        \alpha_\mathrm{steady-state} \geq  \frac{\norm{ \phi^{\iota+1}_\mathrm{reinit} - \phi^{\iota}_\mathrm{reinit}}}{\norm{\phi^{\iota}_\mathrm{reinit}}} 
        \label{eq:steady-state-criterion}
    \end{equation*}
where $\alpha_\mathrm{steady-state}$ is a user-defined steady-state tolerance (set to $10^{-4}$ in this work), and $\iota$ represents an artificial time iteration.
    
\paragraph{Criterion II}
    \begin{equation*}
        \iota \geq N_\tau
        \label{eq:max-step-criterion}
    \end{equation*}
where $N_\tau$ is a user-defined maximum number of artificial time iterations.

Figure \ref{fig:algebraic-reintialization-process-chart} summarizes the steps of the PDE-based reinitialization process. Once the reinitialized phase indicator field replaces the initial phase indicator field (end of Figure \ref{fig:algebraic-reintialization-process-chart}), the solver recomputes the normal vector field with Equation \eqref{eq:projected_normal} and the curvature field with Equation \eqref{eq:curvature_projection}. Finally, the solution of the NS momentum Equation \eqref{eq:ns_weak_form} uses these new solution fields.

\begin{figure}[h!]
    \centering
    \usetikzlibrary{shapes.geometric, arrows}
\usetikzlibrary{positioning}
\definecolor{beige}{HTML}{e9e3cc}
\definecolor{dbeige}{HTML}{7e5b31}

\definecolor{blue}{HTML}{cbe2e6}
\definecolor{dblue}{HTML}{22525b}

\definecolor{interface}{HTML}{ac401b}
\definecolor{dark}{HTML}{464646}

\tikzstyle{startstop} = [rectangle, 
rounded corners=0.75cm, 
minimum width=3cm, 
minimum height=1cm,
text width=4cm,
text centered, 
draw=dbeige, 
fill=beige]
\tikzstyle{process} = [rectangle,
minimum width=3.5cm, 
minimum height=1cm,
text width=4cm, 
text centered, 
draw=dblue, 
fill=blue]
\tikzstyle{decision} = [diamond, 
minimum width=1cm, 
minimum height=1cm, 
text width=3cm,
align=center, 
draw=interface, 
fill=interface!30]
\tikzstyle{arrow} = [thick,
->,
>=stealth]

\begin{tikzpicture}[node distance=2cm]

\node (start) [startstop] {\begin{tabular}[t]{c}Solve for $\phi^{n+1}$ \\Eq. \eqref{eq:vof_weak_form}\end{tabular}};

\node (pro1) [process, below of=start, yshift=-1.5cm] {\begin{tabular}[t]{c}Solve for $\vm{\zeta}^{n+1}$ \\Eq. (\ref{eq:phase_gradient_projection})\\and compute $\vm{n}^{n+1}$\\Eq. \eqref{eq:projected_normal}\end{tabular}};



\node (pro3) [process, right of=pro1, xshift=4.4cm] {\begin{tabular}[t]{c}Solve for $\phi_\mathrm{reinit}^{\iota+1}$ \\Eq. \eqref{eq:algebraic_reinitialization_pde}\end{tabular}};
\node (dec1) [process, below=2.5cm of pro3] {\begin{tabular}[t]{c}Is Criterion\\ I or II  met?\end{tabular}};
\node (stop) [startstop, below of=dec1, yshift=-3cm] {\begin{tabular}[t]{c}Set\\$\phi^{n+1} = \phi_\mathrm{reinit}^{\iota+1}$\end{tabular}};

\draw [arrow] (start) -- (pro1);
\draw [arrow] (pro1) -- (pro3.west);
\draw [arrow] (start.east) -| (pro3.north) node[midway, above] {$\phi^{0}_{\mathrm{reinit}} = \phi^{n+1}$};
\draw [arrow] (pro3.south) -- (dec1);
\draw [arrow] (dec1) -- (stop) node[midway, left] {Yes};
\draw [arrow] (dec1.east)  -- ++(1,0) node[midway, below] {No} |- (pro3.east) node[midway, above] {$\iota\mathrel{+}=1$};

\end{tikzpicture}
    \caption{PDE-based reinitialization process flow diagram}
    \label{fig:algebraic-reintialization-process-chart}
\end{figure}

\subsection{Projection-Based Reinitialization} \label{sec:proj}

The projection-based reinitialization is the simplest of the three reinitialization methods discussed in this article. It projects the phase indicator to a sharper space. The current work studies the method proposed by \citet{Aliabadi2000} in which the reinitialized phase indicator is given by:
\begin{equation}
    \phi_\text{reinit} = 
    \begin{cases}
        c^{1-\alpha}\phi^\alpha \quad &\text{if } 0 \le \phi \le c\\
        1-(1-c)^{1-\alpha}(1-\phi)^\alpha \quad &\text{if } c \le \phi \le 1
    \end{cases} \label{eq:projection-based}
\end{equation}
where $c$ corresponds to the iso-level $\phi=c$ for which the volume is conserved and $\alpha$ is the interface sharpening parameter.
The value of $c$ can be obtained from a volume-conservation routine, as described by~\citet{Aliabadi2000}. The current work, fixes it to $0.5$ from the observation that any other values result in a global displacement of the interface. The interface sharpening parameter $\alpha=1.5$ {is selected} and corresponds to the value proposed by \citet{Aliabadi2000}.

\subsection{Summary and Parameters} \label{sec:reinit_summary}

Table \ref{tab::reinitialization-methods-summary} presents a summary of the three methods presented in~Sections \ref{sec:geometric_reinit} and \ref{sec:mitigation}, highlighting their parameters and the solution profile of the phase indicator in the normal direction to the interface. It shows that the solution of the PDE-based and geometric reinitialization are the same, with $\varepsilon$ controlling the thickness of the interface.

\begin{table}[h]
\renewcommand\arraystretch{1.25}
\centering
\caption{Reinitialization methods summary}
\begin{tabular}{lp{4.5cm}p{4cm}}
\hline
\textbf{Method}           & \textbf{Solution profile} & \textbf{Parameters}  \\ \hline
PDE-based        & {$0.5 -0.5\tanh\pp{{d(\vm{x})}/{(2\varepsilon)}}$} & $\varepsilon, \Delta \tau,$ $ \alpha_\text{steady-state}, N_\tau$ \\
Geometric         & $0.5 -0.5\tanh\pp{{d(\vm{x})}/{(2\varepsilon)}}$& $\varepsilon, d_\text{max}$\\
Projection-based & Piece-wise form given by Equation \eqref{eq:projection-based} & $c, \alpha$ \\ \hline
\end{tabular}
\label{tab::reinitialization-methods-summary}
\end{table}

\newpage

\section{Results} \label{sec:results}
This section studies the results obtained with the geometric reinitialization method on relevant 3D cases: the capillary migration of a droplet in Section \ref{sec:capillary_migration}, the 3D rising bubble benchmark in Section \ref{sec:rising-bubble}, and the Rayleigh-Plateau instability of a jet in Section \ref{sec:rayleigh-plateau}. It also includes a \commentred{sensitivity study to the reinitialization frequency $1/(N_\text{reinit}\Delta t)$} and a critical comparison of the results to those obtained with the PDE- and projection-based methods.

The capillary migration case and the rising bubble benchmark use $\{\mathbb{M}\mathbb{L}\mathbb{T}\}$ dimensions to described the domain sizes and the physical properties, where $\mathbb{M}$ stands for mass, $\mathbb{L}$ for length, and $\mathbb{T}$ for time units.

\subsection{Selection of the Reinitialization Parameter Values}

Table \ref{tab::params} reports the simulation parameters for the three cases. The comparison presented in the following sections results from careful selection of the parameter values for each method. 

For the geometric reinitialization, $\varepsilon$ is proportional to the cell size at the interface. Values of $\varepsilon$ higher than $2h$ lead to better volume conservation; for completeness, Section \ref{sec:capillary_migration} provides a study on the effects of $\varepsilon$. For the maximum reinitialization distance, the current framework considers $d_\mathrm{max} \propto \varepsilon$. Typical values are $d_\mathrm{max} \sim 4\varepsilon$. A higher value $d_\mathrm{max}$ results in additional computation time, but the parameter has no significant effect on the solution, as long as $d_\mathrm{max} \sim 4\varepsilon$.

For the PDE-based approach, the three main parameters are $\varepsilon$, $\Delta\tau$ and $\alpha_{\mathrm{steady-state}}$. They are challenging to select, especially $\varepsilon$, because they are case- and mesh-dependent, so there is no general set that can be re-used. For the value of interface thickness measure $\varepsilon$, \citet{Olsson2007} suggest $\varepsilon \approx h/2$. Higher values ($\varepsilon>h$) lead to poor artificial time convergence to the steady-state or nonphysical oscillations of the phase indicator at the interface. The selected artificial time step $\Delta \tau$ has to enable convergence to steady-state efficiently. A too small value requires a significant number of time-steps ($> 100$) to reach steady-state, while a large value results in a nonphysical solution. \citet{Zahedi2012} suggest $\Delta \tau = \min(\Delta t, h)$. {In the present case, it leads to $\Delta \tau = \Delta t$.} The value of the steady-state criterion $\alpha_{\mathrm{steady-state}}$ has a similar effect. This work consider a value of $\SI{e-4}{}$ to ensure that the error remaining on the steady-state does not affect the results. For the maximum number of artificial time steps, this work considers a high value of $10000$ to ensure convergence to the steady-state. {Typically, the method reaches the steady-state criterion $\SI{e-4}{}$ within $10$ artificial time steps.}

Even if the PDE-based and geometric methods lead to the same tangential hyperbolic solution profile, as presented in Table \ref{tab::reinitialization-methods-summary}, this work considers different values of $\varepsilon$ to report the better results for each approach. 

For the projection-based reinitialization, the iso-level $c$ is $0.5$ to avoid any displacement of the interface, as explained in Section \ref{sec:proj}, and the sharpening parameter $\alpha$ is $1.5$ as proposed by \citet{Aliabadi2000}.

\begin{table}[!htpb]
    \centering
    \footnotesize
    \caption{\centering General and specific parameters for the three cases. CM, RB, and RP stand for Capillary Migration, Rising Bubble, and Rayleigh-Plateau, respectively. Cartesian grids discretize the domain of each case, with $h$ being the side length of the smallest cell.}
    \begin{threeparttable}
    \begin{tabular}{P{0.44\textwidth}cccc}
    \hline
        \multicolumn{5}{c}{General simulation parameters} \\\hline
        \textbf{Parameters}& {\textbf{Dimensions}}& \textbf{CM}& \textbf{RB} & \textbf{RP}  \\ \hline
        Regularization sharpness $\beta$  & - & 20& 20& 20\tnote{*} \\
        Normal projection factor $a$ & - & 1& 1& 4\\
        Curvature projection factor $b$  & - & 1& 1& 1\\ 
        Mesh adaptation frequency & $\ti^{-1}$ & $1/(20\Delta t)$& $1/(5\Delta t)$& $1/(10\Delta t)$ \\
        \hline 
        \multicolumn{5}{c}{Geometric reinitialization} \\\hline
        \textbf{Parameters}& {\textbf{Dimensions}}& \textbf{CM}& \textbf{RB}& \textbf{RP}  \\ \hline
        Interface thickness measure $\varepsilon$ & $\len$& $4h$& $4h$& $3h$ \\
        Max. reinitialization distance $d_\mathrm{max}$ & $\len$ & $4\varepsilon$& $4\varepsilon$& $4\varepsilon$\\
        \hline  
        \multicolumn{5}{c}{PDE-based reinitialization} \\\hline
        \textbf{Parameters}& {\textbf{Dimensions}}& \textbf{CM}& \textbf{RB}& \textbf{RP} \\ \hline
        Interface thickness measure $\varepsilon$ & $\len$& $2h$& $h$ & $h$\\
        Artificial time step $\Delta \tau$ & $\ti$ & $\Delta t$& $\Delta t$& $\Delta t$\\
        Steady-state criterion $\alpha_\mathrm{steady-state}$ & -& $\SI{e-4}{}$&$\SI{e-4}{}$&$\SI{e-4}{}$\\
        Max. number of artificial time step $N_\tau$ & -& $\SI{10000}{}$& $\SI{10000}{}$& $\SI{10000}{}$\\ \hline   
        \multicolumn{5}{c}{Projection-based reinitialization} \\\hline
        \textbf{Parameters}& {\textbf{Dimensions}}& \textbf{CM}& \textbf{RB}& \textbf{RP} \\ \hline
        Iso-level $c$ & -& $0.5$& $0.5$& $0.5$ \\
        Interface sharpening parameter $\alpha$ & - & $1.5$& $1.5$& $1.5$\\
        Initial interface thickness measure $\varepsilon$ & $\len$& $3h$& $3h$& $3h$ \\
        \hline   
    \end{tabular}
    \begin{tablenotes}
        \item[*]The Rayleigh-Plateau case deactivates the approximation $\hat{H}_\Gamma$ described by Equation \eqref{eq:heaviside} for the PDE-based method, else poor volume conservation and slower spatial convergence rates are observed. Instead, the phase-indicator was used directly.
    \end{tablenotes}
    \end{threeparttable}
    
    \label{tab::params}
\end{table}
\subsection{3D Capillary Migration} \label{sec:capillary_migration}

This case considers a spherical droplet subjected to a ST coefficient gradient ($\nabla_\Gamma \gamma \neq 0$). In the absence of external force, a creeping flow develops due to the Marangoni effects described by Equation \eqref{eq:surface_tension_force}. The droplet keeps its spherical shape while migrating along the ST coefficient gradient towards the lowest ST region. 

This application case considers a linear variation of the ST coefficient according to the $x$-position such that:
\begin{equation}
    \gamma = \gamma_0 + \gamma'x
\end{equation}
At steady-state, the droplet migrates along the $x$-axis at a velocity of~\cite{Buscaglia2011}:
\begin{equation}
    U_\infty= -\frac{2R}{3\mu_0(2+3\lambda)}\gamma' \label{eq::capillary-migration-solution}
\end{equation}
where $R$ is the radius of the droplet, $\lambda = {\mu_1}/{\mu_0}$ is the ratio of dynamic viscosities between the inside (fluid 1) and outside (fluid 0) of the droplet. 

\commentblue{In opposition to a purely advection case in which the velocity field would be imposed, the droplet migration results from the tangential motion of the interface. Hence, capturing the migration in a numerical framework requires an accurate representation of the interface throughout the simulation time. In addition, due to the development of a creeping flow, it allows to evaluate the effects of the methods on quasi-static interfaces and assess if errors accumulate.}


\subsubsection{Description of the Case}
Figure \ref{fig::capillary-migration-schematic} presents the schematic of the case and Table \ref{tab::capillary-migration-physical-properties} lists the selected properties, based on the work of \citet{Buscaglia2011}. The bubble radius is $R=0.25\len$ and the domain is a cube with a side length of $L=12R$. For all boundaries, the case considers no-slip conditions for the velocity and no-flux conditions for the phase indicator. The initial condition imposes the droplet in the center of the domain, according to Equations \eqref{eq:initial-condition} and \eqref{eq:initial-condition-d}, with both fluids at rest:
\begin{align}
    \phi_0(\vm{x}) &= 0.5 - 0.5 \tanh\left( \frac{d_0(\vm{x})}{2\varepsilon}\right) \label{eq:initial-condition}\\
    d_0(\vm{x}) &= \| \vm{x}-\vm{x}_0\| - R \label{eq:initial-condition-d}
\end{align}
where $d_0$ is the signed distance function of a sphere of radius $R$ centered at $\vm{x}_0$ according to Figure \ref{fig::capillary-migration-schematic}. The case considers a migration time of $3\ti$.
\begin{figure}[h!]
    \centering
    \def\svgwidth{0.5\textwidth}
          \input{Figures/capillary-migration-schematic.tex}
    \caption{2D cross-section view in the $x$-$y$ plane of the initial state of the capillary migration case}
    \label{fig::capillary-migration-schematic}
\end{figure}

\begin{table}[!htpb]
    \centering
    \caption{Physical properties of the capillary migration case}
    \begin{tabular}{lcc}
        \hline 
        \textbf{Parameters}& {\textbf{Dimensions}}& \textbf{Values} \\ \hline
        {Density ratio ($\rho_\mathrm{0}/\rho_\mathrm{1}$)} & {-} & {1.0} \\
        
        Density of fluid 0 ($\rho_0$) & $\m\len^{-3}$&  1.0 \\

        {Viscosity ratio ($\mu_\mathrm{0}/\mu_\mathrm{1}$)} & {-} & {1.0} \\
        Dynamic viscosity of fluid 0 ($\mu_0$) & $\m\len^{-1}\ti^{-1}$&  1.0 \\

        Reference ST coefficient ($\gamma_0$) & $\m\ti^{-2}$ & 3.0\\ 
        {ST coefficient gradient ($\gamma'$)} & $\m\ti^{-2}\len^{-1}$ & {-1.0} \\ \hline   
    \end{tabular}
    \label{tab::capillary-migration-physical-properties}
\end{table}

\subsubsection{Metrics of Interest}
The analysis focuses on the following metrics of interest: \commentblue{the volume $V$ and surface area $A_\Gamma$ of the droplet}, its migration velocity along the $x$-axis, and the radius distribution along the interface at the end of the simulation. 
\commentblue{The volume $V$ and surface area $A_\Gamma$ of the bubble are, respectively:}
\begin{align}
    V &= \int_{\Omega_1}1\mathrm{d}\Omega \\
    A_\Gamma &= \int_\Gamma 1 dS
\end{align}
\commentblue{The analytical solution predicts a spherical droplet of constant radius $R$, volume $V_\mathrm{exact}=\dfrac{4\pi R^3}{3}$ and surface $A_{\Gamma,\mathrm{exact}}=4\pi R^2$. This study focuses on the maximum deviation in time from the analytical volume and surface area, which corresponds to the $L^\infty$ norms of the relative errors:}
\begin{align}
    \left\| \frac{V-V_\mathrm{exact}}{V_\mathrm{exact}}\right\|_{L^\infty} &= \max_{t\in(0,3\mathbb{T}]}\left| \frac{V(t)-V_\mathrm{exact}}{V_\mathrm{exact}}\right|\\
    \left\| \frac{A_\Gamma-A_\mathrm{exact}}{A_{\Gamma,\mathrm{exact}}}\right\|_{L^\infty} &= \max_{t\in(0,3\mathbb{T}]}\left| \frac{A_\Gamma(t)-A_{\Gamma,\mathrm{exact}}}{A_\mathrm{exact}}\right|
\end{align}
The definition of the migration velocity is:
\begin{equation}
    \text{Migration velocity} = \frac{\int_{\Omega}{u_x \hat{H}_\Gamma} \mathrm{d}\Omega}{\int_{\Omega}\hat{H}_\Gamma\mathrm{d}\Omega}  \label{eq:migration-velocity}
\end{equation}
\commentblue{Equation \eqref{eq::capillary-migration-solution} corresponds to the analytical value of the migration velocity.}
The radius distribution along the interface is:
\begin{align}
    r(\vm{x}) &= \|\vm{x} - \vm{x}_b\| \quad\forall \vm{x}\in\Gamma \\
    \vm{x}_b &= \frac{\int_{\Omega}\vm{x}\hat{H}_\Gamma d\Omega}{\int_{\Omega}\hat{H}_\Gamma d\Omega} \label{eq::barycenter_vector}
\end{align}
where $\vm{x}_b$ is the position of the droplet barycenter computed according to Equation \eqref{eq::barycenter_vector}.  

\subsubsection{Simulation Settings}
This application first assesses \commentblue{the effect of the interface thickness $\varepsilon$ on the spatial convergence of the volume and surface for the geometric reinitialization applied at each time step. The current study considers three values of $\varepsilon$: $2h$, $4h$, and $8h$. For the spatial convergence assessment, three adaptively refined hexahedral grids discretize the domain. Table \ref{tab::capillary-migration-refinements} reports the refinement levels of each mesh along with the selected time steps. The latter are constant throughout the simulation and to enable reasonable reinitialization Courant-Friedrichs-Lewy condition (CFL) values, they are higher than the capillary limit described by \cite{Brackbill1992}: } 
\begin{equation}
    \Delta t_\gamma = \sqrt{\frac{(\rho_\mathrm{0}+\rho_\mathrm{1})h^3}{4\pi\gamma}} \label{eq::capillary-time-step}
\end{equation}
where $h$ is {the minimum cell size}, defined in this work as the side length of the smallest cell. Experience and the results presented in this section show that this choice does not lead to any stability issue. The CFL values stabilize at approximately 0.01.

\commentblue{Then, the analysis compares the results obtained with the geometric method to those of the two reference reinitialization methods (PDE- and projection-based)} and investigates three reinitialization frequencies: $f=1/(N_\text{reinit}\Delta t)$, with $N_\text{reinit} \in \{1,10,100\}$. \commentblue{This study considers the medium refinement level of Table \ref{tab::capillary-migration-refinements} for the three reinitialization methods and frequencies.} 
\begin{table}[h]
    \centering
    \caption{Refinement levels for the capillary migration convergence study. ${R=0.25\mathbb{L}}$. }
    \footnotesize
    \begin{tabular}{P{65pt}C{85pt}C{85pt}C{65pt}}
        \hline 
        \textbf{Ref. level}&  \textbf{Max. cell size} & \textbf{Min. cell size} & \textbf{Time step} \\ \hline
        Coarse&     $\SI{5.00e-1}R$&    $\SI{6.25e-2}R$ & $0.026\mathbb{T}$\\
        Medium&     $\SI{2.50e-1}R$&    $\SI{3.13e-2}R$ & $0.013\mathbb{T}$\\
        Fine  &     $\SI{1.25e-1}R$&   $\SI{1.56e-2}R$ & $0.007\mathbb{T}$\\\hline
    \end{tabular}
    \label{tab::capillary-migration-refinements}
\end{table}

\subsubsection{Results}
\begin{figure}[h]  
    \begin{subfigure}[b]{1\textwidth}
        \begin{minipage}{0.48\textwidth}
            \centering
            \includegraphics[width=1\textwidth]{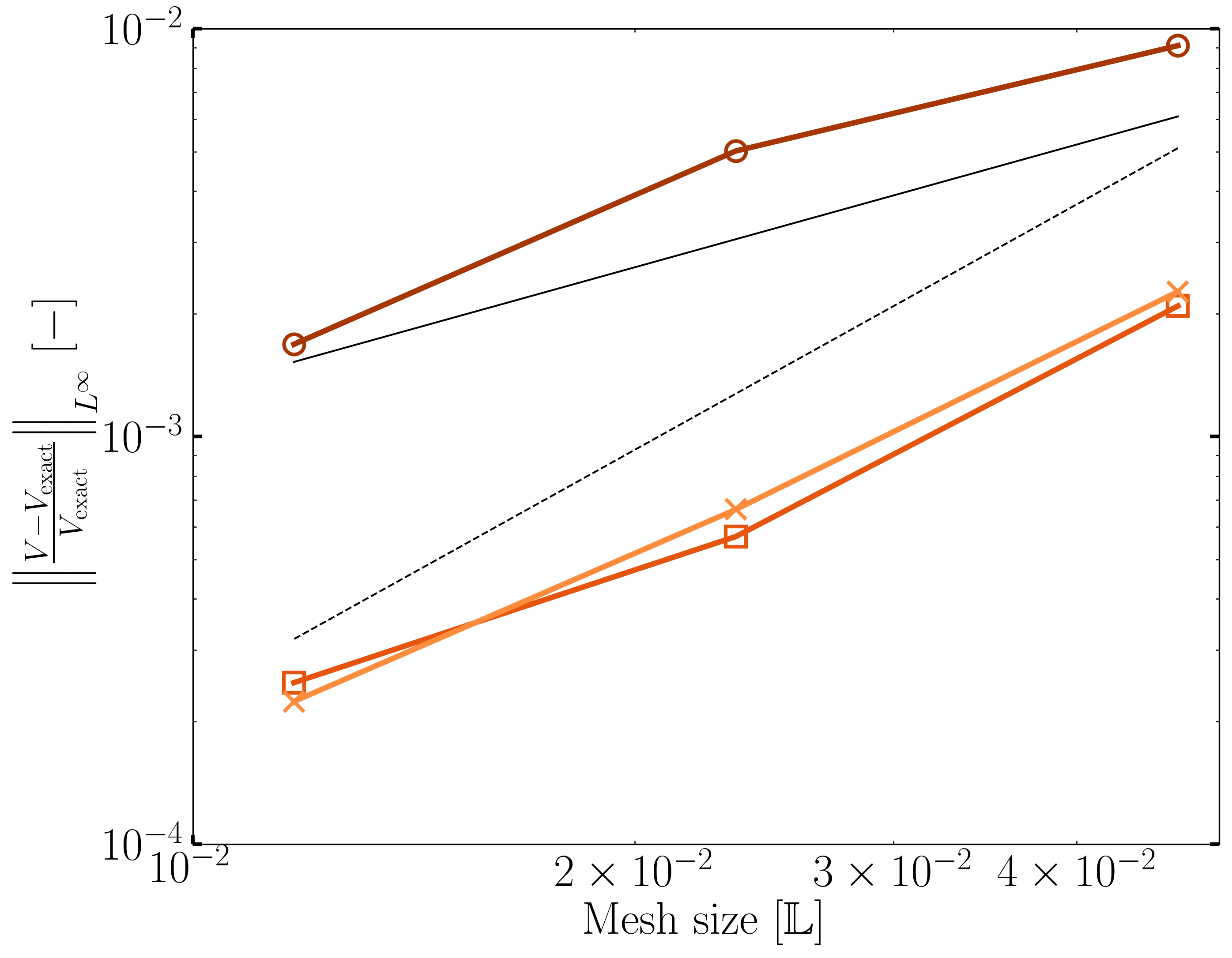}
        \end{minipage}
        \hfill
        \begin{minipage}{0.48\textwidth}
            \centering
            \includegraphics[width=1\textwidth]{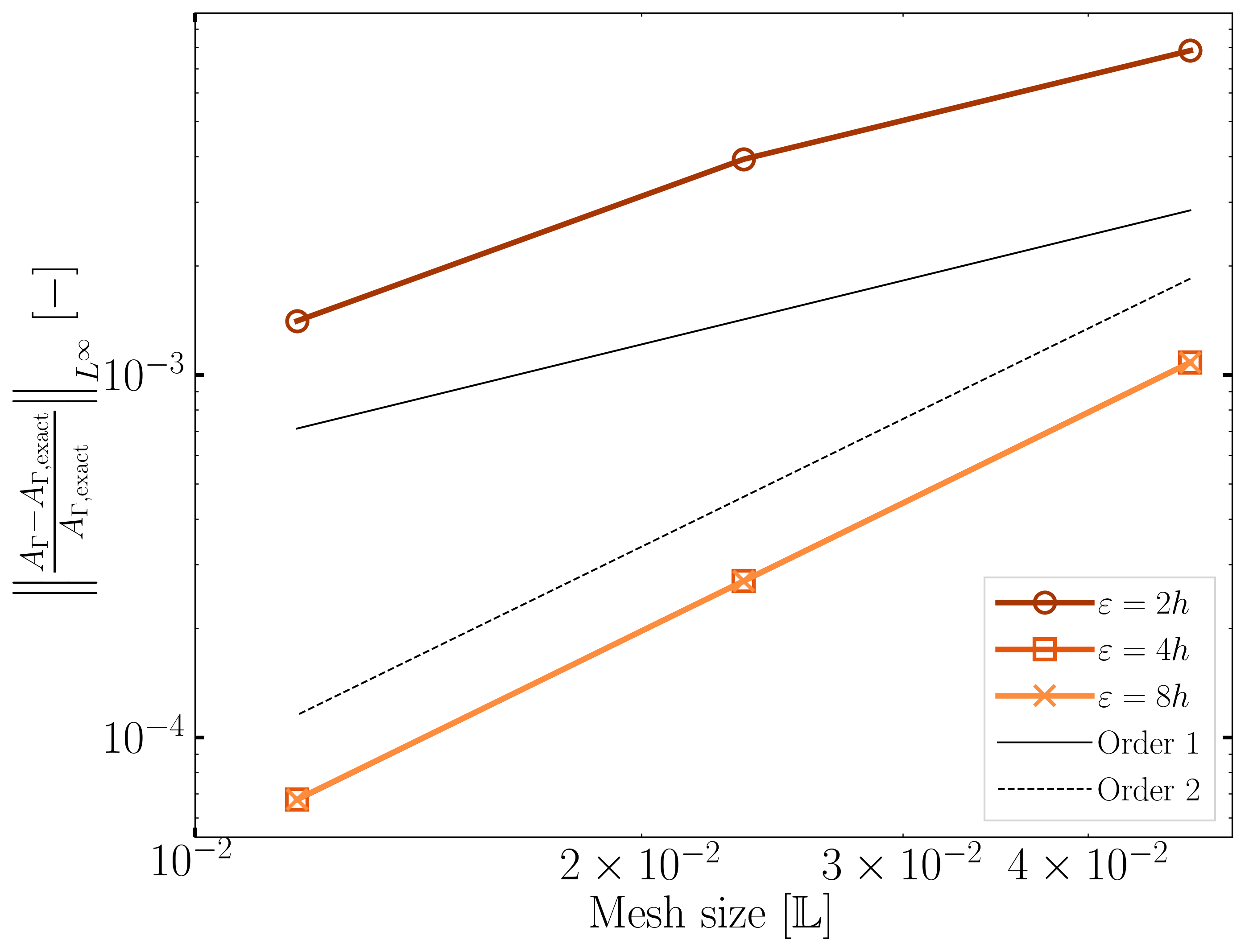}
        \end{minipage}
        
    \end{subfigure}
    \caption{Effects of the interface thickness $\varepsilon$ on the spatial convergence of the errors on the volume (left panel) and surface (right panel) of the droplet  obtained with the geometric method. Higher values of $\varepsilon$ decrease the error norms and lead to better convergence rates.}
    \label{fig::epsilon-study}
\end{figure}

\commentblue{Figure \ref{fig::epsilon-study} presents the spatial convergence of the $L_\infty$ norm in time of the relative error on the volume and surface area for the three values of the interface thickness: $\varepsilon = \{2h, 4h, 8h\}$. For $\varepsilon = 4h$ and $8h$, the volume and surface errors collapse on each other, with a reduction of approximately one order of magnitude compared to $\varepsilon = 2h$. Additionally, the rates of convergence for those two values are closer to a second order, especially for the surface area. The results highlight that the geometric method removes the restriction on $\varepsilon$, without jeopardizing the volume conservation, nor the surface accuracy, even for quasi-static interfaces as this case. }

\begin{figure}[h]
    \begin{subfigure}[b]{0.95\textwidth}
        \centering
        \includegraphics[width=0.7\textwidth]{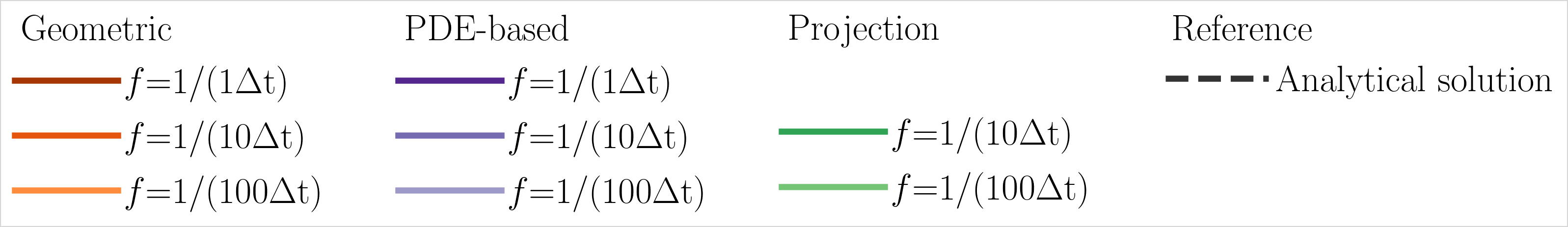}
    \end{subfigure}    
    \begin{subfigure}[b]{1\textwidth}
        \begin{minipage}{0.48\textwidth}
            \centering
            \includegraphics[width=1\textwidth]{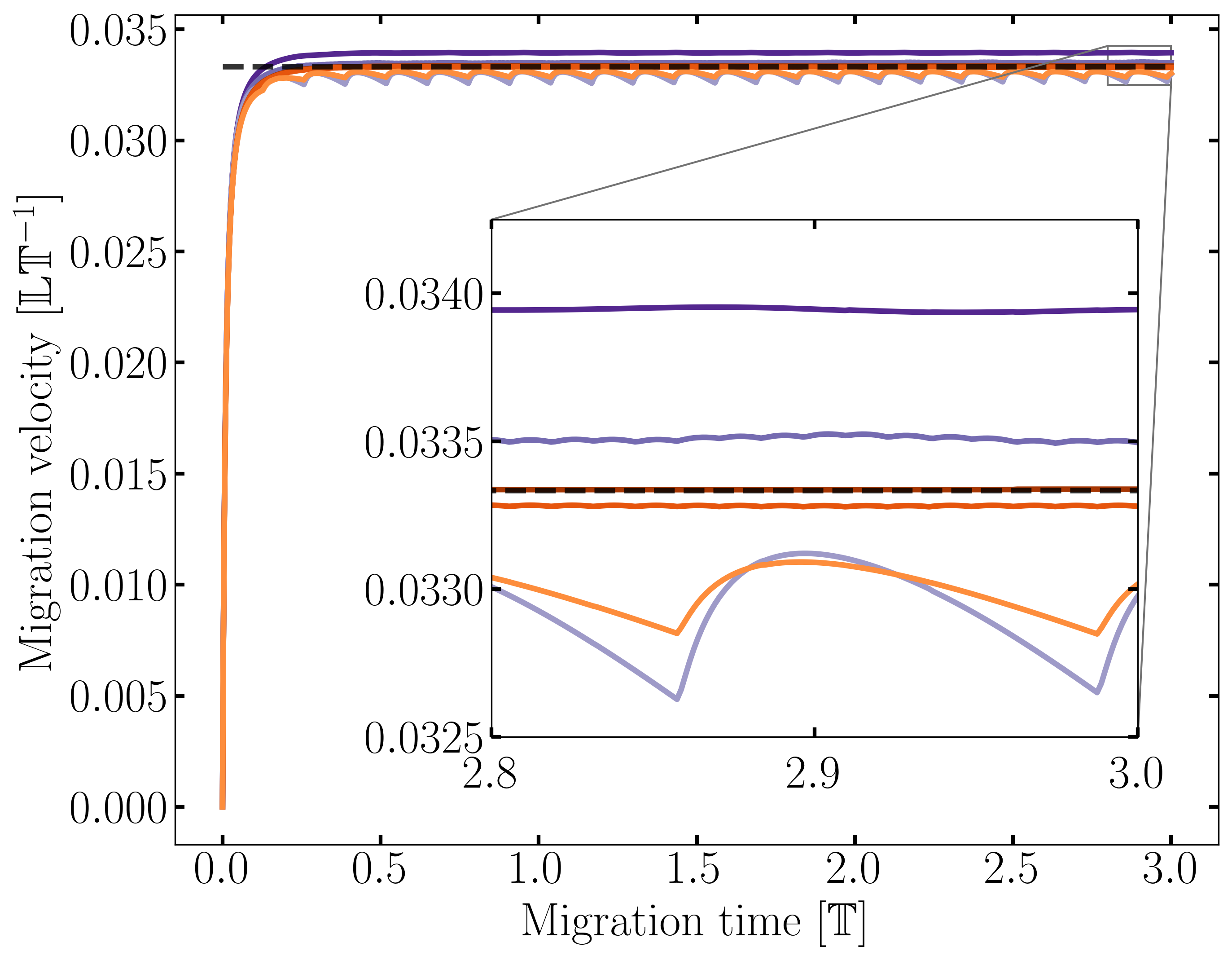}
        \end{minipage}
        \begin{minipage}{0.48\textwidth}
            \centering
            \includegraphics[width=1\textwidth]{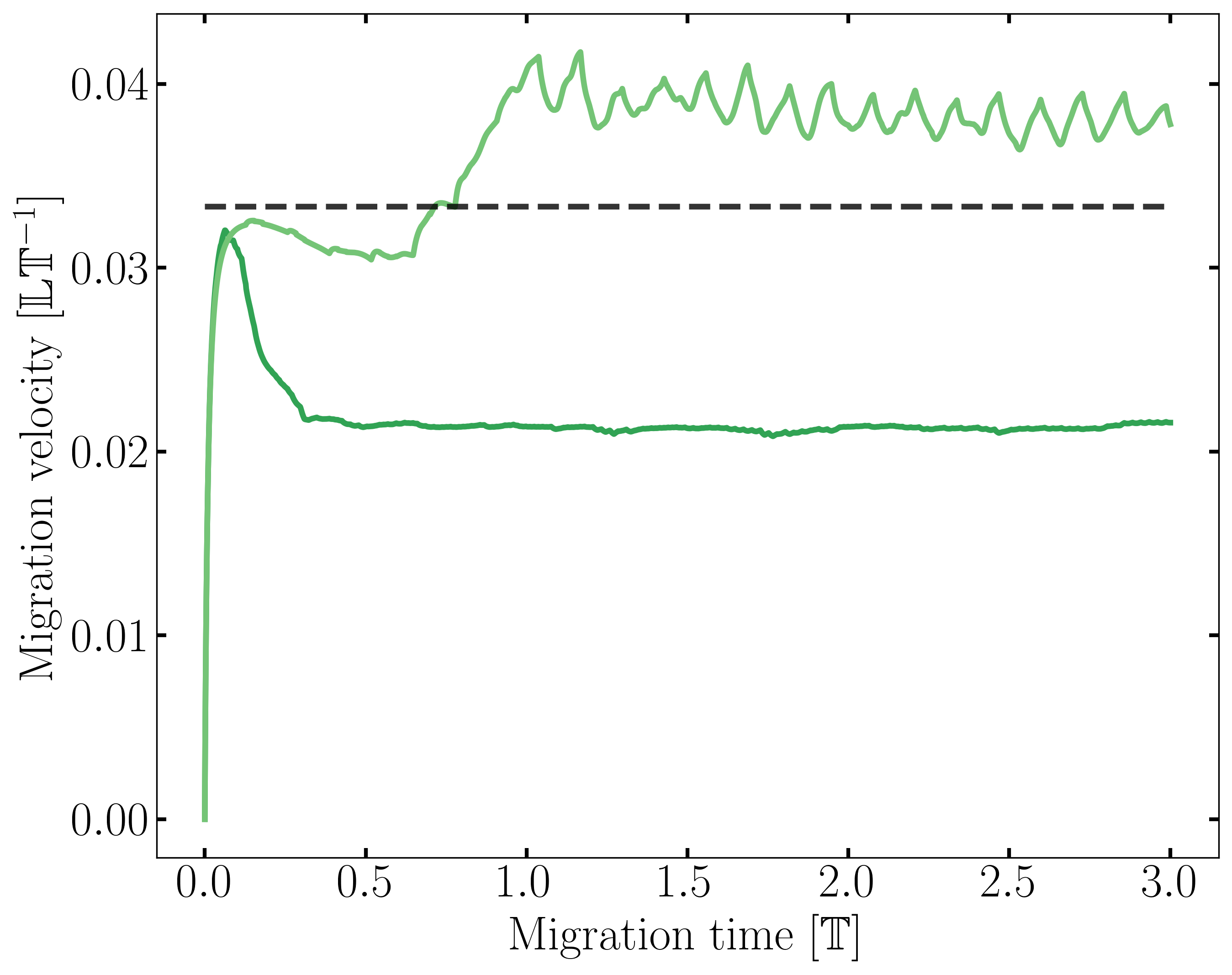}
        \end{minipage}
    \end{subfigure}
    \caption{Comparison of the migration velocity for the different reinitialization methods at different reinitialization frequencies. The PDE-based and geometric methods recover the analytical velocity described by Equation \eqref{eq::capillary-migration-solution} (left panel), while the projection-based approach fails to capture the migration dynamics (right panel).}
    \label{fig::migration-velocity}
\end{figure}

\begin{figure}
    \centering
    \begin{subfigure}[b]{0.95\textwidth}
        \centering
        \includegraphics[width=0.7\textwidth]{Figures/capillary-migration/legend.png}
    \end{subfigure}
    \begin{subfigure}[b]{1\textwidth}
        \centering
        \begin{minipage}{0.49\textwidth}
            \includegraphics[width=1\textwidth]{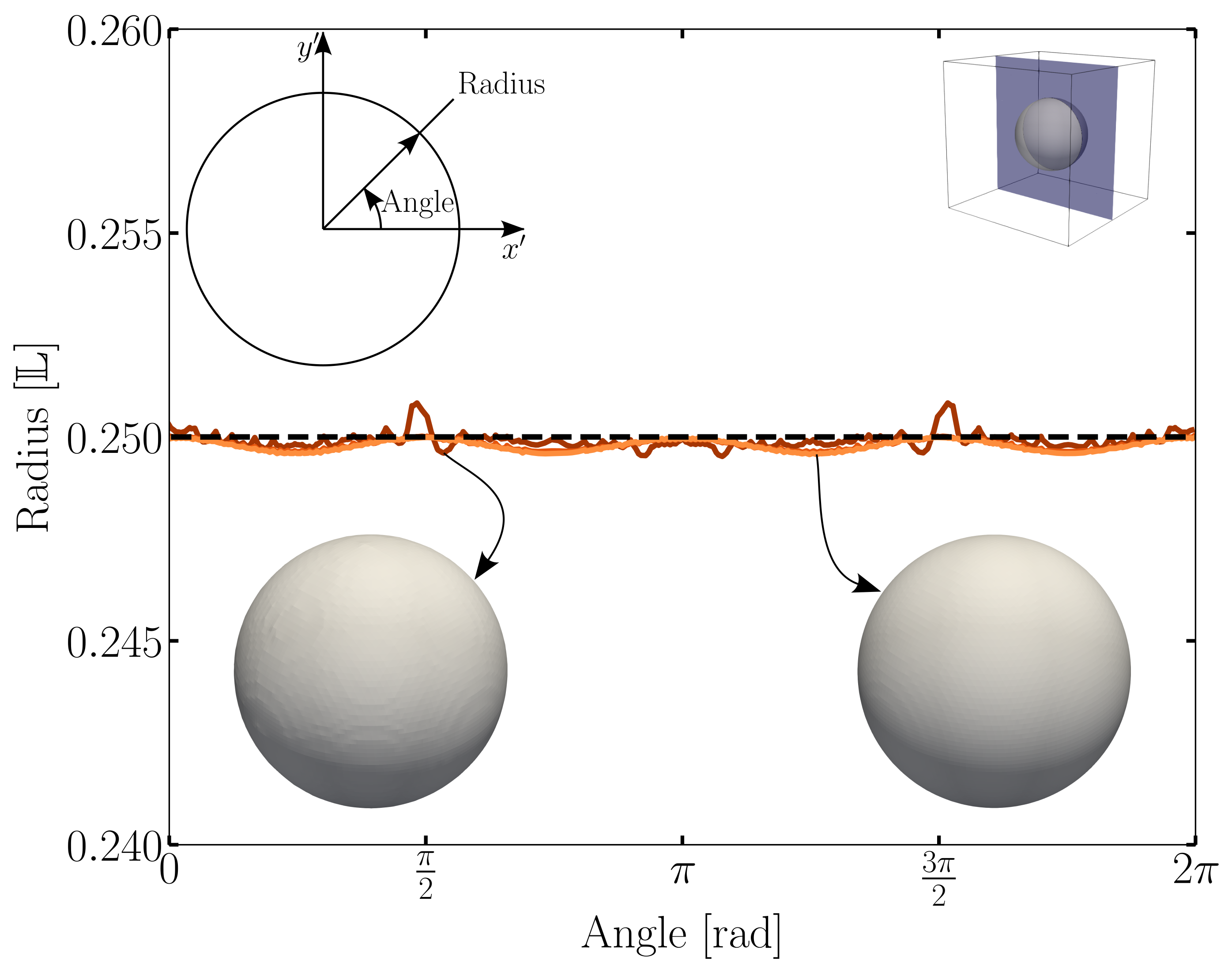}
        \end{minipage}
        \hfill
        \begin{minipage}{0.49\textwidth}
            \centering
            \includegraphics[width=1\textwidth]{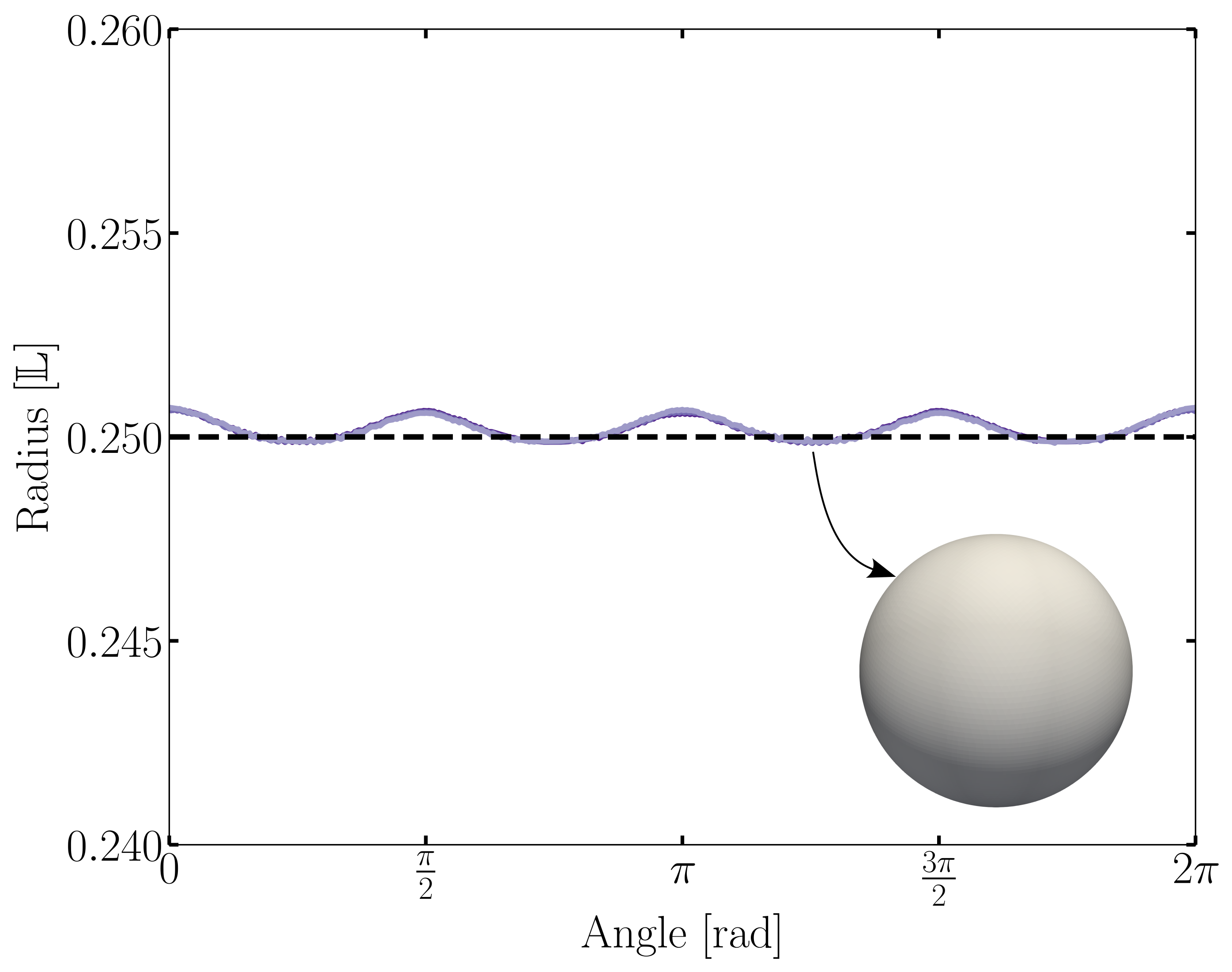}
        \end{minipage}
        \begin{minipage}{0.49\textwidth}
            \centering
            \includegraphics[width=1\textwidth]{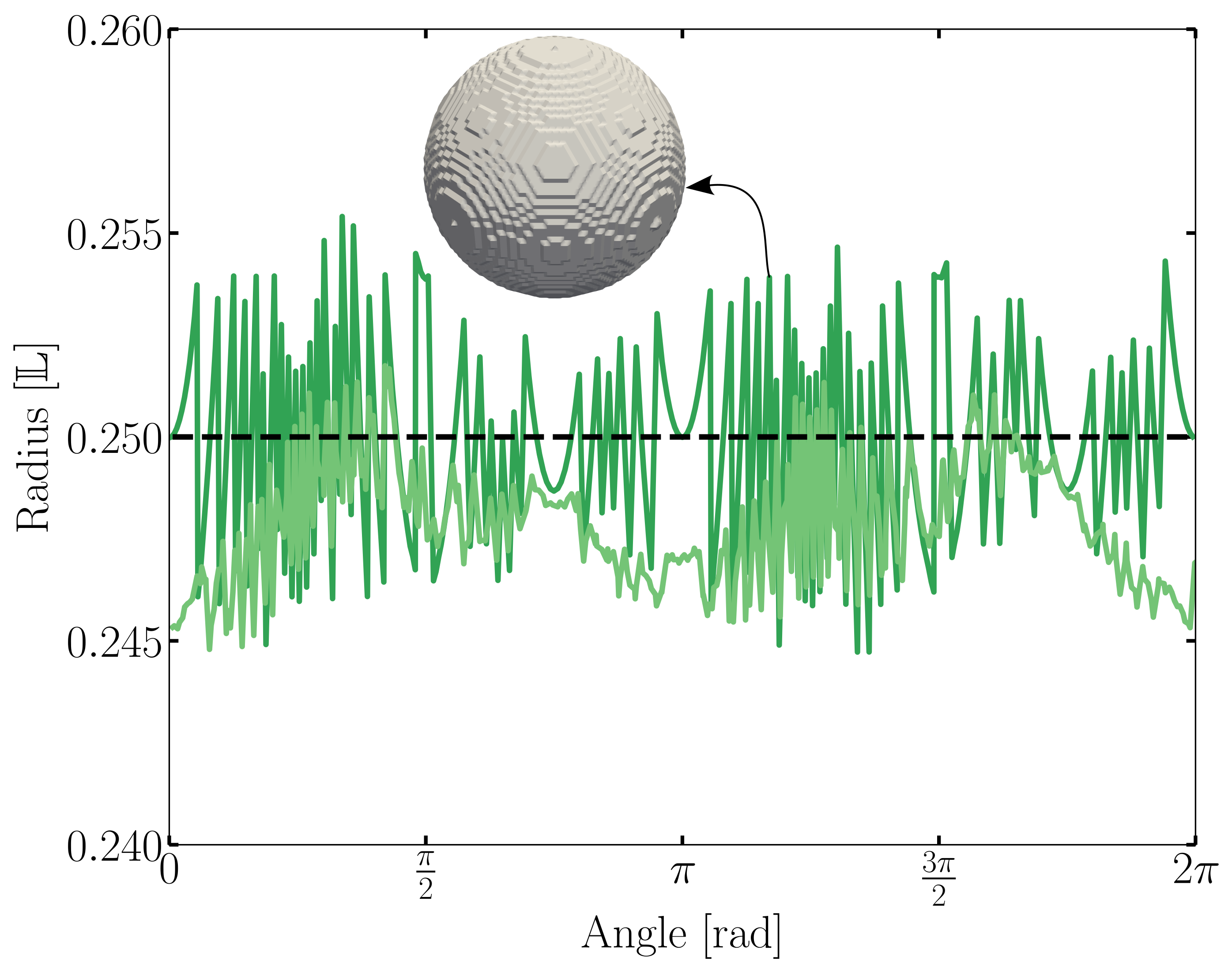}
        \end{minipage}
    \end{subfigure}
    \caption{Comparison of the final radius distribution in the $x$-$y$ plane with respect to the azimuthal angle for the different reinitialization methods at different reinitialization frequencies. On the left panel, the geometric reinitialization results capture expected shape for the lower frequencies. However, at the highest frequency, oscillations appear at the surface of the droplet. On the right panel, the PDE-based reinitialization results indicate a smooth quasi-spherical droplet. Lastly on the bottom panel, the projection-based results display a poor conservation of the spherical shape.}
    \label{fig::capillary-migration-radius}
\end{figure}

Figure \ref{fig::migration-velocity} presents the evolution of the migration velocity and Figure \ref{fig::capillary-migration-radius} displays the radius distribution in the $xy$-plane with respect to the azimuthal angle at the end of the simulation time. The results of geometric, \mbox{PDE-,} and projection-based reinitializations are respectively in orange, purple, and green. A darker shade of the color indicates a higher reinitialization frequency. All figures for the results use the same color scheme for all plots. Overlays of the iso-surface $\phi=0.5$ give an insight of the overall shape of the droplet at the end of the simulation.

The PDE-based and the geometric reinitialization result in migration velocities in good agreement with the reference value for all frequencies. Both methods leads to velocity variations due to the reinitialization frequency at $f=1/(100\Delta t)$. It highlights the importance of the reinitialization in this case: as the phase indicator diffuses between two reinitialization steps, the bubble slows down. When applied, the reinitialization method recovers a sharper indicator field and the bubble tends toward the expected migration velocity. \commentblue{The migration velocity obtained with the geometric reinitialization is less sensitive to the frequency than the PDE-based results.}

The final radius distribution for the PDE-based method, presented in the right plot of Figure \ref{fig::capillary-migration-radius}, oscillates smoothly around a value close to the expected sphere radius for all frequencies. The geometric method (Figure \ref{fig::capillary-migration-radius}, left plot) results in the same behavior for $f \in \{1/(10\Delta t),1/(100\Delta t)\}$ \commentblue{with oscillations of smaller amplitude}, while the reinitialization frequency of $1/\Delta t$ leads to sharper oscillations of similar amplitude as the PDE-based results. Additionally, these variations are asymmetric with respect to the front and tail of the droplet. 
For this frequency ($1/\Delta t$), the reinitialization CFL is approximately $0.01$ for the PDE-based and geometric methods. It leads to a negative impact on the droplet shape for the geometric method, while it has no effect on the PDE-based results, as highlighted in Figure \ref{fig::capillary-migration-radius}. In this specific test case, the geometric method does not perform as well at low reinitialization CFL.

Figure \ref{fig::capillary-migration-radius} reveals that the projection-based reinitialization leads to strong oscillations of the radius and it fails to capture the migration of the droplet for all frequencies, as presented in Figure \ref{fig::migration-velocity}. 

\subsection{3D Rising Bubble Benchmark} \label{sec:rising-bubble}
The rising bubble benchmark originates from the work of \citet{Hysing2009} in 2D. The current application case focuses on the 3D extension of the benchmark proposed by \citet{adelsberger2014}.

\subsubsection{Description of the Case}
The benchmark simulates an incompressible bubble of radius $R=0.25\len$ rising in the center of a cuboid column due to buoyancy, as illustrated in Figure \ref{fig:rising-bubble-schematic}. The selected set of physical properties follows the one proposed by \citet{adelsberger2014}, denoted Case 1 in the reference work. Table \ref{tab::rising-drop-physical-properties} lists them along with the two relevant dimensionless numbers: the Reynolds number $\mathrm{Re}= {\rho_0 U_\mathrm{g} D_\mathrm{b}}/{\mu_0}$ and the Eötvös number ${\mathrm{Eo}}= {\rho_0 U_\mathrm{g}^2 D_\mathrm{b}}/{\gamma}$, where $U_\mathrm{g} = \sqrt{gD_\mathrm{b}}$ is the reference velocity and $D_\mathrm{b}=2R$ is the diameter of the bubble. 
\begin{figure}[h!]
    \centering
    \def\svgwidth{0.5\textwidth}
          \input{Figures/rising-bubble-schematic.tex}
    \caption{2D cross-section view in the x-y plane of the initial state of the rising bubble case}
    \label{fig:rising-bubble-schematic}
\end{figure}

\begin{table}[!htpb]
    \centering
    \caption{Physical properties of the rising bubble case}
    \begin{tabular}{lcc}
        \hline 
        \textbf{Parameters}& {\textbf{Dimensions}}& \textbf{Values} \\ \hline
        {Reynolds number (${\mathrm{Re}}$)} & {-} & {35} \\
        {Eötvös number (${\mathrm{Eo}}$)} & {-} & {10} \\
        {Density ratio ($\rho_\mathrm{0}/\rho_\mathrm{1}$)} & {-} & {10} \\
        
        Density of fluid 0 ($\rho_0$) & $\m\len^{-3}$&  1000 \\

        {Dynamic viscosity ratio ($\mu_\mathrm{0}/\mu_\mathrm{1}$)} & {-} & {10} \\
        Dynamic viscosity of fluid 0 ($\mu_0$) & $\mathbb{M}\mathbb{L}^{-1}\mathbb{T}^{-1}$&  10 \\

        ST coefficient ($\gamma$) & $\mathbb{M}\mathbb{T}^{-2}$ & 24.5 \\ 
        {Gravitational acceleration magnitude ($g$)} & $\mathbb{L}\mathbb{T}^{-2}$ & {0.98} \\ \hline   
    \end{tabular}
    \label{tab::rising-drop-physical-properties}
\end{table}

        



The problem formulation imposes all walls to no-slip conditions for the NS equations \cite{Turek2019,adelsberger2014} and no-flux for the phase indicator transport. Both fluids are initially at rest. The initial condition of the phase indicator follows Equations \eqref{eq:initial-condition} and \eqref{eq:initial-condition-d} where $\vm{x}_0$ corresponds to the center of the sphere of radius $R$ illustrated in Figure \ref{fig:rising-bubble-schematic}.
The case considers a rising time of $3\ti$.

\subsubsection{Metrics of Interest} \label{sec:rising-bubble-metrics}
The metrics of interest are the barycenter height and the rise velocity along the rising axis ($y$-axis), based on the benchmark of \citet{adelsberger2014}. The current work considers the following definitions:
\begin{align}
    \text{Barycenter height} &= \frac{\int_{\Omega}y\hat{H}_\Gamma d\Omega}{\int_{\Omega}\hat{H}_\Gamma d\Omega} \label{eq:barycenter}\\
    \text{Rise velocity} &= \frac{\int_{\Omega}{u_y}\hat{H}_\Gamma d\Omega}{\int_{\Omega}\hat{H}_\Gamma d\Omega}  \label{eq:rising-velocity}
\end{align}
where $u_y$ is the $y$ component of the velocity vector $\vm{u}$. The rise velocity corresponds to the mean velocity in the $y$ direction inside the bubble.

This case also monitors the relative volume evolution $V/V_0$ and the sphericity ${A_\mathrm{eq}}/{A_\Gamma}$ of the bubble, computed according to:
\begin{align}
    \frac{V}{V_0} &=\frac{\int_{\Omega_\mathrm{1}}d\Omega}{\left . \int_{\Omega_\mathrm{1}}d\Omega\right|_{t=0}} \label{eq:volume}\\
    \frac{A_\mathrm{eq}}{A_\Gamma} &= \frac{\pi^{\frac{1}{3}}(6V)^{\frac{2}3{}}}{\int_{\Gamma}dS} \label{eq:sphericity}
\end{align}
where $V$ is the volume at the time $t$, $V_0$ is the initial volume, $A_\mathrm{eq}$ is the area of a sphere with a volume equivalent to $V$ and $A_\Gamma = \int_{\Gamma}dS$ is the surface area of the bubble. 

The work of \citet{adelsberger2014} provides reference solutions for the barycenter height, rise velocity, and sphericity from three different solvers: DROPS, NaSt3DGPF, and OpenFoam. \citet{Turek2019} also report simulation results of the solver FeatFlow for the rise velocity and sphericity. Table \ref{tab:reference-codes} presents a summary of the methods used in the reference solvers.

\begin{landscape}
\begin{table}[]
\caption{Reference solvers specifications for the rising bubble case, based on \cite{adelsberger2014}.}
\label{tab:reference-codes}
\footnotesize

\begin{tabular}{M{25pt}|M{120pt}|M{75pt}|M{78pt}|M{50pt}|l|M{50pt}|} 
\cline{2-7} 
 &
  \textbf{Developer} &
  \textbf{Spatial \mbox{discretization}} &
  \textbf{Time \mbox{discretization}} &
  \textbf{Interface solver}&
  \textbf{STF modeling} & \textbf{Reinitia-lization method}\\ \hline
\multicolumn{1}{|l|}{\begin{tabular}[c]{@{}l@{}}\textbf{DROPS}\\ \cite{adelsberger2014, gross2011, drops} \end{tabular}} &
  \begin{tabular}[c]{@{}l@{}}Institute of Geometry\\ and Applied Mathematics\\ (IGPM),\\ RWTH Aachen\end{tabular} &
  XFEM &
  Implicit $\theta$-scheme &
  LS &
  \begin{tabular}[c]{@{}l@{}}- CSF\\ - Curvature approximated \\ with weak formulation of\\ Laplace-Beltrami-Operator\end{tabular} &
  Fast-marching method\\  \hline
\multicolumn{1}{|l|}{\begin{tabular}[c]{@{}l@{}}\textbf{FeatFlow}\\ \cite{Turek2019,turek1998}\end{tabular}} &
  \begin{tabular}[c]{@{}l@{}}Institute of Applied\\ Mathematics (Chair III),\\ TU Dortmund\end{tabular} &
  FEM &
  Not specified &
  LS &
  \begin{tabular}[c]{@{}l@{}}- Semi-implicit treatment\\ - Laplace-Beltrami\\ transformation\end{tabular} &
   Not \mbox{specified}\\  \hline
\multicolumn{1}{|l|}{\begin{tabular}[c]{@{}l@{}}\textbf{NaSt3DGPF}\\ \cite{adelsberger2014, croce2010, engel_nast3dgpf}\end{tabular}} &
  \begin{tabular}[c]{@{}l@{}}Institute for Numerical\\ Simulation (INS),\\ University of Bonn\end{tabular} &
  Finite differences &
  2nd order Adams-Bashforth scheme&
  LS &
  CSF &
  PDE-based\\  \hline
\multicolumn{1}{|l|}{\begin{tabular}[c]{@{}l@{}}\textbf{OpenFOAM}\\ \cite{adelsberger2014, scheufler2023}\end{tabular}} &
  Multiple contributors (open source) &
  Finite volumes &
  Implicit Euler scheme&
  VOF &
  CSF &
  None\\ \hline
\end{tabular}
\end{table}
\end{landscape}

\subsubsection{Simulation Settings}
For each reinitialization method, the study investigates three reinitialization frequencies: $f=1/(N_\text{reinit}\Delta t)$, with $N_\text{reinit} \in \{1,10,100\}$. 

An adaptively refined Cartesian grid discretizes the domain. The minimum and maximum element sizes are $\SI{7.81e-3}{}\len$ and $\SI{6.25e-2}{}\len$ respectively. The selected time-step respects the capillary time step limit defined in Equation \eqref{eq::capillary-time-step}. For this case, it results in a constant time step of $\Delta t = \SI{1.3e-3}\ti$. Hence, the CFL is not constant, and stabilizes at the end of the simulation at {averaged values of 0.082, 0.069, and 0.075} for the geometric, PDE-based, and projection-based methods, respectively.

\commentblue{The geometric and projection-based methods use the same specific parameters as in the capillary migration case. For the PDE-based approach, the value of the interface thickness measure $\varepsilon$ changes, yielding to better results in terms of the shape, volume conservation, sphericity and flow-based quantities, than if the same value as in the capillary migration case is used. It highlights that the parameters selection for the PDE-based approach is case dependent.}

\subsubsection{Results}
\begin{figure}
    \centering
    \begin{subfigure}[b]{1\textwidth}
        \centering
        \begin{minipage}{0.32\textwidth}
            \centering
            \vspace{25pt}
            \includegraphics[width=1\textwidth,trim={12cm 0pt 12cm 2cm},clip]{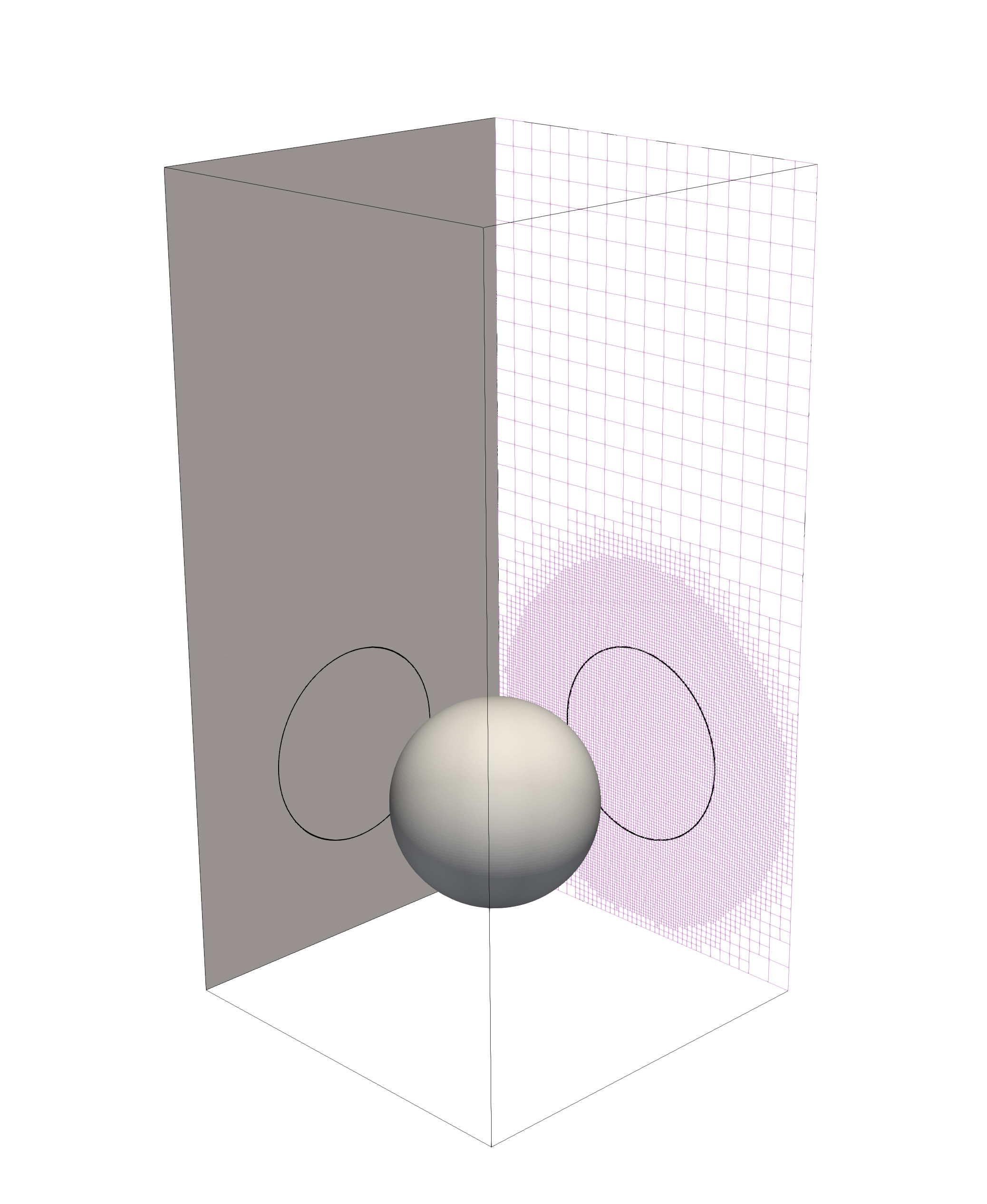}
            \caption{$t=0.0\mathbb{T}$}
        \end{minipage}
        \hfill
        \begin{minipage}{0.32\textwidth}
            \centering
            \begin{tikzpicture}
                \node[inner sep=0pt] at (0,0) {\includegraphics[width=1\textwidth,trim={12cm 0pt 12cm 2cm},clip]{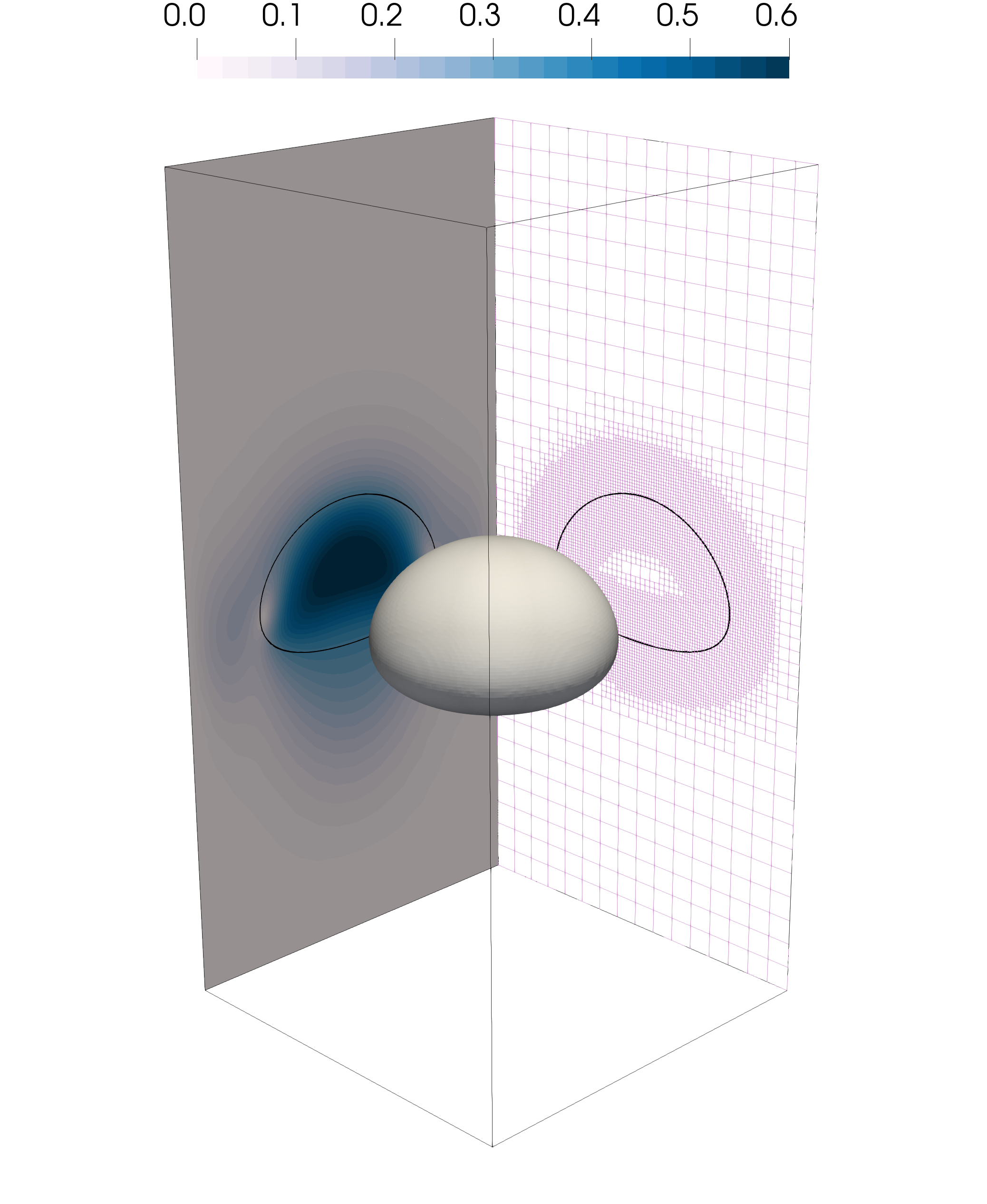}};
                \node[inner sep=0pt] at (-1.95,4) {\footnotesize 0.0};
                \node[inner sep=0pt] at (-1.3,4) {\footnotesize 0.1};
                \node[inner sep=0pt] at (-0.65,4) {\footnotesize 0.2};
                \node[inner sep=0pt] at (0,4) {\footnotesize 0.3};
                \node[inner sep=0pt] at (0.65,4) {\footnotesize 0.4};
                \node[inner sep=0pt] at (1.3,4) {\footnotesize 0.5};
                \node[inner sep=0pt] at (1.95,4) {\footnotesize 0.6};
                \node[inner sep=0pt] at (0,4.5) {\footnotesize Velocity $[\mathbb{LT}^{-1}]$};
            \end{tikzpicture}
            \caption{$t=1.5\mathbb{T}$}
        \end{minipage}
        \hfill
        \begin{minipage}{0.32\textwidth}
            \centering
            \vspace{25pt}
            \includegraphics[width=1\textwidth,trim={12cm 0pt 12cm 2cm},clip]{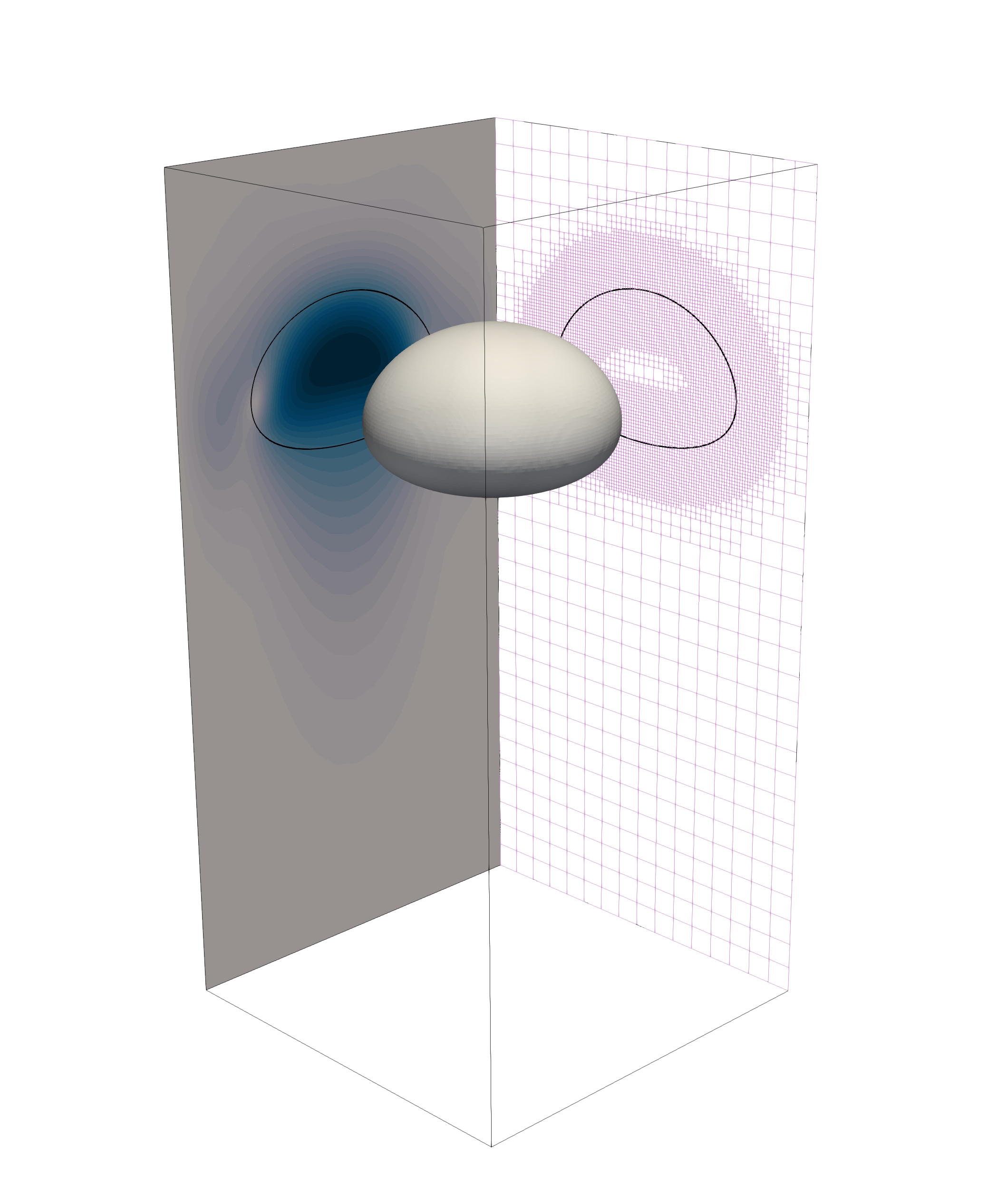}
            \caption{$t=3.0\mathbb{T}$}
        \end{minipage}
    \end{subfigure}
    \caption{Snapshots of the bubble shape, velocity field and mesh for the geometric reinitialization applied at each time step. The velocity field corresponds to the velocity in the $y$-$z$ plane passing through the center of the bubble. The adaptively refined mesh in the $x$-$z$ plane passing through the center of the bubble is shown on the right part of each panel in light pink.}
    \label{fig:bubble-snapshots}
\end{figure}

Figure \ref{fig:bubble-snapshots} shows snapshots of the bubble shape, velocity field, and mesh, obtained with the geometric method at $f=1/\Delta t$. The right part of each snapshot shows the adaptively refined mesh, a new capability of the geometric method presented in this work, unlike in the original works of \citet{Mut2006} and \citet{Ausas2011}. 

\begin{figure}
    \centering
    \begin{subfigure}[b]{0.95\textwidth}
        \centering
        \includegraphics[width=0.525\textwidth]{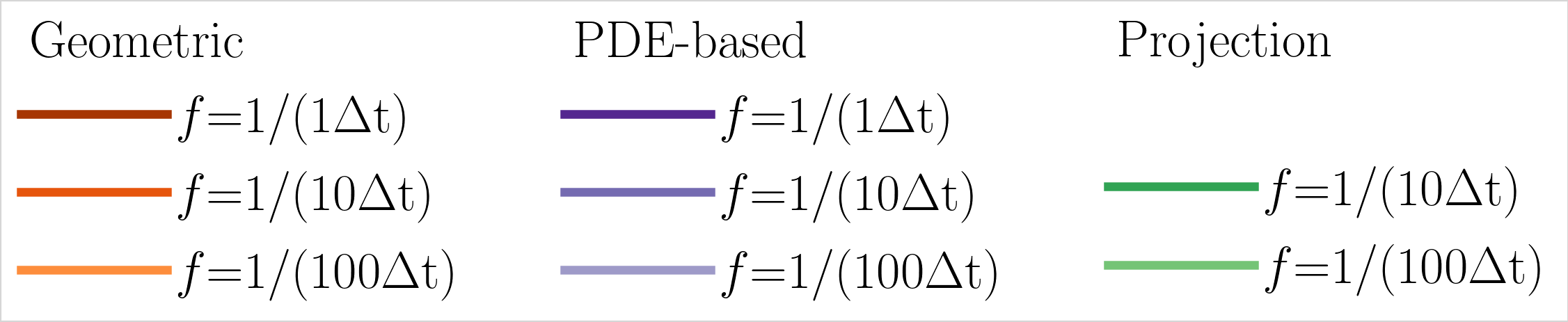}
    \end{subfigure}
    
    \begin{subfigure}[b]{1\textwidth}
        \centering
        \begin{minipage}{0.49\textwidth}
            \includegraphics[width=1\textwidth]{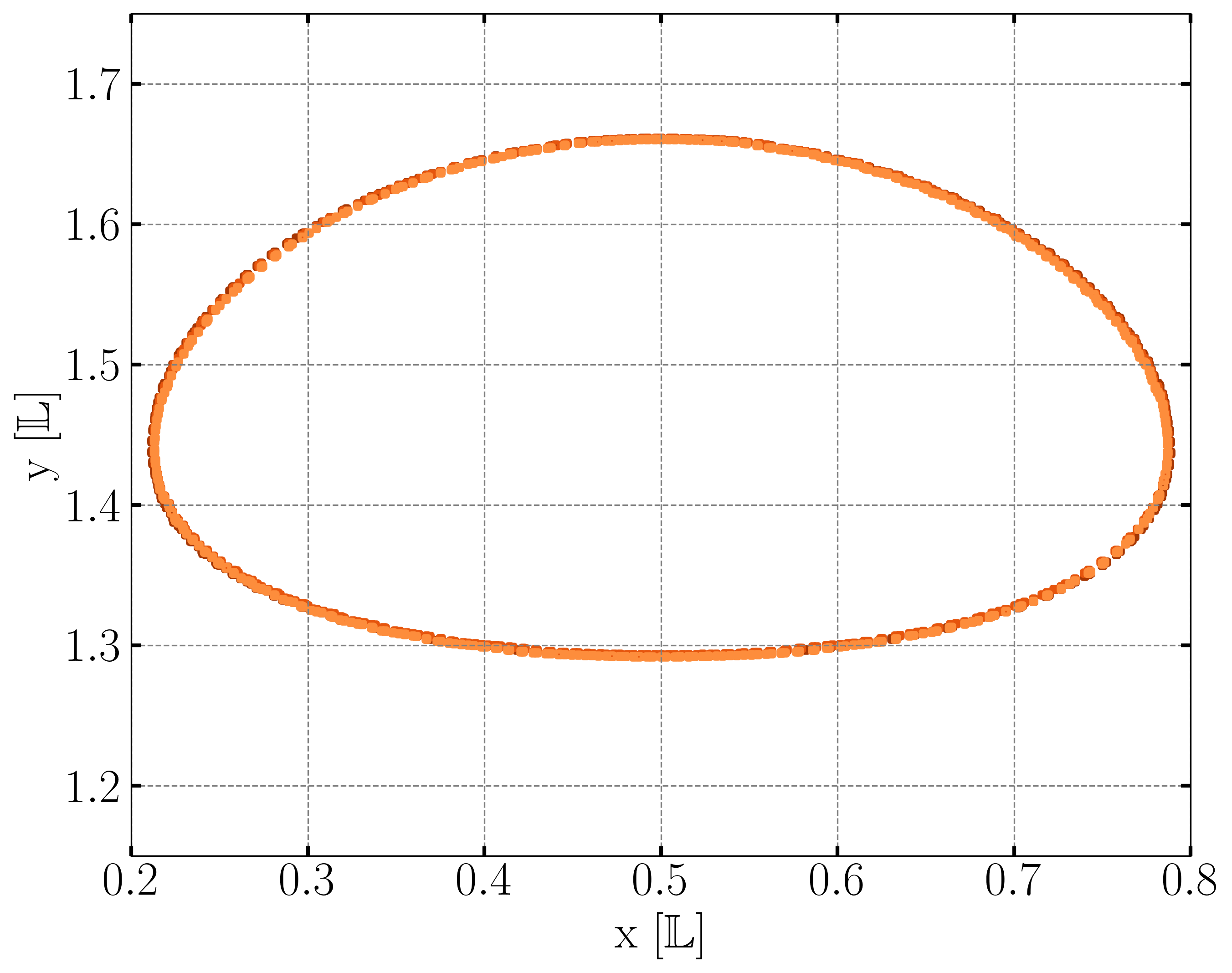}
        \end{minipage}
        \hfill
        \begin{minipage}{0.49\textwidth}
            \centering
            \includegraphics[width=1\textwidth]{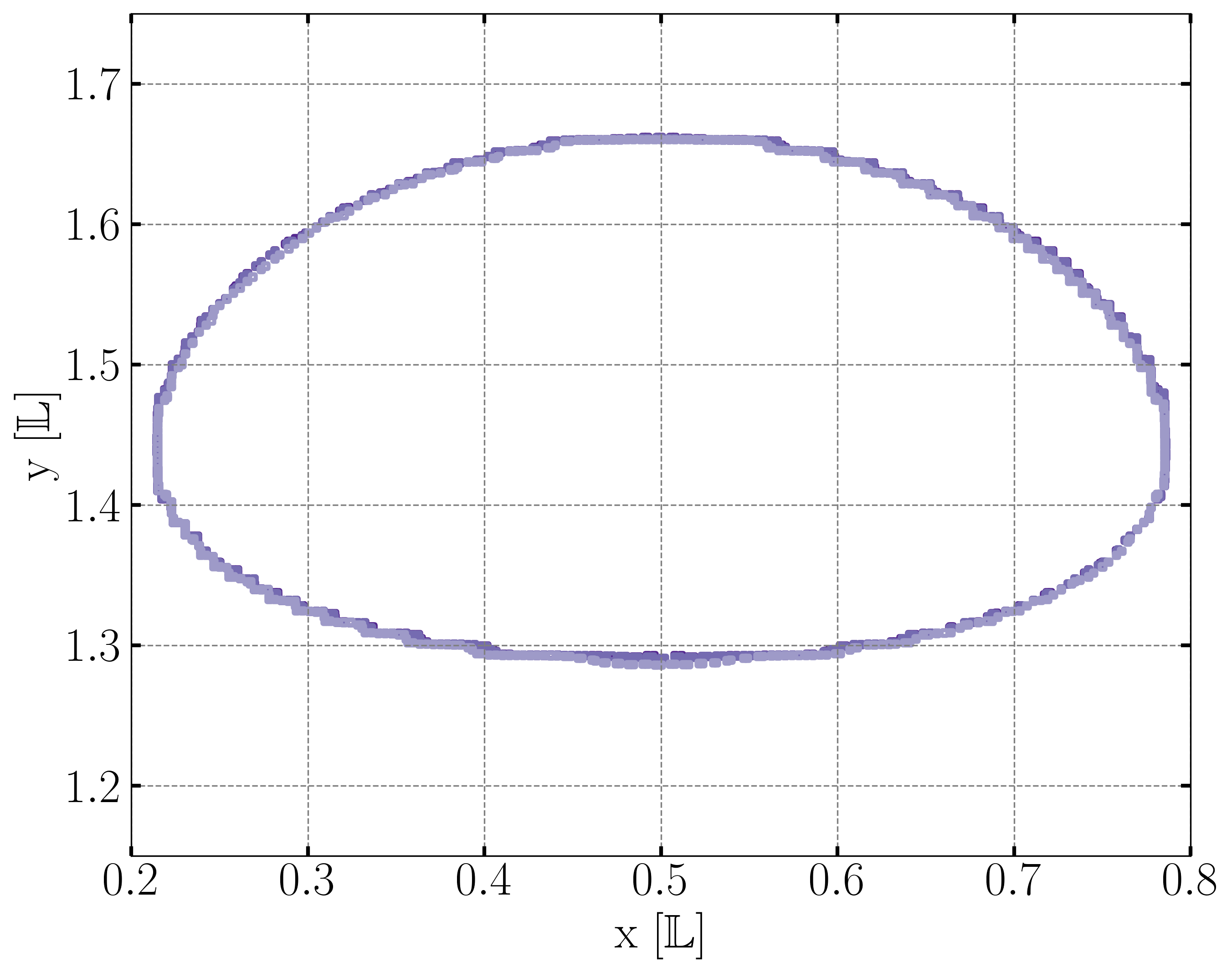}
        \end{minipage}
        \begin{minipage}{0.49\textwidth}
            \centering
            \includegraphics[width=1\textwidth]{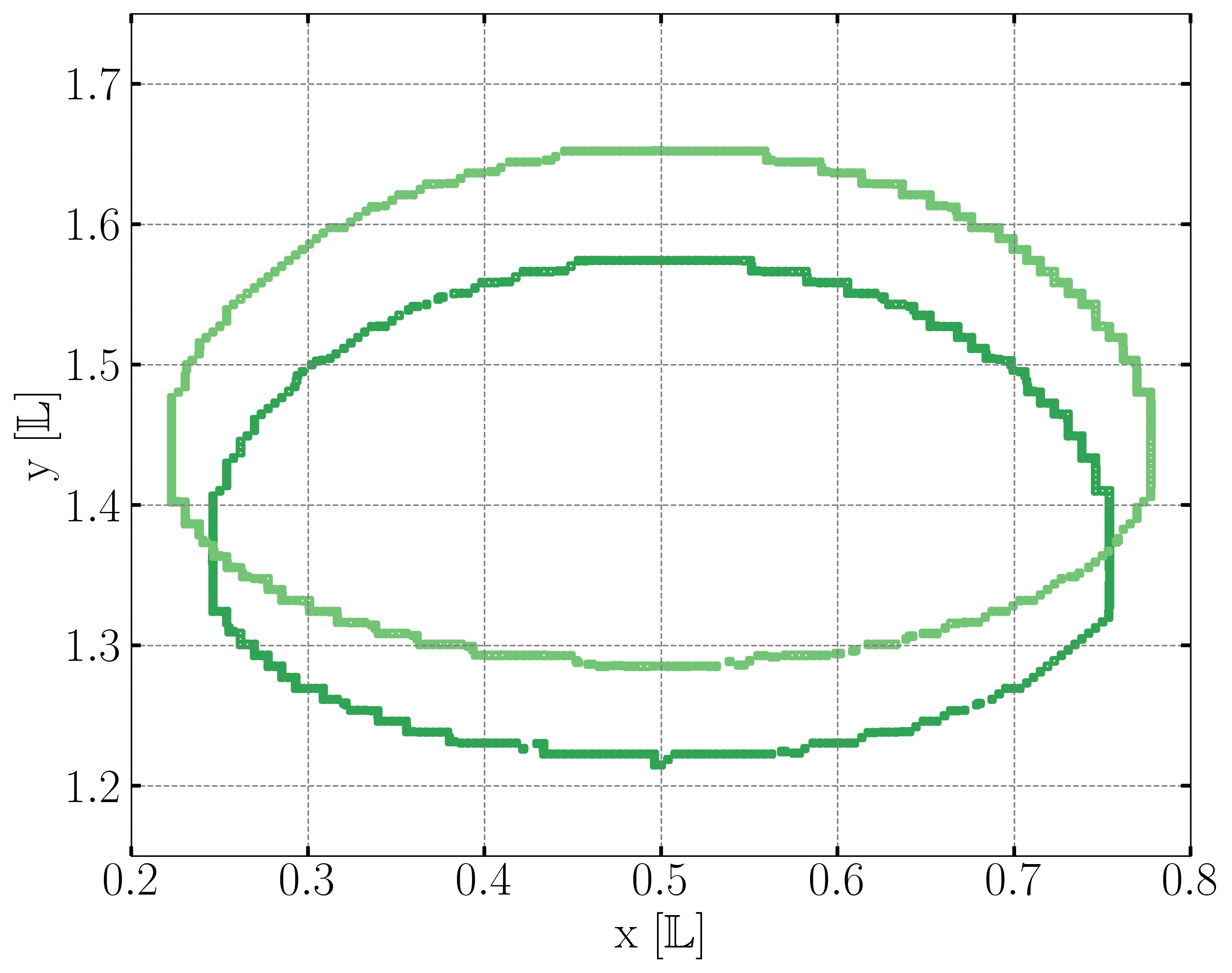}
        \end{minipage}
    \end{subfigure}
    \caption{Comparison of the final bubble interface contours for different reinitialization methods at different reinitialization frequencies. The contours are extracted from the slice along the $xy$-plane passing by the center of the domain. On the upper panels, the geometric (left) and PDE-based (right) methods display very little influence of the reinitialization frequency. In contrast, on the lower panel, the projection-based method results in two significantly different bubble shapes.}
    \label{fig:bubble-contours-case1}
\end{figure}

\begin{figure}[h!]

    \begin{subfigure}[b]{0.95\textwidth}
        \centering
        \includegraphics[width=0.7\textwidth]{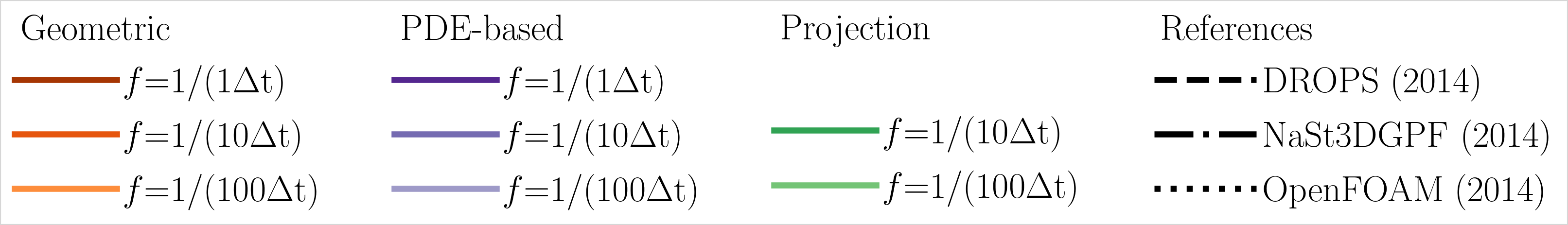}
    \end{subfigure}    
    \begin{subfigure}[b]{0.95\textwidth}
        \begin{minipage}{0.90\textwidth}
            \centering
            \includegraphics[width=1\textwidth]{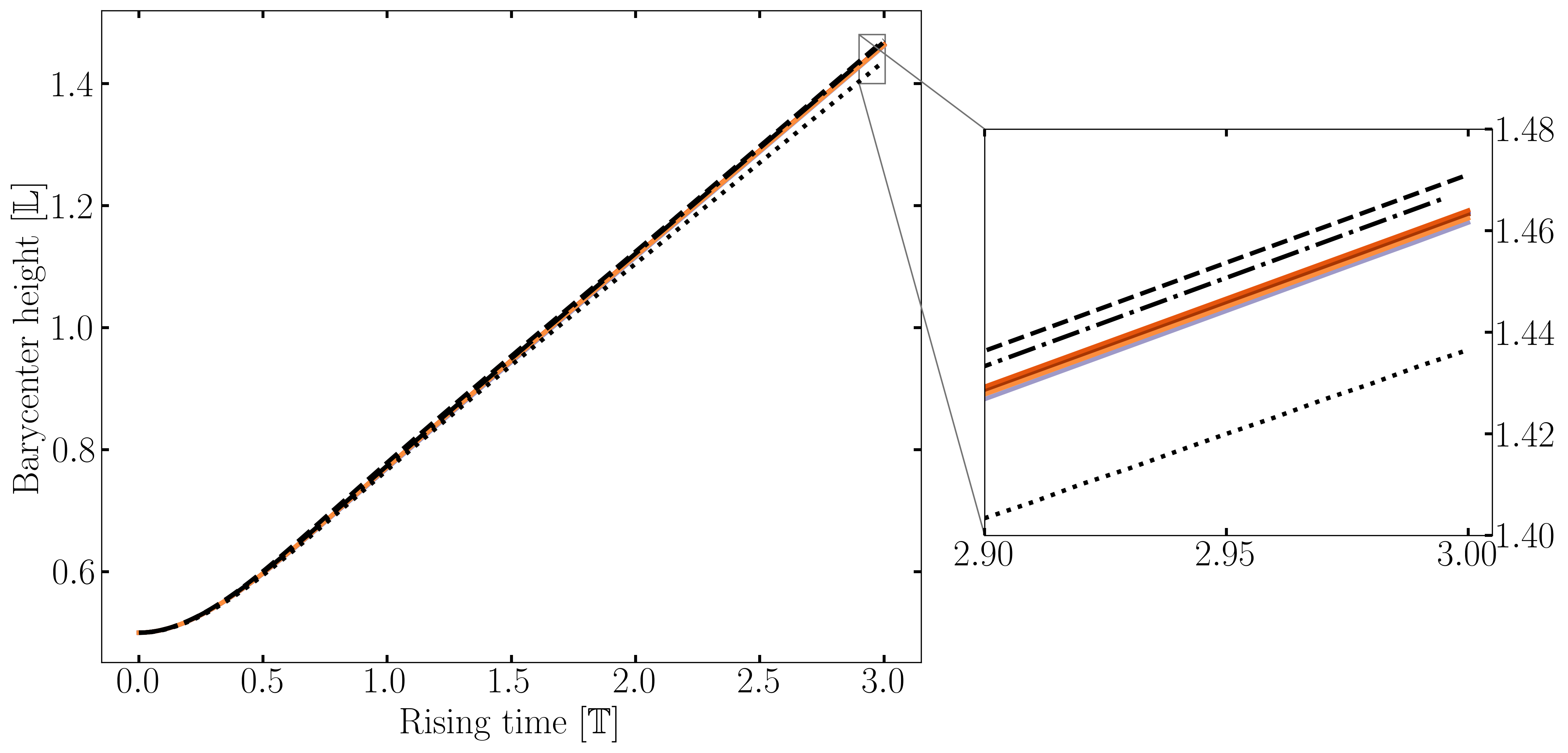}
        \end{minipage}
        \begin{minipage}{0.90\textwidth}
            \centering
            \includegraphics[width=1\textwidth]{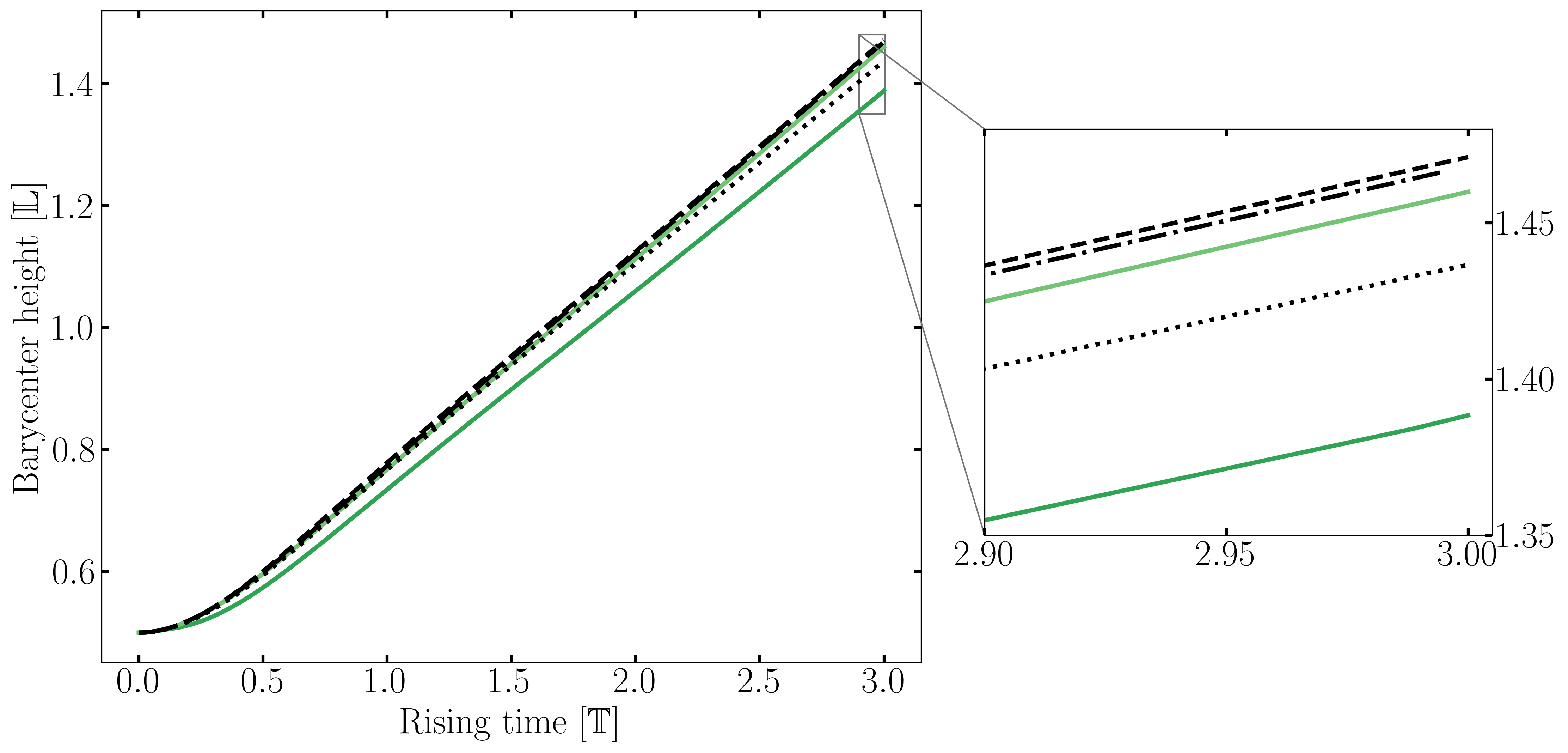}
        \end{minipage}
    \end{subfigure}
    \caption{Comparison of the barycenter evolution for different reinitialization methods at different reinitialization frequencies with reference values \cite{Turek2019, adelsberger2014}. On the top panel, geometric and PDE-based methods capture trends similar to references (with the exception of OpenFOAM \cite{adelsberger2014}). On bottom panel, projection-based results display a larger dependency to the reinitialization frequency.}
    \label{fig:barycenter-case1}
\end{figure}

\begin{figure}[h!]

    \begin{subfigure}[b]{0.95\textwidth}
        \centering
        \includegraphics[width=0.7\textwidth]{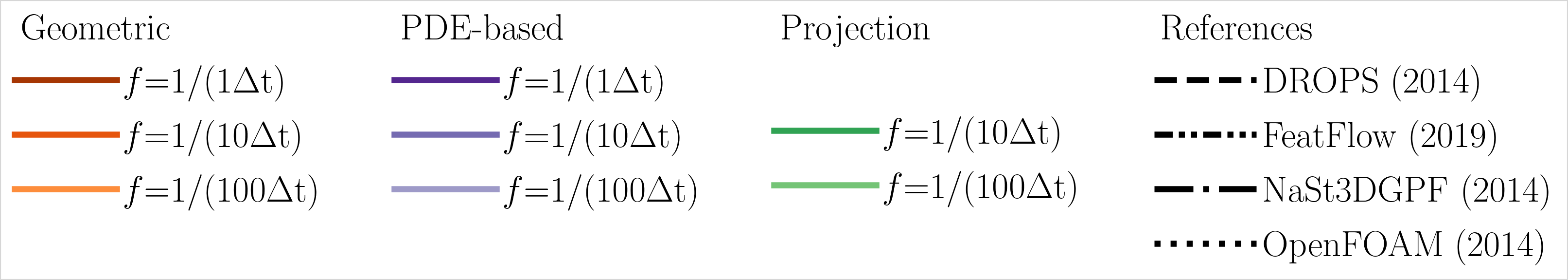}
    \end{subfigure}    
    \begin{subfigure}[b]{0.95\textwidth}
        \begin{minipage}{0.48\textwidth}
            \centering
            \includegraphics[width=1\textwidth]{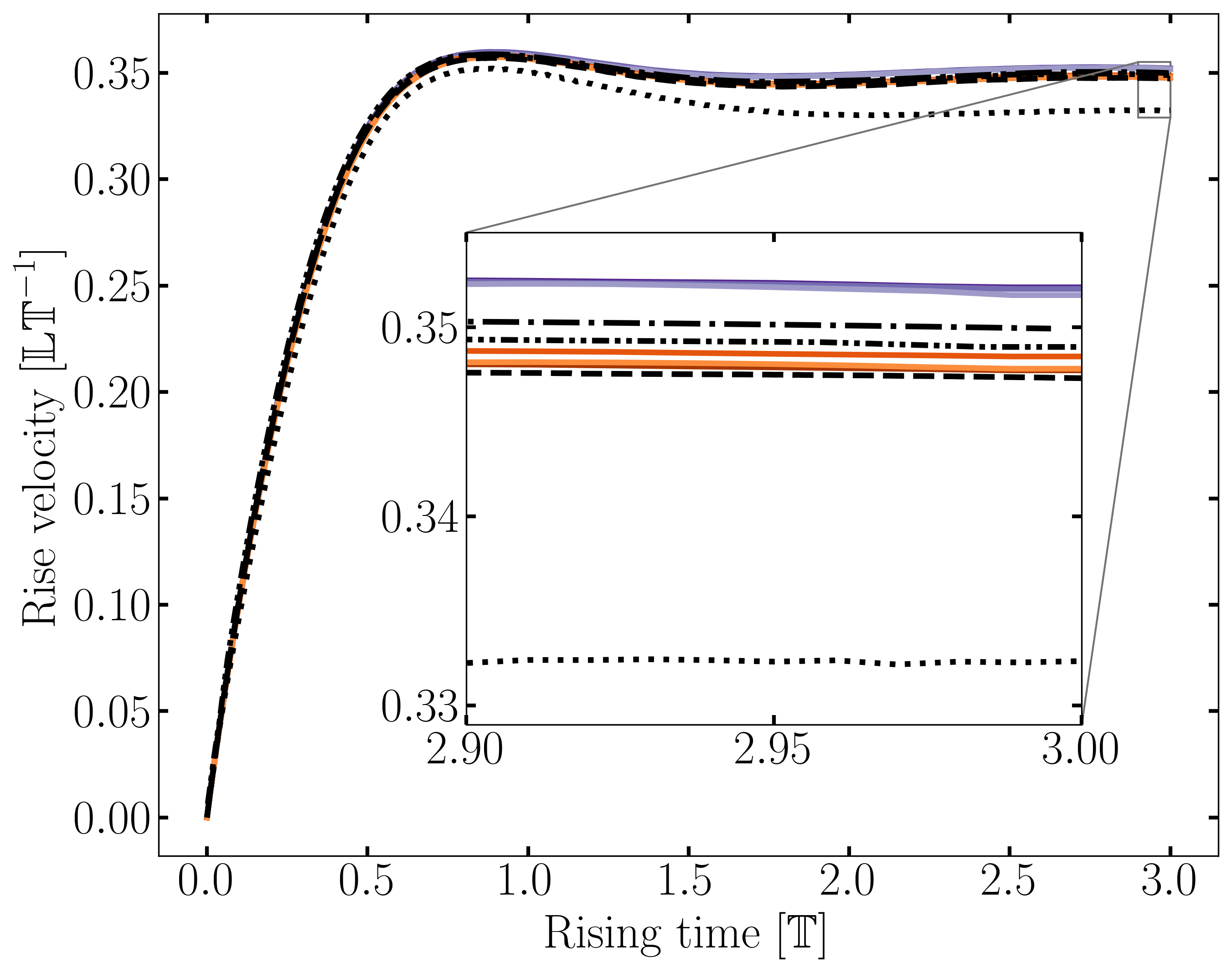}
        \end{minipage}
        \begin{minipage}{0.48\textwidth}
            \centering
            \includegraphics[width=1\textwidth]{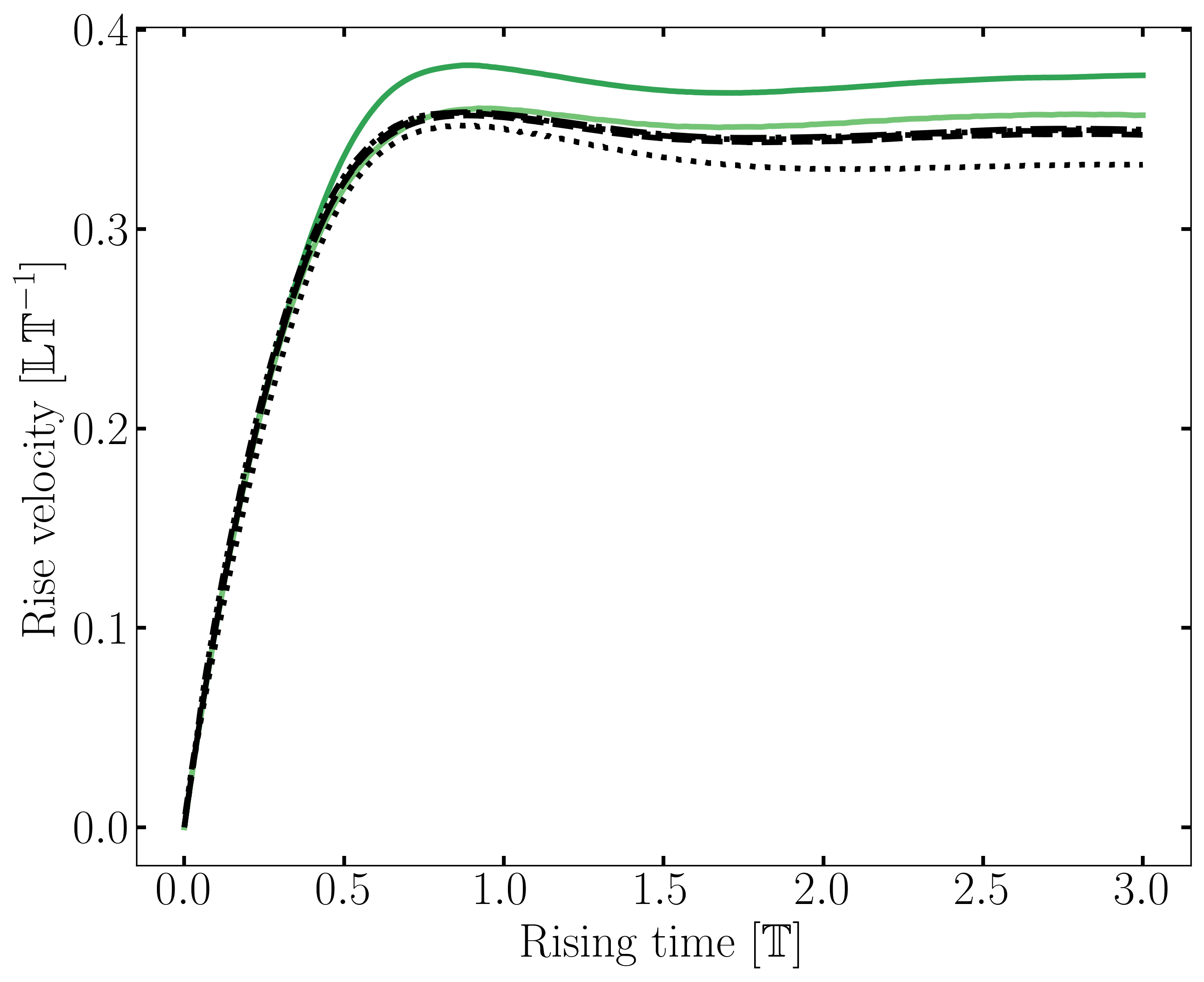}
        \end{minipage}
    \end{subfigure}
    \caption{Comparison of the rise velocity for different reinitialization methods at different reinitialization frequencies with reference values \cite{Turek2019, adelsberger2014}. The geometric and PDE-based methods (left panel) capture similar trends to those of the references (with the exception of OpenFOAM \cite{adelsberger2014}). In contrast, the projection based method (right panel), overestimates the rise velocity.}
    \label{fig:velocity-case1}
\end{figure}

\begin{figure}[h!]
    \centering
    \begin{subfigure}[b]{0.95\textwidth}
        \centering
        \includegraphics[width=0.525\textwidth]{Figures/rising-bubble-3d/legend_no_ref.png}
    \end{subfigure}
    \begin{subfigure}[b]{0.95\textwidth}
        \begin{minipage}{0.48\textwidth}
            \centering
            \includegraphics[width=1\textwidth]{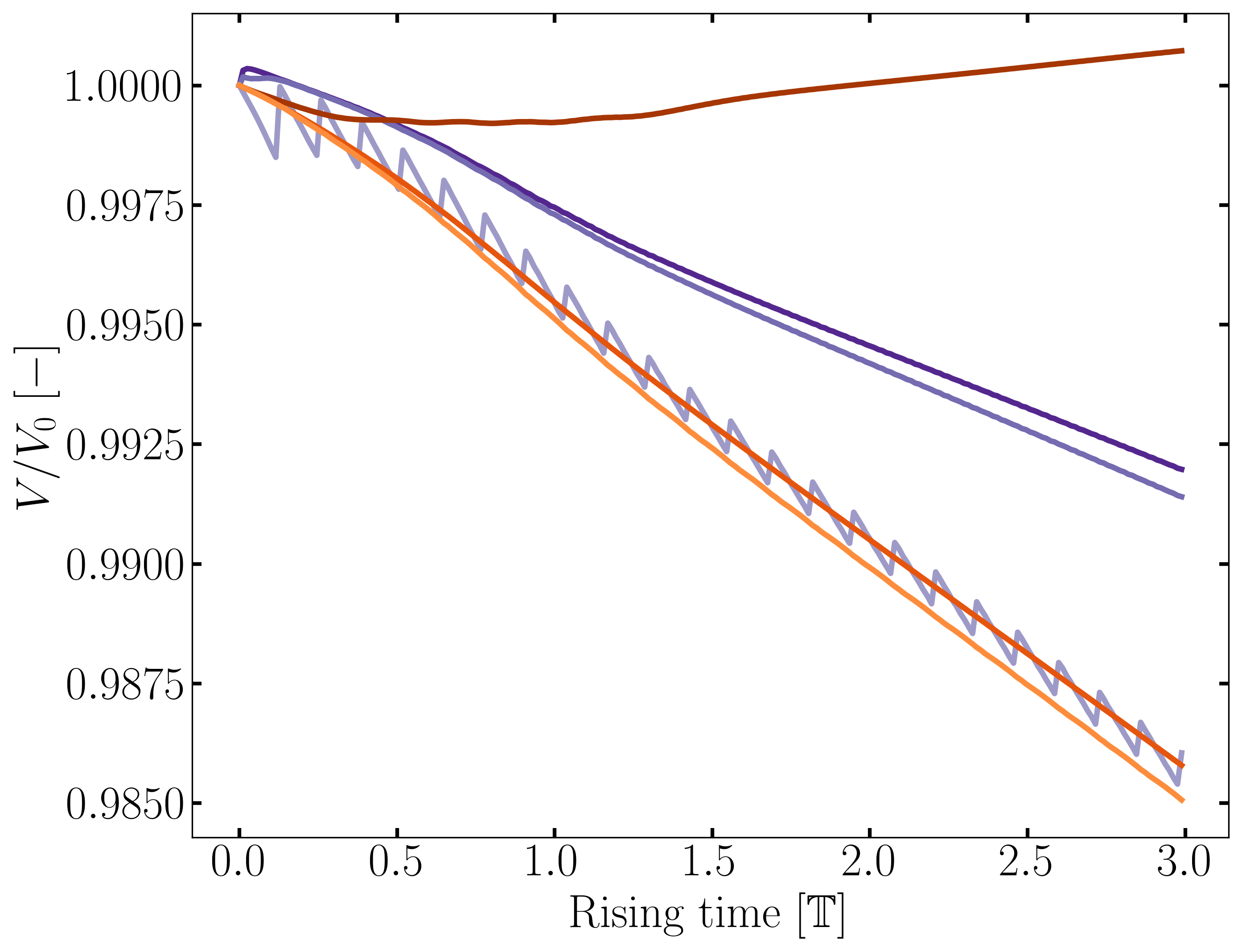}
        \end{minipage}
        \begin{minipage}{0.48\textwidth}
            \centering
            \includegraphics[width=1\textwidth]{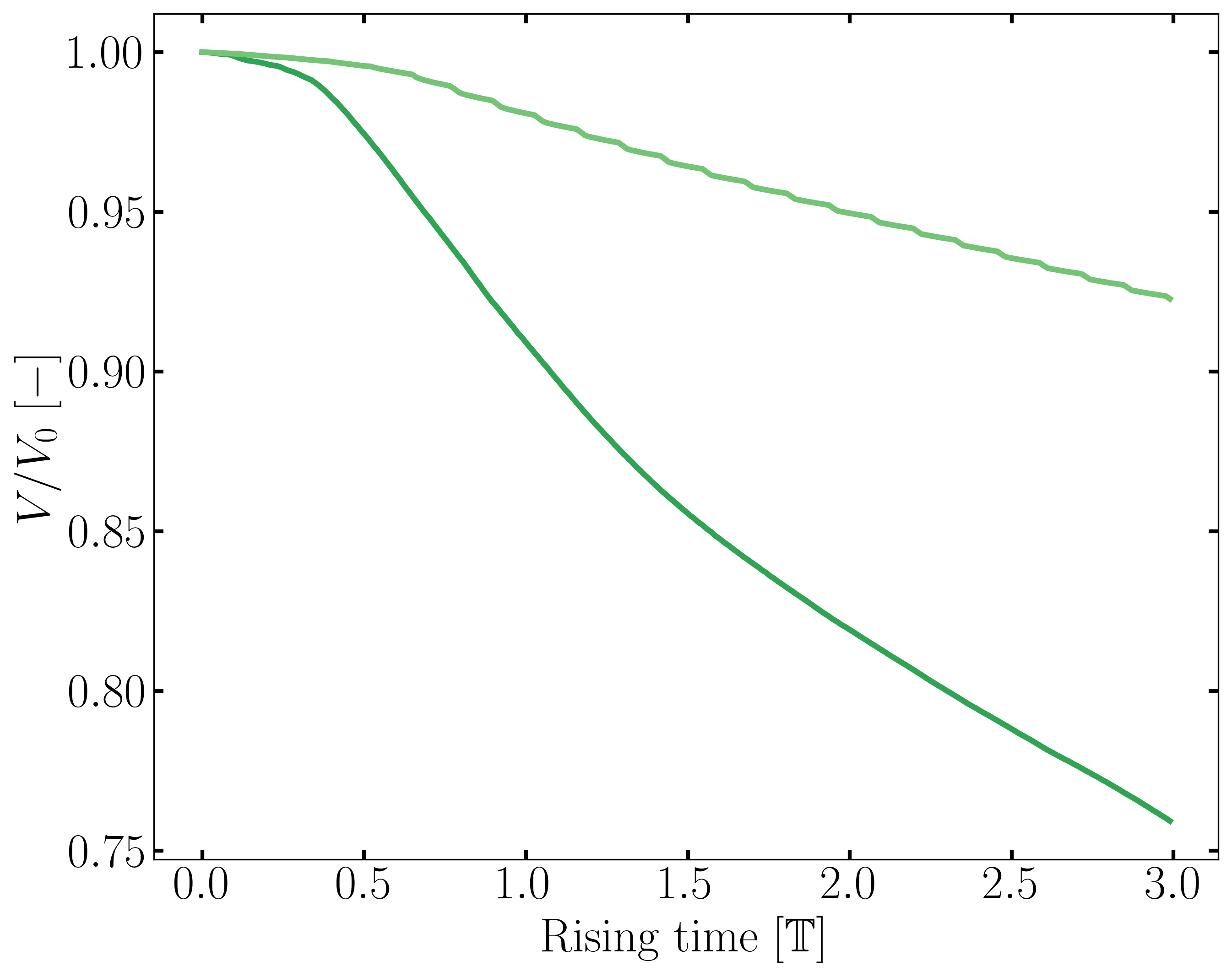}
        \end{minipage}
    \end{subfigure}
    \caption{Comparison of relative geometrical volume evolution for different reinitialization methods at different reinitialization frequencies. On the left panel, geometric and PDE-based reinitialization methods show good volumetric conservation (under 2\% loss). In contrast, projection-based reinitialization (right panel) displays poor volume conservation (more than 25\% loss for the highest frequency).}
    \label{fig:mass-case1}
\end{figure}

\begin{figure}[h!]
    \centering
    \begin{subfigure}[b]{0.95\textwidth}
        \centering
        \includegraphics[width=0.7\textwidth]{Figures/rising-bubble-3d/legend_with_featflow.png}
    \end{subfigure}
    \begin{subfigure}[b]{0.95\textwidth}
        \begin{minipage}{0.48\textwidth}
            \centering
            \includegraphics[width=1\textwidth]{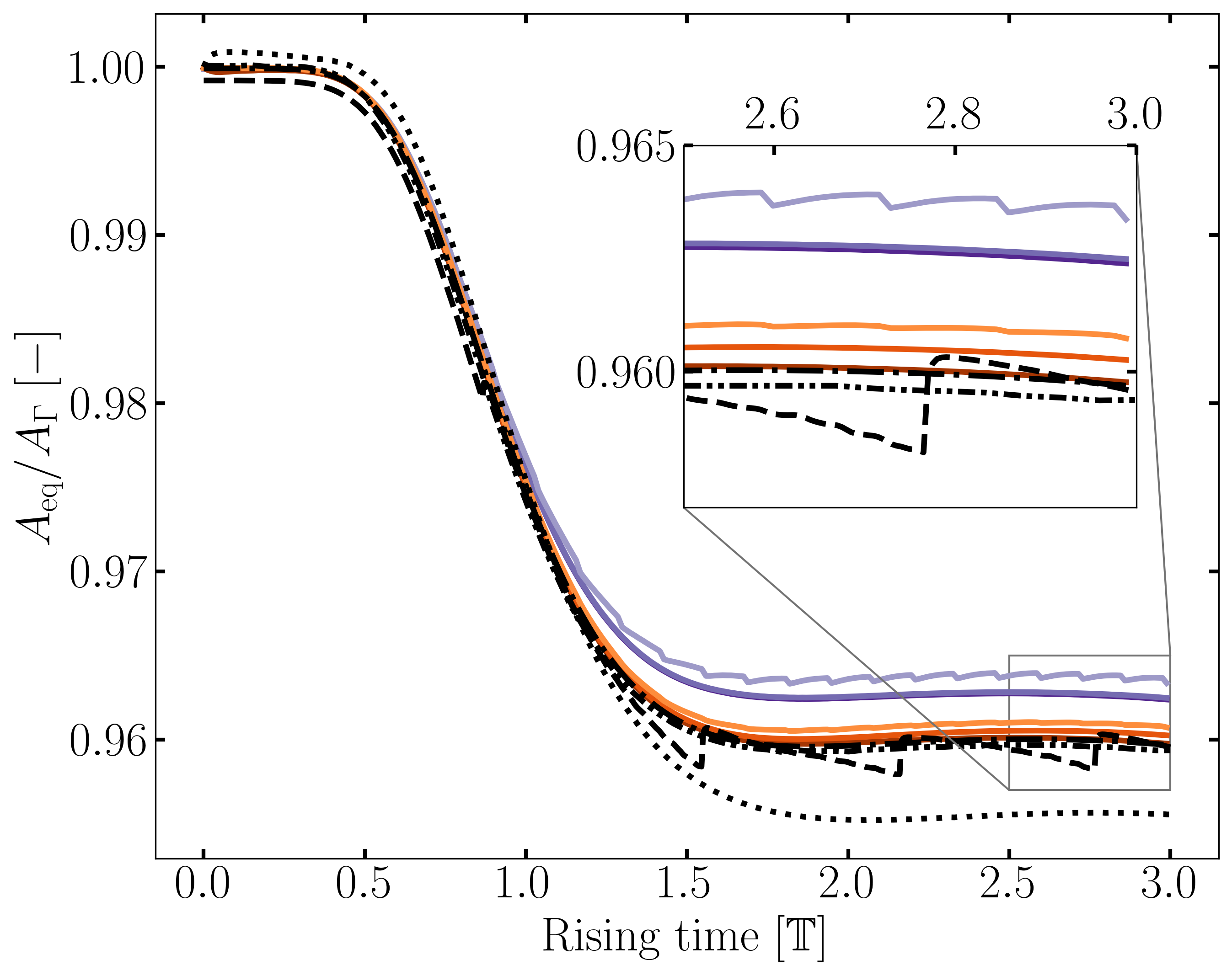}
        \end{minipage}
        \begin{minipage}{0.48\textwidth}
            \centering
            \includegraphics[width=1\textwidth]{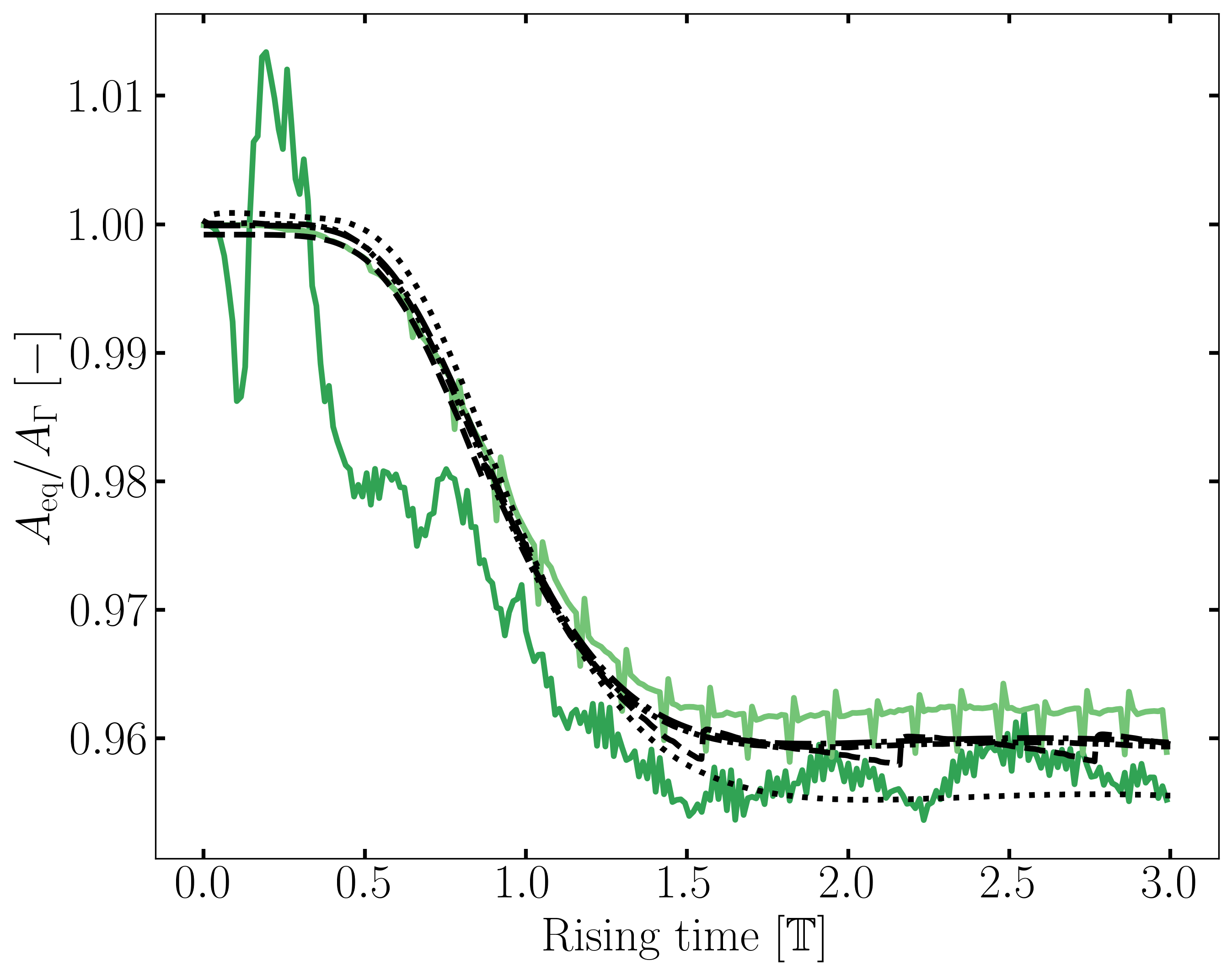}
        \end{minipage}
    \end{subfigure}
    \caption{Comparison of the sphericity evolution for the different reinitialization methods at different reinitialization frequencies with reference values \cite{Turek2019, adelsberger2014}. On the left panel, the geometric reinitialization results capture an evolution similar to references (with the exception of OpenFOAM \cite{adelsberger2014}) while, the PDE-based reinitialization results stabilize at a higher value. On the right panel, the projection-based reinitialization leads to nonphysical oscillations.}
    \label{fig:sphericity-case1}
\end{figure}

Figures \ref{fig:bubble-contours-case1} to \ref{fig:sphericity-case1} present the final bubble shape, as well as the time evolution of the barycenter height, rise velocity, volume conservation, and sphericity. The reference curves correspond to the results for the finest discretization reported in \cite{Turek2019,adelsberger2014}.

\paragraph{Bubble shape and barycenter position}
Figure \ref{fig:bubble-contours-case1} shows the bubble contour in the $xy$-plane at the end of the simulation time for each method. 

For the geometric approach, the reinitialization frequency has no significant effect on the bubble shape and position: for all frequencies, the contours are superposed.  The evolution of the barycenter height showed in Figure \ref{fig:barycenter-case1} agrees well the reference results of DROPS and NaStDGPF.

The PDE-based reinitialization method leads to the same results quality as the geometric approach. However, the shapes obtained with the PDE-based reinitialization follow a staircase pattern. The latter is caused by the choice of a small value of $\varepsilon=h$ .

The projection-based method shows the strongest effect of the reinitialization frequency on all the metrics of interest. Only the results for the frequencies of $1/(10\Delta t)$ and $1/(100\Delta t)$ are presented because the solver fails with a reinitialization frequency of $1/\Delta t$. The bubble contours in Figure \ref{fig:bubble-contours-case1} show a strong staircase pattern for both frequencies, with significant differences in size and position. The corresponding barycenter evolution presented in Figure \ref{fig:barycenter-case1} for $f=1/(10\Delta t)$ is not comparable to the reference curves, while the results for $f=1/(100\Delta t)$ are in better agreement.

\paragraph{Rise velocity}
Figure \ref{fig:velocity-case1} shows the evolution in time of the rise velocity. For the geometric and PDE-based approaches, the rise velocity evolution agrees well with the reference curves, especially to the ones of DROPS, NaSt3DGPF, and FeatFlow. There is no significant effect of the frequency. The projection-based results for the lowest frequency of $f=1/(100\Delta t)$ are in better agreement with the reference values than the ones with a frequency of $1/(10\Delta t)$.

\paragraph{Relative volume evolution} The relative volume evolution is presented in Figure \ref{fig:mass-case1}.
For the geometric method, the relative volume evolution remains close to $1$ for $f=1/\Delta t$, while the results for $f\in\{1/(10\Delta t), 1/(100\Delta t)\}$ present a volume loss of up to approximately $2\%$. The PDE-based reinitialization results in a volume loss of less than $1\%$ for $f=1/(\Delta t)$. At the lowest frequency of $f=1/(100\Delta t)$, the volume loss is approximately $\approx1.5\%$. However, at this frequency, the evolution features jumps at regular time intervals corresponding to the application of the reinitialization. This variation in the volume indicates that the interface moves at each reinitialization step.
The projection-based results reveal a significant volume loss: the final bubble volume is approximately $75\%$ of its initial volume for $f=1/(10\Delta t)$ and $92\%$ for $f=1/(100\Delta t)$. 

For all methods, there is no steady-state reached for the volume: it keeps decreasing for all frequencies, even when the rise velocity in Figure \ref{fig:velocity-case1} reaches a constant value. However, for the geometric method at $f=1/\Delta t$, the rate of volume loss is significantly lower than all other methods.

\paragraph{Sphericity}
Figure \ref{fig:sphericity-case1} shows the evolution in time of the sphericity. For the geometric reinitialization, the evolution is smooth, indicating that the method does not alter significantly the interfacial dynamics. Additionally, the geometric results are in better agreement with the reference data, compared to the PDE-based method. For the latter, the sphericity is higher than the reference values, indicating that the shape of the bubble is closer to a sphere. Additionally, it features jumps resulting from the reinitialization frequency of $f=1/(100\Delta t)$. It also is consistent with the observation that PDE-based method moves the interface.

For the projection-based method, large oscillations dominate the sphericity evolution, and the results are not comparable to the reference curves. The behavior of the sphericity is assumed to be linked to the staircase shape reported in Figure \ref{fig:bubble-contours-case1}, where the interface follows the underlying Cartesian mesh. For the lower frequency of $f=1/(100\Delta t)$, the sphericity evolution is in the expected range, despite the presence of smaller oscillations.

\clearpage
\subsection{3D Rayleigh-Plateau Instability} \label{sec:rayleigh-plateau}
The Rayleigh-Plateau instability is a case where the ST force has a destabilizing effect, in opposition to the two previous applications. For a capillary liquid jet, the instability acts to reduce the surface energy of the jet, leading to its breakup into droplets~\cite{Denner2022}. It is a 3D case with complex interfacial deformations and the current work uses it to assess the spatial convergence of the reinitialization methods. Due to the poor performances of the projection-based method in the previous cases and this one, this study focuses only on the PDE-based and geometric approaches.

\subsubsection{Case Description}
\begin{figure}[h!]
    \centering
    \def\svgwidth{0.9\textwidth}
          \input{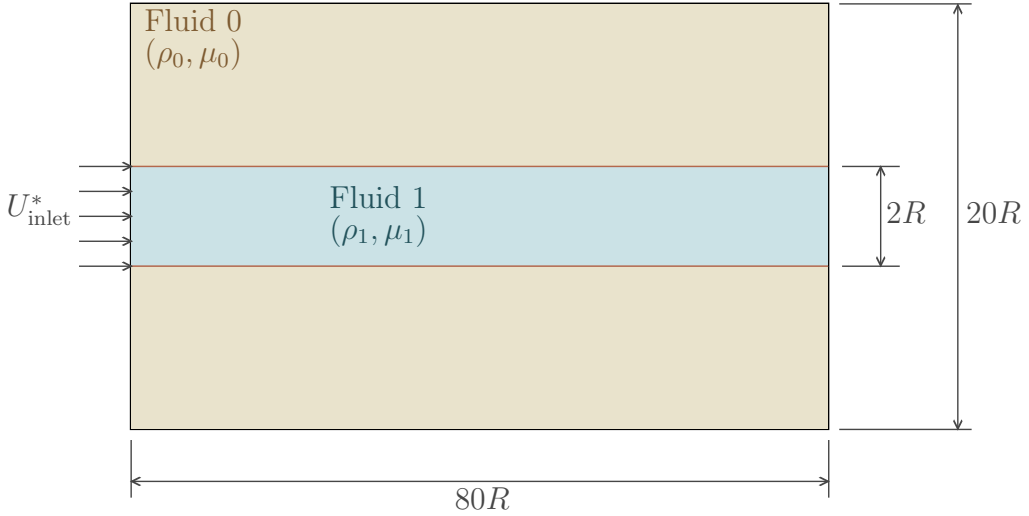}
    \caption{2D cross-section view in the $x$-$y$ plane of  the initial state of the Rayleigh-Plateau case, where Fluid 1 corresponds to the liquid jet}
    \label{fig::rayleigh-plateau-schematic}
\end{figure}

Figure \ref{fig::rayleigh-plateau-schematic} shows the schematic of the case. The domain is a rectangular prism of dimensions $80R \times20R \times20R$, where $R$ is the radius of the jet at the inlet. The jet develops along the $x$-axis of the domain.

Based on the work of \citet{Denner2022}, the jet undergoes a periodic perturbation through the $x$-component of inlet velocity $U^*_\mathrm{inlet}$, imposed at the left side of the domain, according to:
\begin{equation}
    U^*_\mathrm{inlet} = 
    \begin{cases}
        0 &\quad \forall\vm{x}\in \Omega_0\\
        U_\mathrm{inlet}\left(1+\delta_0\sin\left(\dfrac{k_\mathrm{jet} U_\mathrm{inlet}t}{R}\right)\right) &\quad \forall\vm{x}\in \Omega_1
    \end{cases}
\end{equation}
where $U_\mathrm{inlet}$ is the unperturbed inlet velocity, $\delta_0$ is the dimensionless excitation amplitude, and $k_\mathrm{jet} $ is the dimensionless wavenumber. The case considers $R=\SI{1.145e-3}{\meter}$, $\delta_0 = 0.3$, $k_\mathrm{jet} =0.7$, and the physical properties listed in Table \ref{tab::rayleigh-plateau-physical-properties}. It corresponds to an Ohnesorge number $\pp{\mathrm{Oh}=\mu_1/\sqrt{\gamma\rho_1 R}}$ of 0.1. The selected Weber number $\pp{\mathrm{We}=\rho_1RU_\mathrm{inlet}^2/\gamma}$ is 50, which results in $U_\mathrm{inlet}=\SI{1.569}{\meter\per\second}$. 

For the NS equations, the case imposes the Dirichlet boundary condition $\vm{u}=[U^*_\mathrm{inlet},0,0]^{\transpose{}}$ at the left boundary, and outlet conditions at the other boundaries. For the phase indicator, it imposes a Dirichlet condition at the left boundary according to:
\begin{align}
    \phi_\mathrm{jet}(\vm{x}) &= 0.5 - 0.5 \tanh\left( \frac{d_\mathrm{jet}(\vm{x})}{2\varepsilon}\right) \label{eq:inlet-condition}\\
    d_\mathrm{jet}(\vm{x}) &= \sqrt{(y-y_0)^2+(z-z_0)^2} - R \label{eq:inlet-condition-d}
\end{align}
where $d_\mathrm{jet}$ is the signed distance from the unperturbed jet interface and $\vm{x_0}=\pp{x_0,y_0,z_0} = \pp{0,0,0}$ is the center of the jet at the left boundary. It considers no-flux conditions at the remaining boundaries. 

The initial condition imposes a uniform velocity of $\vm{u}_0 = [U_\mathrm{intlet},0,0]^{\transpose{}}$ in the jet, and null elsewhere. For the phase indicator, it follows Equations \eqref{eq:inlet-condition} and \eqref{eq:inlet-condition-d}. The simulation time is $\SI{0.08}{\second}$.

\begin{table}[!htpb]
    \centering
    \caption{Physical properties of the Rayleigh-Plateau instability case}
    \begin{tabular}{lcc}
        \hline 
        \textbf{Parameters}& {\textbf{Units}}& \textbf{Values} \\ \hline
        {Density ratio ($\rho_\mathrm{0}/\rho_\mathrm{1}$)} & {-} & {$\SI{1e-3}{}$} \\
        
        Density of fluid 0 ($\rho_0$) & $\SI{}{\kilo\gram\per\cubic\meter}$&  1.196 \\

        {Dynamic viscosity ratio ($\mu_\mathrm{0}/\mu_\mathrm{1}$)} & {-} & {$\SI{1e-2}{}$} \\
        Dynamic viscosity of fluid 0 ($\mu_0$) & $\SI{}{\kilo\gram\per\meter\per\second}$&  $\SI{3.038e-4}{}$ \\

        ST coefficient ($\gamma$) & $\SI{}{\kilo\gram\per\square\second}$ & $\SI{6.740e-2}{}$\\ \hline
    \end{tabular}
    \label{tab::rayleigh-plateau-physical-properties}
\end{table}

\subsubsection{Metric of Interest}
The surface energy $E_\Gamma$ of the jet drives the development of the instability~\cite{deGennes2005}:
\begin{equation}
    E_\Gamma = \gamma A_\Gamma \label{eq::surface-energy}
\end{equation}
where $A_\Gamma$ is the surface area of the jet. Hence, the metric of interest is the relative surface area of the jet $A_\Gamma/A_{\Gamma,0}$, with $A_{\Gamma,0}$ corresponding to the surface area of the jet at $t=\SI{0}{\second}$. The surface area computation follows the method described in Section \ref{sec:rising-bubble-metrics}. This case also monitors the relative volume evolution, computed according to Equation \eqref{eq:volume}.

\subsubsection{Simulation Settings}
To assess the spatial convergence, this study monitors $A_\Gamma/A_{\Gamma,0}$ and $V/V_0$ for four adaptively refined Cartesian meshes, reported in Table \ref{tab::rayleigh-plateau-refinements}.  It considers a constant CFL of $0.25$ for all meshes, which also ensures the respect of the capillary time-step limit  (Equation \eqref{eq::capillary-time-step}). 

The study restricts itself to only one reinitialization frequency of ${f=1/(10\Delta t)}$. This choice aims to limit the growth of parasitic capillary waves, described by \citet{Denner2017}, without introducing additional artificial diffusion or limiters.  Additionally, as denoted at the bottom of Table \ref{tab::params}, this study deactivates the approximation $\hat{H}_\Gamma$ described by Equation \eqref{eq:heaviside} for the PDE-based method to report better volume conservation and spatial convergence. It considers instead $\hat{H}_\Gamma(\vm{x}) = \phi(\vm{x})$.

\begin{table}[h]
    \centering
    \caption{Refinement levels for the Rayleigh-Plateau instability convergence study. ${R=\SI{1.145e-3}{\meter}}$. The approximate number of cells is reported for the initial jet and changes for the PDE-based and geometric methods due to difference values of $\varepsilon$ ($h$ and $3h$, respectively).}
    \footnotesize
    \begin{tabular}{P{45pt}C{65pt}C{65pt}C{45pt}C{45pt}}
        \hline 
        \textbf{Ref. level}& {\textbf{Approx. no. cells (PDE)}}& {\textbf{Approx. no. cells (Geo.)}}&  \textbf{Max. cell size} & \textbf{Min. cell size}\\ \hline
        Coarse&\SI{3.6e5}{}&\SI{7.1e5}{}& $1.25R$&$0.16R$\\
        Medium&\SI{1.5e6}{}&\SI{2.5e6}{}& $0.63R$&$0.08R$\\
        Fine&\SI{6.4e6}{}&\SI{1.1e7}{}& $0.31R$&$0.04R$\\
        Extra-fine&\SI{3.0e7}{}&\SI{4.6e7}{}& $0.16R$&$0.02R$\\\hline
    \end{tabular}
    \label{tab::rayleigh-plateau-refinements}
\end{table}

\subsubsection{Results}

\begin{figure}
        \begin{subfigure}[b]{1\textwidth}
            \centering
            \includegraphics[width=1.
            \textwidth]{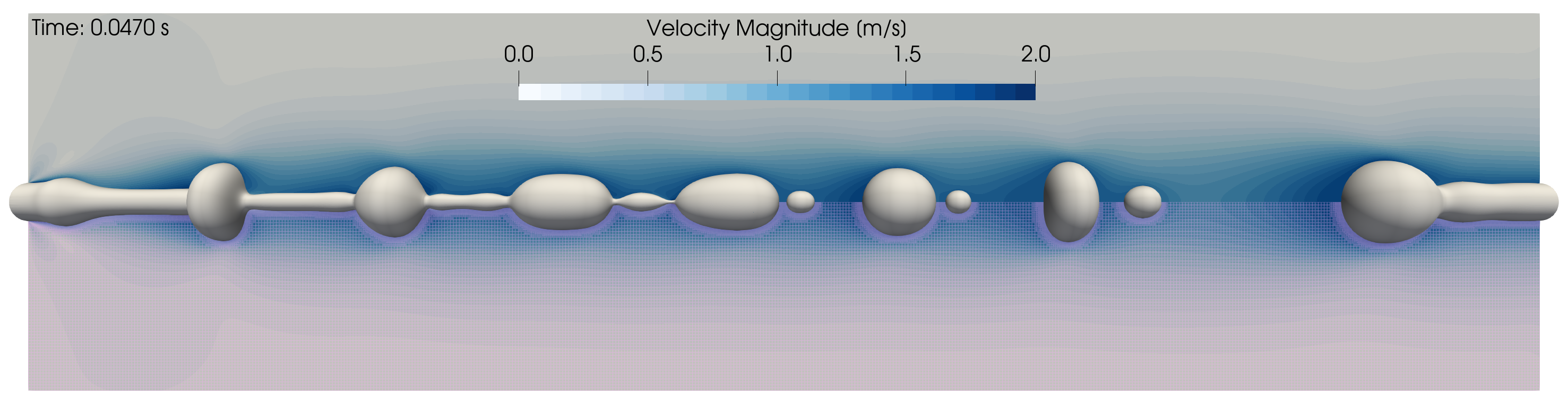}
            \caption{Iso-surface $\phi=0.5$ representing the interface}
            \label{fig:rayleigh-plateau-snapshot-isosurface}
        \end{subfigure}
        \begin{subfigure}[b]{1\textwidth}
            \centering
            \begin{tikzpicture}
                \node[inner sep=0pt] at (0,0) {\includegraphics[width=1.\textwidth]{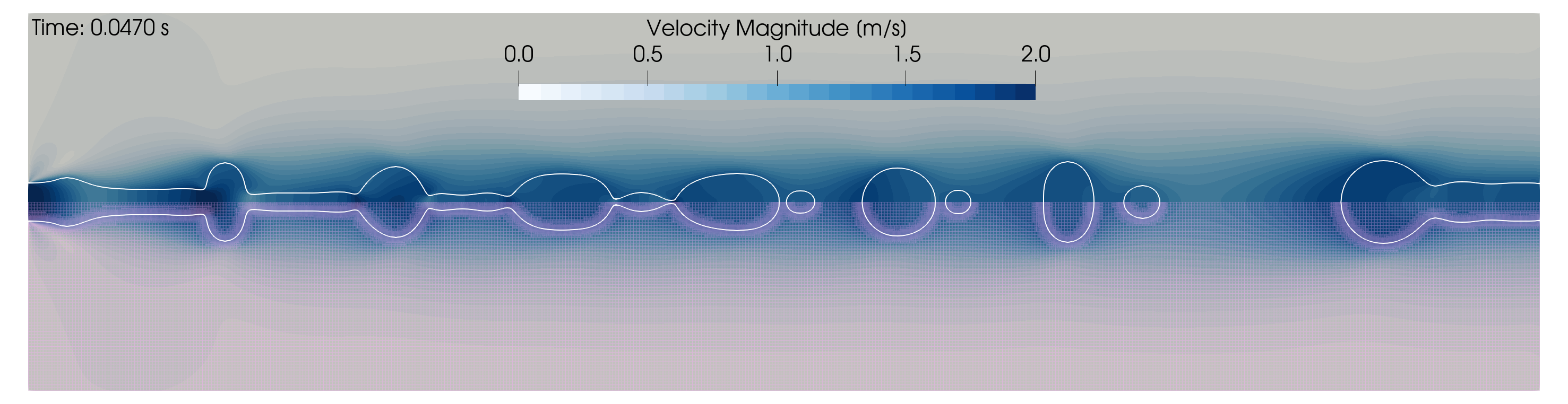}};
                \node[inner sep=0pt] at (0,-4) {\includegraphics[width=0.96\textwidth]{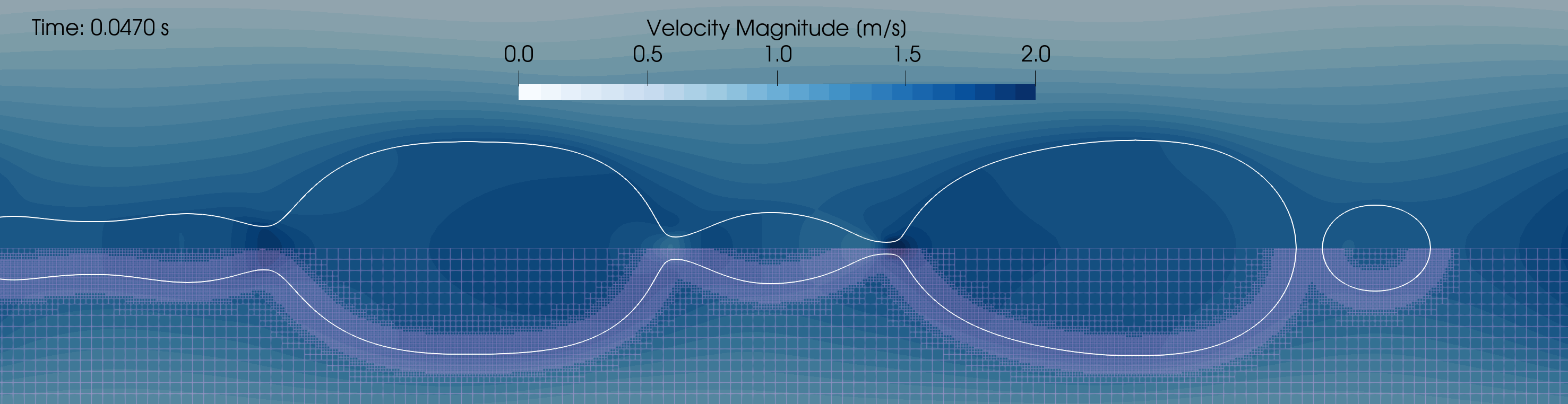}};
                \draw (-3.0,-0.5) -- (-6.55,-2.3);
                \draw (0.5,-0.5) -- (6.55,-2.3);
                \draw (-3.0,-0.5) rectangle (0.5,0.5);
            \end{tikzpicture}
            \caption{Zoom on the breaking}
            \label{fig:rayleigh-plateau-snapshot-zoom}
        \end{subfigure}
            
            
        \caption{Geometry of the jet at $t=\SI{0.0470}{\second}$ for the geometric method with the extra-fine mesh. The velocity field corresponds to the velocity in the $x$-$y$ plane. The adaptively refined mesh is shown on the bottom half of each panel in light pink.}
        \label{fig:rayleigh-plateau-snapshot}
        
    \end{figure}

\commentblue{Figure \ref{fig:rayleigh-plateau-snapshot} presents different snapshots of the jet at $t=\SI{0.0470}{\second}$, obtained with the geometric method on the extra-fine mesh. The bottom half of each snapshot shows the adaptively refined mesh. The top snapshot shows the 3D interface as well as the velocity field in the $x$-$y$ plane going through the center of the jet. It demonstrates the ability of the geometric method to handle complex flows featuring breaking and satellite droplets. The middle and bottom snapshots show the 2D contour of the interface in the same $x$-$y$ plane. They highlight the details of the necking of the jet just before breaking, which are well captured by the solver due to the adaptive mesh~refinement. }

\begin{figure}
    \centering
    \begin{subfigure}[b]{0.95\textwidth}
        \centering
        \includegraphics[width=0.35\textwidth]{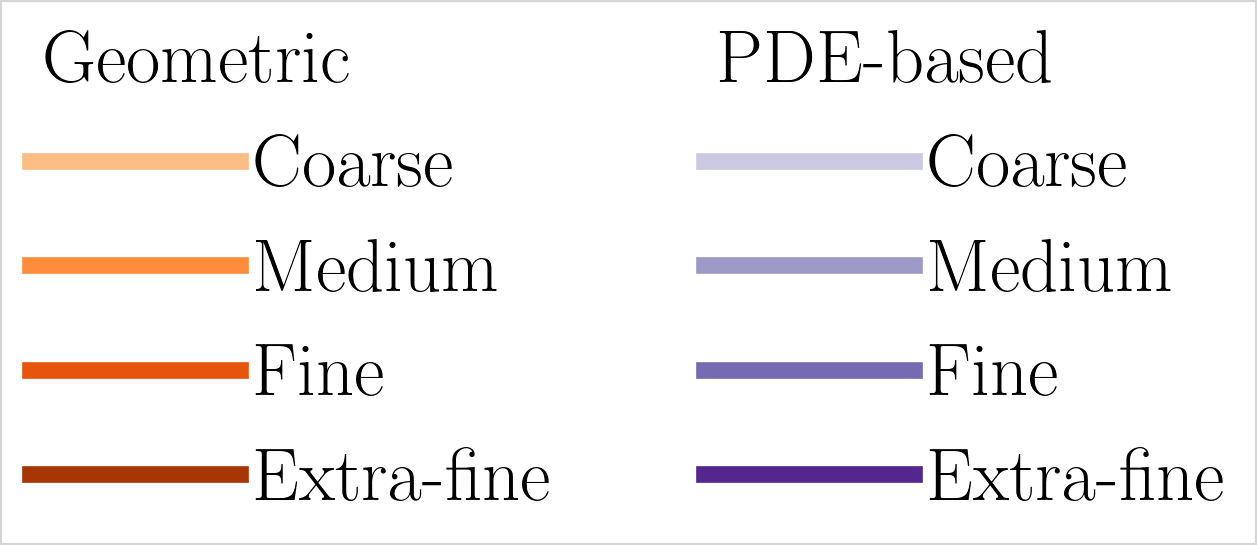}
    \end{subfigure}
    \begin{subfigure}[b]{1\textwidth}
        \centering
        \begin{minipage}{0.48\textwidth}
            \includegraphics[width=1\textwidth]{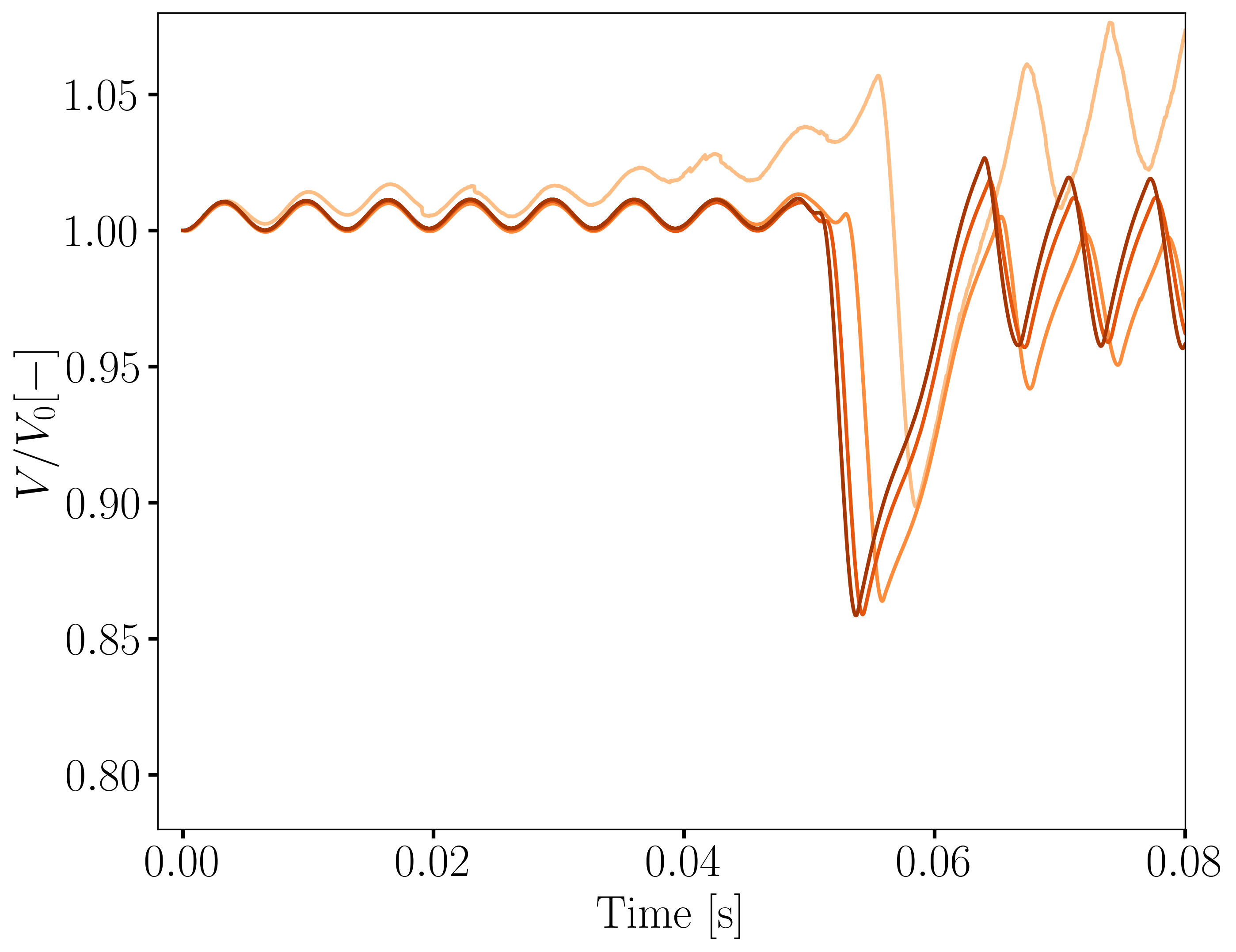}
        \end{minipage}
        \hfill
        \begin{minipage}{0.48\textwidth}
            \centering
            \includegraphics[width=1\textwidth]{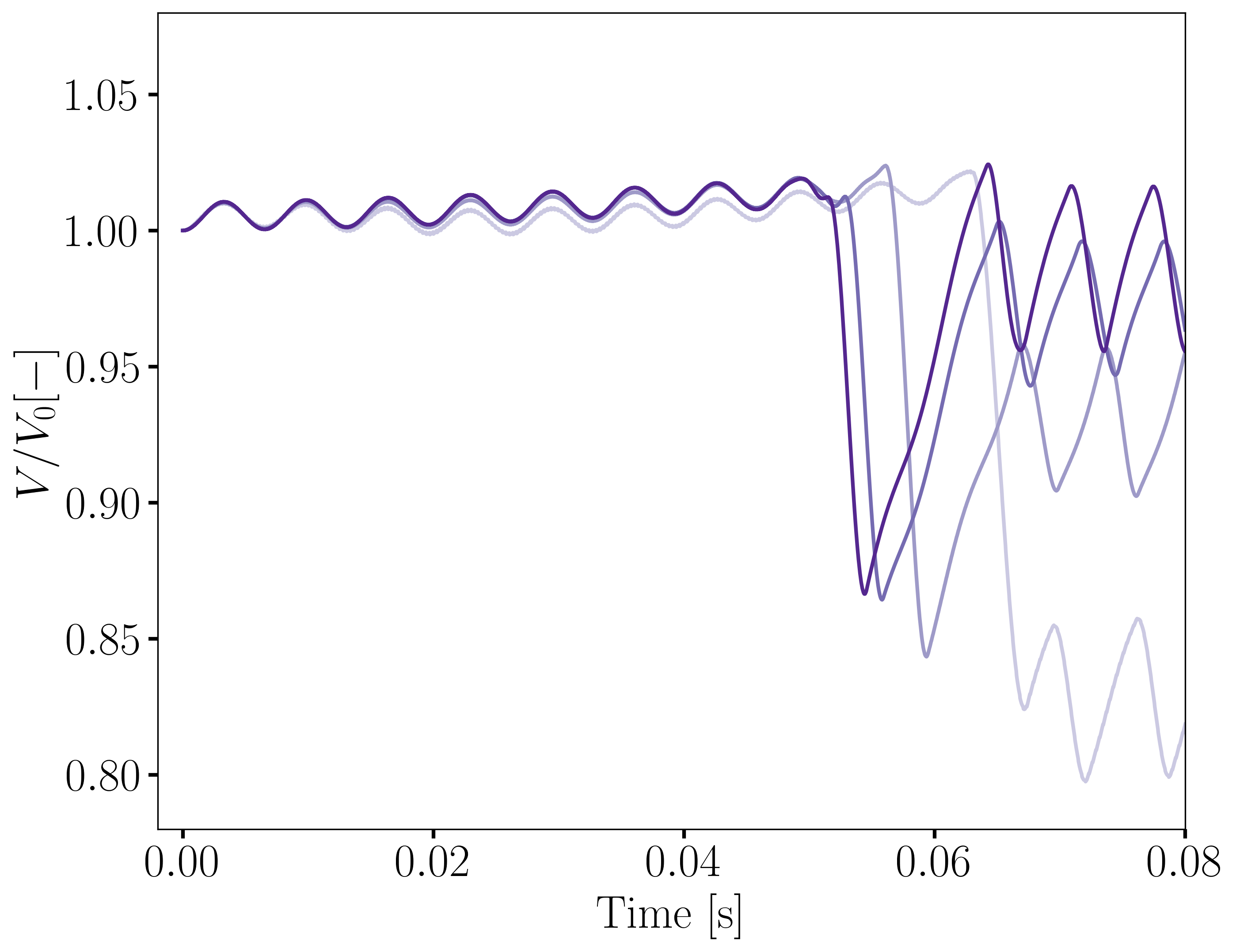}
        \end{minipage}
        \begin{minipage}{0.48\textwidth}
            \includegraphics[width=1\textwidth]{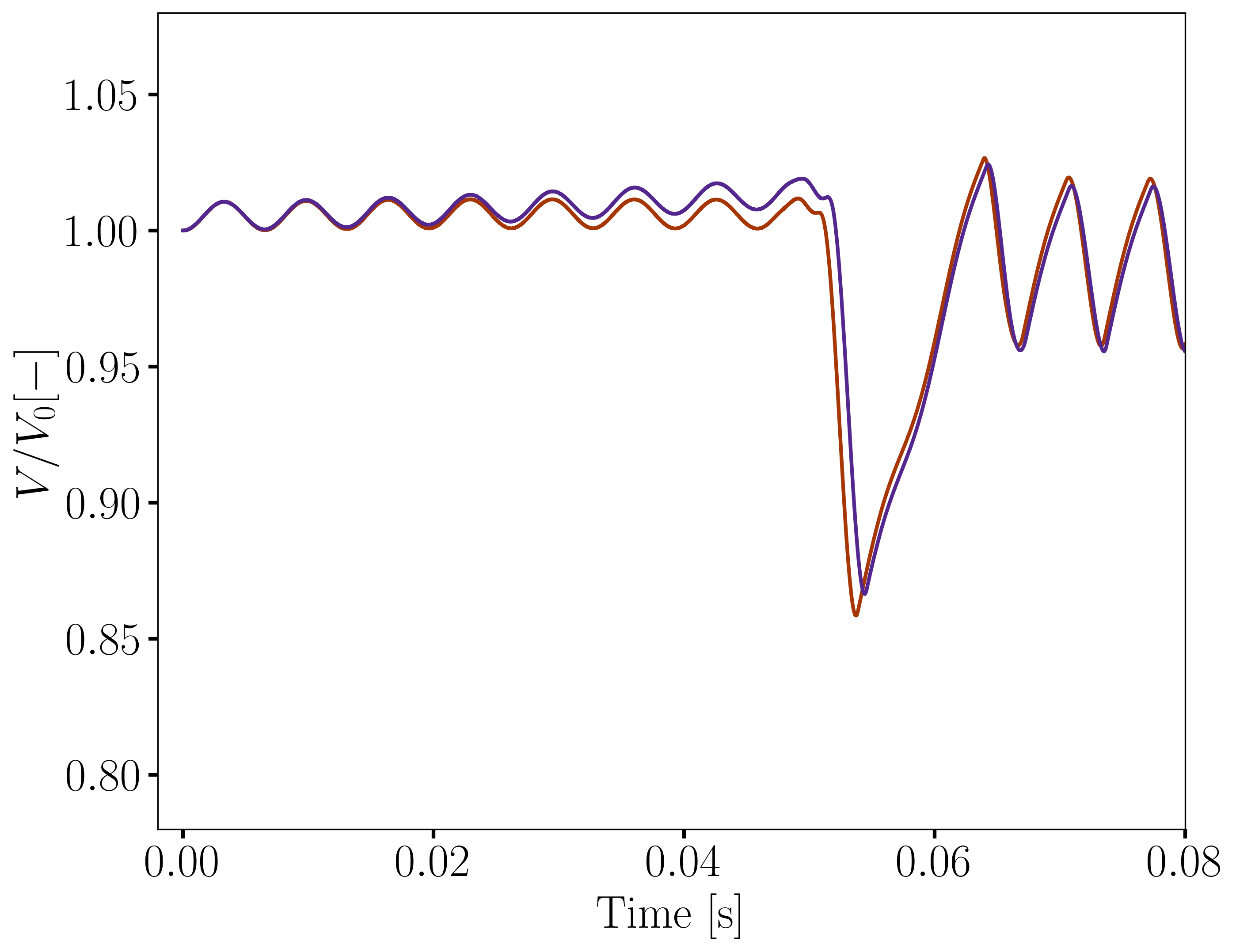}
        \end{minipage}
    \end{subfigure}
    \caption{Spatial convergence of the relative volume for the Rayleigh-Plateau case. On the top-left panel, the geometric reinitialization leads to the convergence of the relative volume to a value oscillating  around $1$ for $t<\SI{0.05}{\second}$. Then, for $t>\SI{0.05}{\second}$, the convergence is slower. On the top-right panel, the PDE-based method results in similar trends, with a relative volume oscillating slightly above 1 for $t<\SI{0.05}{\second}$ and a slower convergence rate than the geometric approach for $t>\SI{0.05}{\second}$. The bottom panel presents the comparison of the two methods results for the extra-fine meshes, highlighting that both methods converge toward similar solutions.}
    \label{fig::rayleigh-plateau-volume}
\end{figure}
\begin{figure}
    \centering
    \begin{subfigure}[b]{0.95\textwidth}
        \centering
        \includegraphics[width=0.35\textwidth]{Figures/rayleigh-plateau/legend.png}
    \end{subfigure}
    \begin{subfigure}[b]{1\textwidth}
        \centering
        \begin{minipage}{0.48\textwidth}
            \includegraphics[width=1\textwidth]{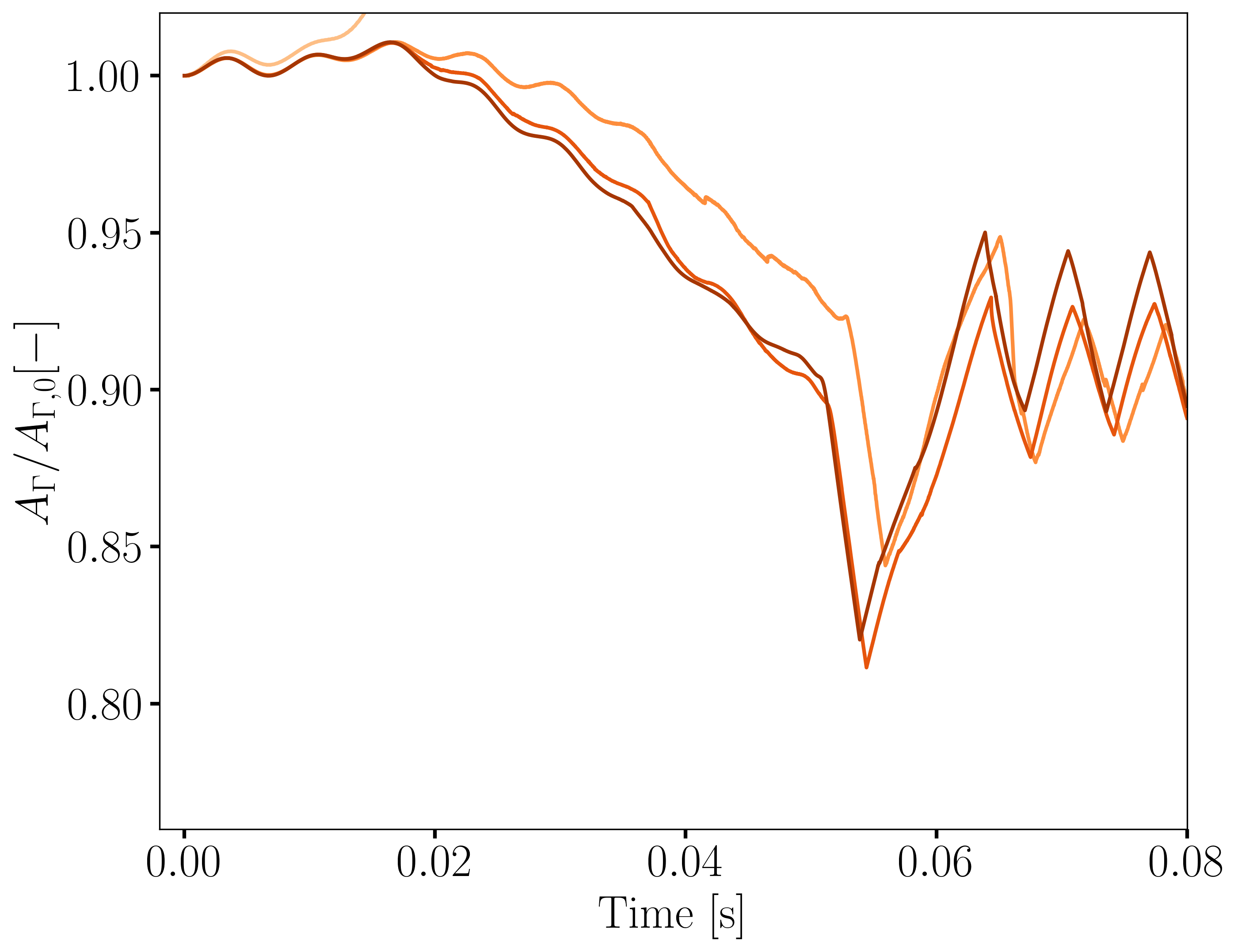}
        \end{minipage}
        \hfill
        \begin{minipage}{0.48\textwidth}
            \centering
            \includegraphics[width=1\textwidth]{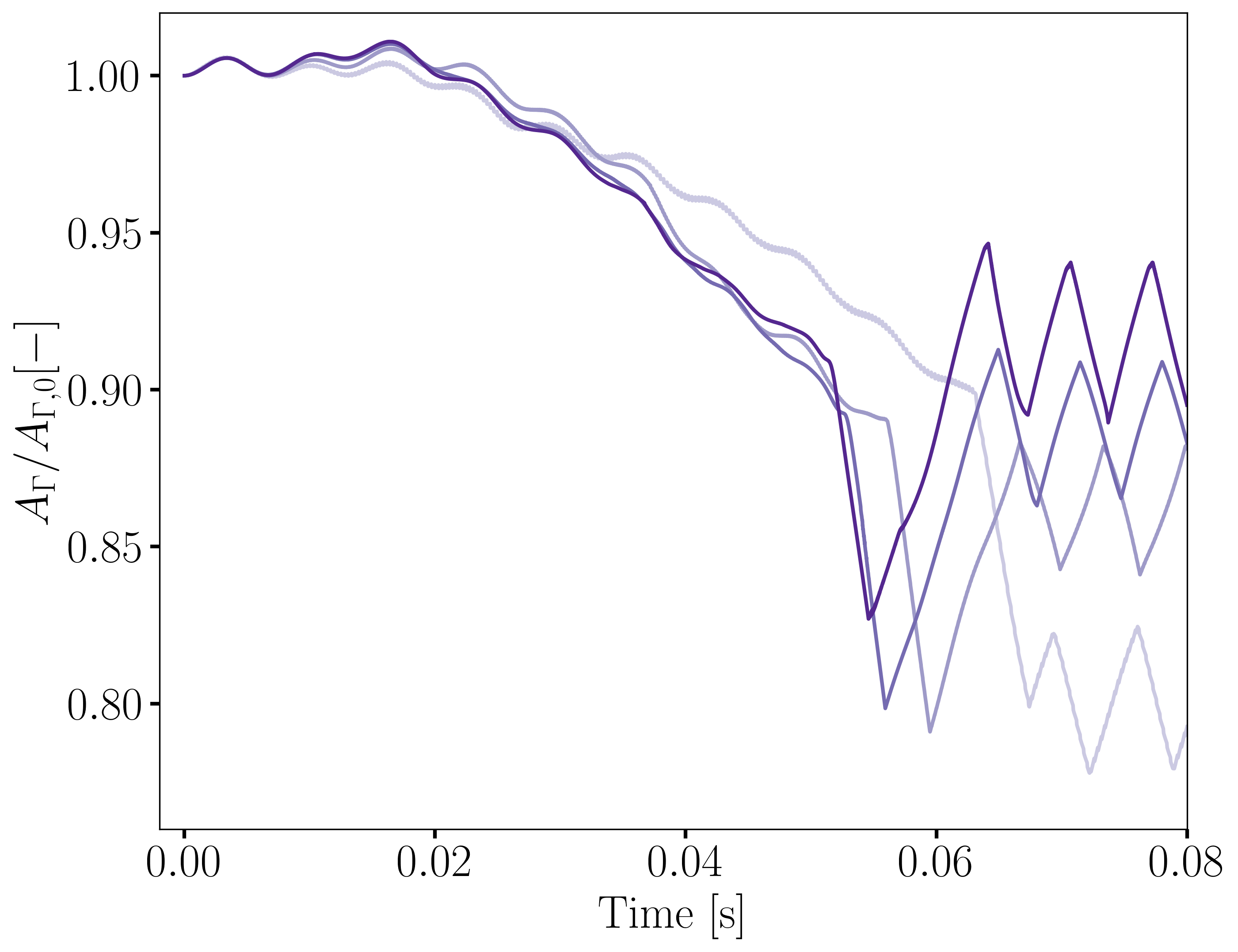}
        \end{minipage}
        \begin{minipage}{0.48\textwidth}
            \includegraphics[width=1\textwidth]{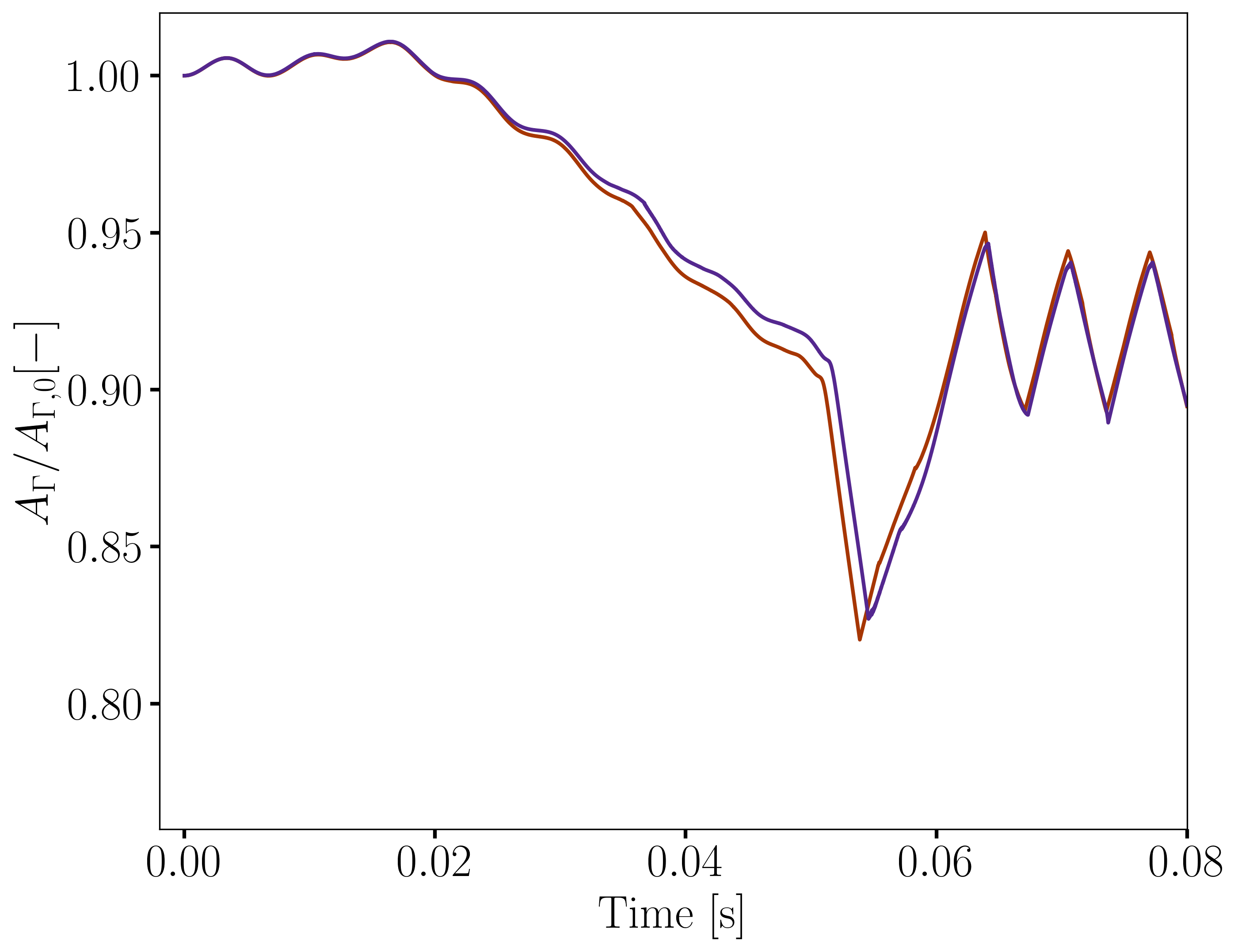}
        \end{minipage}
    \end{subfigure}
    \caption{Spatial convergence of the relative surface for the Rayleigh-Plateau case. On the top-left panel, the geometric reinitialization leads to the convergence of the relative surface, with a slower rate for $t>\SI{0.05}{\second}$. On the top-right panel, the PDE-based method results in similar trends, however the difference between the two finest meshes is larger for $t>\SI{0.05}{\second}$ than the geometric approach. The bottom panel presents the comparison of the two methods for the extra-fine meshes: they result in relative surface evolutions in great agreement.}
    \label{fig::rayleigh-plateau-surface}
\end{figure}

Figures \ref{fig::rayleigh-plateau-volume} and \ref{fig::rayleigh-plateau-surface} present the spatial convergence of the relative volume and surface, respectively. For both methods, the results reveal two regimes with a transition around $\SI{0.05}{\second}$. The first one presents small oscillations, corresponding to the imposed perturbations of the jet flow rate at the inlet. The second regimes features stronger and sharper oscillations, which corresponds to droplets exiting the computational domain.

For the geometric method, the relative volume converges in space to a periodic oscillation around $V/V_0 \approx 1$ in the first regime, which is the expected result if the solver conserves the volume. Its mesh convergence in the second regime is slower, indicating a slower convergence of the shape of the droplets generated by the breakup. Indeed, the relative surface evolution converges for the fine and extra-fine meshes until approximately $\SI{0.02}{\second}$. This time corresponds to the beginning of the necking of the jet leading to the first breakup. The slower mesh convergence following the breakup indicates that the jet and droplet shapes change with the spatial discretization when the Rayleigh-Plateau instability grows stronger, even if the volume does not change significantly. The results for the coarse mesh is significantly different due to propagation of parasitic capillary waves, and the plot leaves out part of the evolution for the sake of clarity. 

The relative volume obtained with PDE-based reinitialization tends to converge towards a similar solution as the geometric approach, but at a slower rate in the second regime. Additionally, it leads to a volume increase in the first regime, which is not observed with the geometric method. For the surface evolution, the difference between the fine and extra-fine meshes grows with time, also indicating that the method reaches convergence at a slower rate for the jet and droplet shapes than the relative volume.


\section{Conclusion}

This work develops and thoroughly assesses a geometric reinitialization for conservative level-set simulations of capillary flows. Building on the geometric approach of \citet{Ausas2011} developed in the context of classical level-set frameworks, the proposed method is an extension of the former to the conservative level-set method within an open-source finite element framework that allows for fully distributed and adaptively refined, two- and three-dimensional meshes. This extension enables the application of the geometric reinitialization method to large-scale three-dimensional problems with strongly deforming interfaces, including breakup, which were outside the scope of the original work of \citet{Ausas2011}.

The analysis assesses the geometric reinitialization on three 3D capillary-flow problems of increasing complexity, comprising Marangoni- and buoyancy-driven flows as well as the development of a capillary instability leading to interface breakup. Across these cases, the reinitialization method robustly and accurately recovers expected interface topology-based metrics (e.g., the volume, surface area and sphericity), and relevant flow quantities. The analysis of sensitive geometric metrics further demonstrates consistent conservation of the enclosed volume and accurate representation of the interface geometry, even under severe deformation.

Comparison to the well established PDE-based reinitialization method of \citet{Olsson2007} demonstrates that the geometric approach provides a robust alternative which does not restrict the interface thickness to small values, nor significantly alters the interface to conserve the volume. By tackling these challenges, the computations of the normal and curvature are carried out using the smooth and regular phase indicator field; they do not require any additional complex algorithms. It is an advantage over the more recent extensions of the PDE-based method such as the ones proposed by \citet{McCaslin2014}, \citet{Chiodi2017}, and \citet{Janodet2022}. Importantly, the geometric method has only two parameters, $\varepsilon$ and $d_\mathrm{max}$, and is robust across parameter choices and cases. This reduced parameter sensitivity constitutes an important advantage for predictive simulations in which suitable reinitialization parameters cannot be calibrated a priori.

The study of the reinitialization frequency shows that reinitialization at every time step yields the best volume conservation for both the geometric and PDE-based approaches, whereas the investigated flow-based quantities exhibit only weak sensitivity to this parameter. Consequently, reducing the reinitialization frequency provides a suitable option for significantly improving computational performances when small losses/gains in volume are acceptable.

Future work will focus on applying the geometric method to complex multiphysics flows such as the laser powder bed fusion process, which includes at the same time static regions of the powder bed and extreme interface deformations, driven by capillary- and evaporation-induced pressure jumps, as well as strong Marangoni stresses, making the choice of reinitialization method critical for accurately capturing the underlying physics.

\section{Acknowledgment}
The authors acknowledge the technical support and computing time provided by the Digital Research Alliance of Canada. HPL and AA acknowledge financial support from the Natural Sciences and Engineering Research Council of Canada (NSERC) and the Fonds de recherche du Québec – Nature et technologies (FRQNT). MSF acknowledges financial support by the European Research Council through the ERC Starting Grant ExcelAM under award number 101117579. BB acknowledges financial support from the Natural Sciences and Engineering Research Council of Canada
(NSERC) through the RGPIN-2020-04510 Discovery Grant and the funding from the Multiphysics Multiphase Intensification Automatization Workbench (MMIAOW) Canadian Research Chair Level 2 in computer-assisted design and scale-up of alternative energy vectors for sustainable chemical processes (CRC-2022-00340).

\newpage


\bibliographystyle{elsarticle-num-names}
\bibliography{references}






\end{document}